\documentclass[useAMS,usenatbib]{mnras}

\usepackage{newtxtext,newtxmath}

\usepackage[T1]{fontenc}
\usepackage{ae,aecompl}
\usepackage[draft]{graphicx}   
\usepackage{amsmath}    
\usepackage{amssymb}    
\usepackage{blkarray}
\usepackage{bigstrut}
\usepackage{layouts}
\usepackage[math]{cellspace}
\usepackage{breqn}
\usepackage{bm}
\usepackage{xpatch}
\usepackage{setspace}
\usepackage{slashed}
\usepackage{subcaption}
\usepackage{mathtools}

\usepackage[noend]{algpseudocode}

\newcommand{\vectb}[1]{\boldsymbol{#1}}
\newcommand{\mat}[1]{{\mathbf{#1}}}
\newcommand{\G}{\mat{G}}
\newcommand{\GH}{\mat{G}^{H}}
\newcommand{\B}{\mat{B}}

\newcommand{\E}{\mat{E}}
\newcommand{\EH}{\mat{E}^{H}}
\newcommand{\K}{\mat{K}}
\newcommand{\KH}{\mat{K}^{H}}
\newcommand{\V}{\mat{V}}

\newcommand{\MM}{\mat{M}}

\newcommand{\Br}{\mat{r}}
\newcommand{\Beps}{\bm{\epsilon}}
\newcommand{\CM}{CM}
\newcommand{\CU}{CU}
\newcommand{\Csky}{CV}
\newcommand{\Cnoise}{C$\sigma$}
\newcommand{\DMA}{DMA}
\newcommand{\DUA}{DUA}
\renewcommand{\P}{\mathcal{P}}
\newcommand{\g}{\bm{g}}
\newcommand{\C}{\bm{C}}
\newcommand{\thetab}{\bm{\theta}}
\DeclarePairedDelimiter\ceil{\lceil}{\rceil}

\makeatletter
{\itemsep\z@}{\itemsep=2ex plus2pt}{}{} 
\makeatother

\title[Solution intervals considered harmful]{Solution intervals considered harmful: on the optimality of radio interferometric gain solutions}

\author[Sob et al.]{
U. M. Sob$^{1}$\thanks{E-mail: mulricharmel@gmail.com (UMS)},
H. L. Bester$^{2,1}$,
O. M. Smirnov$^{1,2}$,
J. S. Kenyon$^{1,2}$,
C. Russeeawon$^{1}$
\\
$^{1}$Department of Physics and Electronics, Rhodes University, Grahamstown, 6140, South Africa\\
$^{2}$SARAO, 2 Fir Street, Black River Park, Observatory, 7925, South Africa
}

\date{Accepted XXX. Received YYY; in original form ZZZ}

\pubyear{2020}

\begin{document}
\label{firstpage}
\pagerange{\pageref{firstpage}--\pageref{lastpage}}
\maketitle

\begin{abstract}
Solution intervals are often used to improve the signal-to-noise ratio during radio interferometric gain calibration.
This work investigates how factors such as the noise level, intrinsic gain variability, degree of model incompleteness, and the presence of radio frequency interference impact the selection of solution intervals for calibration. 
We perform different interferometric simulations to demonstrate how these factors, in combination with the choice of solution intervals, affect calibration and imaging outputs and discuss practical guidelines for choosing optimal solution intervals. 
Furthermore, we present an algorithm capable of automatically selecting suitable solution intervals during calibration. 
By applying the algorithm to both simulated and real data, we show that it can successfully choose solution intervals that strike a good balance between capturing intrinsic gain variability and not fitting noise as long as the data are not too inhomogeneously flagged.
Furthermore, we elaborate on several practical aspects that emphasize the need to develop regularised calibration algorithms that do not require solution intervals. 
\end{abstract}




\section{Introduction}\label{intro}
Radio interferometry (RI) attempts to enhance the sensitivity and angular resolution of radio images by correlating measurements from multiple radio telescopes. 
The measured radio interferometric data are generally corrupted by various instrumental and atmospheric effects that have to be removed through a process known as radio interferometric gain calibration \citep{Hamaker1996,RIME1} or simply calibration. 
Calibration is usually performed in stages known as first, second and third generation calibration, abbreviated as 1GC, 2GC and 3GC, respectively \citep{noordam2010meqtrees}. 
Since neither the sky nor the propagation errors (also called antenna gains) are known prior to observing some astrophysical field of interest (viz. the target field), target scans are usually interspersed with calibrator scans i.e. high signal-to-noise ratio (SNR) observations of a well modelled object or part of the sky. 
During 1GC, the known properties of the calibrator fields are used to determine the dominant factors influencing the antenna gains.
These are then transferred to the target field by, for example, interpolating between calibrator scans and correcting the data by applying the inverse of the interpolated gains (sometimes referred to as calibrator transfer). 
The corrected data is the starting point for self-calibration.

Self-calibration (see \citet{cornwell1981wilkinson} for example), or 2GC, is an iterative process which alternates between calibration and imaging to successively refine the model of the target field. 
Since the model is not known a priori, there is always a risk of absorbing certain features of the model into the calibration solutions during calibration (or absorbing calibration artefacts into the model during imaging).
The risk increases with decreasing SNR per degree of freedom (DoF), where the number of DoFs involved during calibration is determined by the gain parametrisation.
A common way to try and decrease the number of DoFs for the problem is to use the notion of time and frequency solution intervals during which the gains are assumed to be constant.

Solution intervals essentially parametrise gains as a sum of boxcar functions in time and frequency (see Appendix \ref{sec:appendixC}). 
Since this decreases the number of DoFs, it reduces the risk of biasing the model during calibration.
However, it can also result in significant errors in the gain estimates if the underlying gains vary substantially within a solution interval.
Thus, an optimal solution interval minimises the number of DoFs, while still allowing for the intrinsic variability of the gains.
As far as we are aware, an automatic and statistically rigorous way to determine optimal solution intervals does not exist.
Instead, they are often set empirically by exploiting expertise and intuition gained from experience, possibly requiring the self-calibration procedure to be repeated multiple times. 
This is already impractical for precursor telescopes such as MeerKAT \citep{jonas2018meerkat} and will become increasingly so as data rates increase with the next generation of radio telescopes (such as the Square Kilometre Array (SKA) \citet{schilizzi2008square}, for example). 
Thus, it becomes increasingly important to automate as much of the data reduction as possible. 

The problem is exacerbated by the fact that certain key science targets can be extremely faint compared to the foregrounds, necessitating high dynamic range\footnote{Dynamic range is defined as the ratio of the brightest feature to the faintest feature in the image distinguishable from noise.} imaging.
When combined with the characteristic wide fields of view of modern interferometers, this means that direction dependent (DD) self-calibration \citep{RIME3} is sometimes required. 
DD calibration not only increases the number of DoFs but also adds another layer of complexity viz. both signal and intrinsic gain variability can be DD.
Thus, ideally, different directions require different solution intervals. 
A similar problem is encountered when data are very inhomogeneously flagged i.e. when the fraction of flagged data changes significantly as a function of time, frequency or antenna. 
In extreme cases, there is no single choice of solution interval that can adequately capture intrinsic gain variations while allowing for sufficient SNR.

This paper aims to determine which factors are relevant for determining optimal solution intervals and provide a practical way for selecting such intervals when possible. We will also highlight some of the inherent limitations of the solution interval approach and motivate the need for developing more sophisticated regularised approaches to the problem. 
The rest of the paper is organised as follows. In $\S$\ref{cal_prob}, we give a brief overview of radio interferometric calibration and discuss the concept of solution intervals used for gain parametrisation by most calibration packages. This parametrisation is further analysed in $\S$\ref{sol_int} where we discuss the main factors which influence the choice of optimal solution intervals. In $\S$\ref{sec:interval_search}, we describe the details of our optimal solution interval search algorithm. Next in $\S$\ref{multi_chan} and $\S$\ref{real_video_data}, we present successful results after applying the algorithm for direction independent (DI) calibration of both simulated and real data, respectively. In $\S$\ref{meerkatdata}, we discuss how to apply the algorithm in the case of DD calibration and discuss more practical aspects that make the search for optimal solution intervals difficult. We summarise the paper in $\S$\ref{conclusions} and discuss possible extensions to this research.

\emph{About the title:} This work started as a simple question, \emph{what is the statistically optimal gain solution interval?} Well-posed and self-contained, the problem promised to be no more than half a quiet afternoon's work for a crack Bayesian. Several (occasionally very frustrating) years later, we have come to appreciate that solution intervals are simple, powerful, practical, straightforward, and an absolute pain to analyze. These traits are reminiscent of a nowadays archaic, but once ubiquitous programming construct, a critique of which by \citet{dijkstra1968harmful}, \emph{Go To Statement Considered Harmful}, has become a computer science cliche. \\

\emph{Notation:} ${()}^T$, ${()}^*$ and $()^H$ denote the transpose, the complex conjugate and the complex conjugate transpose operators respectively. $\mathbf{I}$ denotes the identity matrix. $\mathrm{CN}$ is used to represent a circular complex Gaussian distribution. An over-tilde is used to denote the vectorisation of a $2\times 2$ matrix into a $4\times 1$ vector, and an over-hat denotes an estimated quantity (obtained as the result of an optimisation problem, for example)

\emph{A note on simulations:} The setups for all the simulations performed for this study are summarised in Table \ref{Table:sim_setups}. Each simulation is referred to as \textbf{Simulation-x} where \textbf{x} is the corresponding identifier in the table. All simulations use the software package MeqTrees \citep{noordam2010meqtrees} to compute and populate a measurement set with the visibilities for the simulation. For simplicity, we will mainly consider the DI model given by Eq. \eqref{eq_rime_simple}. Antenna gains are simulated by drawing realisations from Gaussian processes (GP) with a fixed covariance function and using the hyper-parameters shown in Table \ref{Table:sim_setups}. The individual realisations are assumed to be independent across antennas (more details are given in Appendix~\ref{sec:appendixD}).

\section{Problem overview}\label{cal_prob}
This paper focuses on the importance of solution intervals and the practical difficulties encountered when choosing solution intervals during calibration. This section describes the concept of a solution interval as implemented by most calibration software packages, and examines some key factors to consider when choosing solution intervals.

\subsection{Calibration}\label{cal_nlls}
The radio interferometry measurement equation (RIME) \citep{Hamaker1996,RIME1} describes the transformations a signal undergoes as it travels from its source to our receivers here on earth. Using a generalised RIME model, the corrupted model visibility $\MM_{pq}$ for an antenna pair $pq$ is related to the sky brightness matrix $\B$ via (assuming a discrete source distribution), 
\begin{equation}
\MM_{pq} \,=\, \G_{p} \,\left( \sum_{d}\E_{dp}\K_{dp}\B_{pqd}\KH_{dq}\EH_{dq}\right) \, \GH_{q}\, , 
\label{rime_1}
\end{equation}
where $d$ labels direction, $\G_{p}$ is the DI gain of antenna $p$, $\E_{dp}$ denotes a DD gain (e.g. antenna primary beam pattern) for antenna $p$ in direction $d$, and $\K_{dp}$ represents the geometric phase delay term. Measured visibilities $\V_{pq}$ are inevitably corrupted by noise and radio frequency interference (RFI). Thus, the measurement process takes the form
\begin{equation}
\V_{pq} \,=\, \MM_{pq}\, + \, \Beps_{pq} \,+\, \text{RFI}\, , 
\label{rime_2}
\end{equation} 
where the noise, $\bm{\epsilon}_{pq}$, is assumed to follow a circular complex Gaussian distribution. In general, all the above quantities can be $2\times 2$ complex valued matrices. 
The vectorised noise can be described as
\begin{equation}
    \tilde{\Beps}_{pq}  \sim  \mathrm{CN}(0,\Sigma_{pq})
\end{equation}
where $\Sigma_{pq}$ is a $4\times 4$ diagonal noise covariance matrix. Hence, assuming that all RFI corrupted visibilities have been carefully flagged, maximum likelihood calibration aims at minimising the negative log-likelihood
\begin{equation}
\min_{\vectb{\theta}} \sum_{pq | p < q} \tilde{\Br}_{pq}^H \Sigma_{pq}^{-1} \tilde{\Br}_{pq} \quad \mbox{with} \quad \tilde{\Br}_{pq} = \tilde{\V}_{pq}-\tilde{\MM}_{pq},
\label{eq_rime_2}
\end{equation} 
where the gains are parametrised by $\thetab$. This problem can be solved efficiently as a non-linear least squares (NLLS) problem. In what follows, we use the complex formulation of \cite{smirnov2015radio} which has been implemented in the CubiCal\footnote{\url{https://github.com/ratt-ru/CubiCal}} package. In particular, we will use the complex $2\times 2$ \citep{kenyon2018cubical} and the robust $2\times 2$ solvers \citep{sob2019robust}. As discussed in $\S$\ref{sec:model completeness}, the latter is important when dealing with residual RFI and partially complete sky models.

Sometimes, for example when dealing with the DI calibration problem, it will be convenient to denote the summation term of Eq. \eqref{rime_1} as $\C_{pq}$. In this case, the measurement process can be written succinctly as
\begin{equation}
\V_{pq} = \G_p \C_{pq} \G_q^H + \Beps_{pq} + \text{RFI}.
\label{eq_rime_simple}
\end{equation} 
When there are no DD effects present (or we are ignoring them), the quantity $\C_{pq}$ just contains the model visibilities corresponding to the current sky model used for calibration. 

\subsection{Solution Intervals} \label{gain_parametrization}
Naively solving for an independent gain at each time and frequency usually results in overfitting, as the algorithm will tend to fit noise as well as signal, especially at low SNR.  Solution intervals attempt to reduce the extent of overfitting by imposing time and frequency intervals over which the gains should be constant. In practice, this is often achieved by factoring the effective gain into multiple Jones terms. For example, in a typical 1GC procedure, one would solve for a {\em bandpass} term that has a high frequency resolution but is constant in time, and a proper gain term that has high time resolution, but is constant in frequency. 

In this work, we consider the general case of a single gain term that is assumed constant over some finite time and frequency intervals.
This corresponds to having multiple measurements for each baseline across the chosen intervals. The optimisation problem, Eq. \eqref{eq_rime_2}, is then slightly modified as follows
\begin{equation}
\min_{\vectb{\theta}} \sum_{pqs| p < q} \tilde{\Br}_{pqs}^H \Sigma_{pqs}^{-1} \tilde{\Br}_{pqs} ,
\label{eq_rime_2s}
\end{equation}
where $s$ is an index for all the time and frequency samples that are part of the same solution interval. If the full domain of the problem is a $N_t \times N_\nu$ time/frequency grid, the solution interval approach parametrises the gains as a sum of boxcar functions, i.e.
\begin{equation}
\G_{p}(t, \nu, \thetab_p) = \thetab_{p, ij}\bm{\sqcap}_{t_i, t_{i+\Delta_t}}(t) \bm{\sqcap}_{\nu_j, \nu_{j+\Delta_\nu}}(\nu),
\label{sol_int_param0}
\end{equation} 
where $\thetab_{p, ij}$ is a 2$\times $2 matrix containing the gain values for antenna $p$ in an interval labelled by $ij$, $\Delta_t$ is the resolution of the coarsened time grid, $\Delta_\nu$ is the resolution of the coarsened frequency grid and the boxcar function is defined as
\begin{equation}
\bm{\sqcap}_{a, b}(x) = \left\{ \begin{array}{ccc}
1 & \mbox{if} & x \in [a, b), \\
0 & ~~ & \mbox{otherwise}.
\end{array} \right.
\end{equation}
The units that discretise the full grid are the integration time, $\delta_t$, and the channel width, $\delta_\nu$. Each solution interval is therefore a multiple of $\delta_t$ and $\delta_\nu$. We denote this by $n_t$ and $n_\nu$\footnote{Throughout the paper, when employed without units, the term solution interval refers to the discrete units $n_t$ and $n_\nu$.}, such that
\begin{equation}
n_t = \frac{\Delta_t}{\delta_t} \quad \mbox{and} \quad n_\nu = \frac{\Delta_\nu}{\delta_\nu}.
\end{equation} 
Thus, if the full grid consists of $N = N_tN_\nu$ points, and each solution interval contains $n = n_tn_\nu$ such points, there will be a total of $M_t = \ceil*{\frac{N_t}{n_t}}$\footnote{$\ceil*{.}$ is the ceil function defined such that $\ceil*{x}$ is the smallest integer $\geq\; x$.} time intervals and $M_\nu = \ceil*{\frac{N_\nu}{n_\nu}}$ frequency intervals giving a total of $M = M_tM_\nu$ parameters for each gain (along each direction in the DD case). 

To illustrate the basic problem, let's assume we are calibrating a dataset with a very low SNR. If the chosen solution interval is too small, the data will consist mostly of noise and the algorithm will therefore fit noise instead of signal. As the size of the solution interval is increased, the noise from multiple data points tend to cancel out allowing the algorithm to start tracking the signal and the variance in the inferred gain solution will be reduced. However, as the size of the interval grows, we eventually lose the ability to track rapid gain variations resulting in a biased gain estimate. Thus, the error in the estimated gains has a contribution from two terms: one from the uncertainties in the data, and the second from the bias introduced as a result of approximating gains as a sum of boxcars. Using the definition of the mean squared error (MSE) of a parameter, $\vectb{\theta}$, and its estimate, $\hat{\vectb{\theta}}$, we have
\begin{eqnarray}
 \text{MSE} \, &=& \,  \mathop{\mathbb{E}}(\hat{\vectb{\theta}} \,-\, \vectb{\theta})^{2}, \label{mse1} \\
        &=&\, \mathop{\mathbb{E}}\left(\hat{\vectb{\theta}} - \mathop{\mathbb{E}}[\hat{\vectb{\theta}}]\right)^{2} \, +\, \left(\mathop{\mathbb{E}}[\hat{\vectb{\theta}}] - \vectb{\theta}\right)^{2},\label{mse3}\\
        &=& \, \mathrm{Var}(\hat{\vectb{\theta}})\, +\, \mathrm{Bias}(\hat{\vectb{\theta}}, \vectb{\theta})^{2},\label{mse2}
\end{eqnarray}
where $\mathop{\mathbb{E}}$ is the expectation operator with respect to the probability distribution governing $\hat{\vectb{\theta}}$. The first term of Eq. \eqref{mse2} is the variance of the estimated parameter and represents the error in the estimated parameter due to the uncertainties in the measured data. This corresponds to the error in the gains because of the noise in the visibilities, unmodelled sources, and RFI for the calibration problem. The second term is called the $\mathrm{Bias}$, and it results from the information lost when we approximate the gains as a sum of boxcars when using solution intervals. 

\textbf{Simulation-i} provides a general summary of the effects of solution intervals on calibration outputs. We simulate a field containing 100 point sources using a single channel MeerKAT array configuration. The sources have random positions, and their fluxes follow a power law distribution with a peak flux of 0.5 Jy. The simulated visibilities are corrupted with gains from a GP (see Table \ref{Table:sim_setups}). The data are repeatedly calibrated with different time intervals using both complete and incomplete sky models. 

\begin{figure*}
\captionsetup{font=footnotesize,labelfont=footnotesize}
\begin{subfigure}{0.48\textwidth}
\centering
  \includegraphics[width=8cm,height=6cm,keepaspectratio]{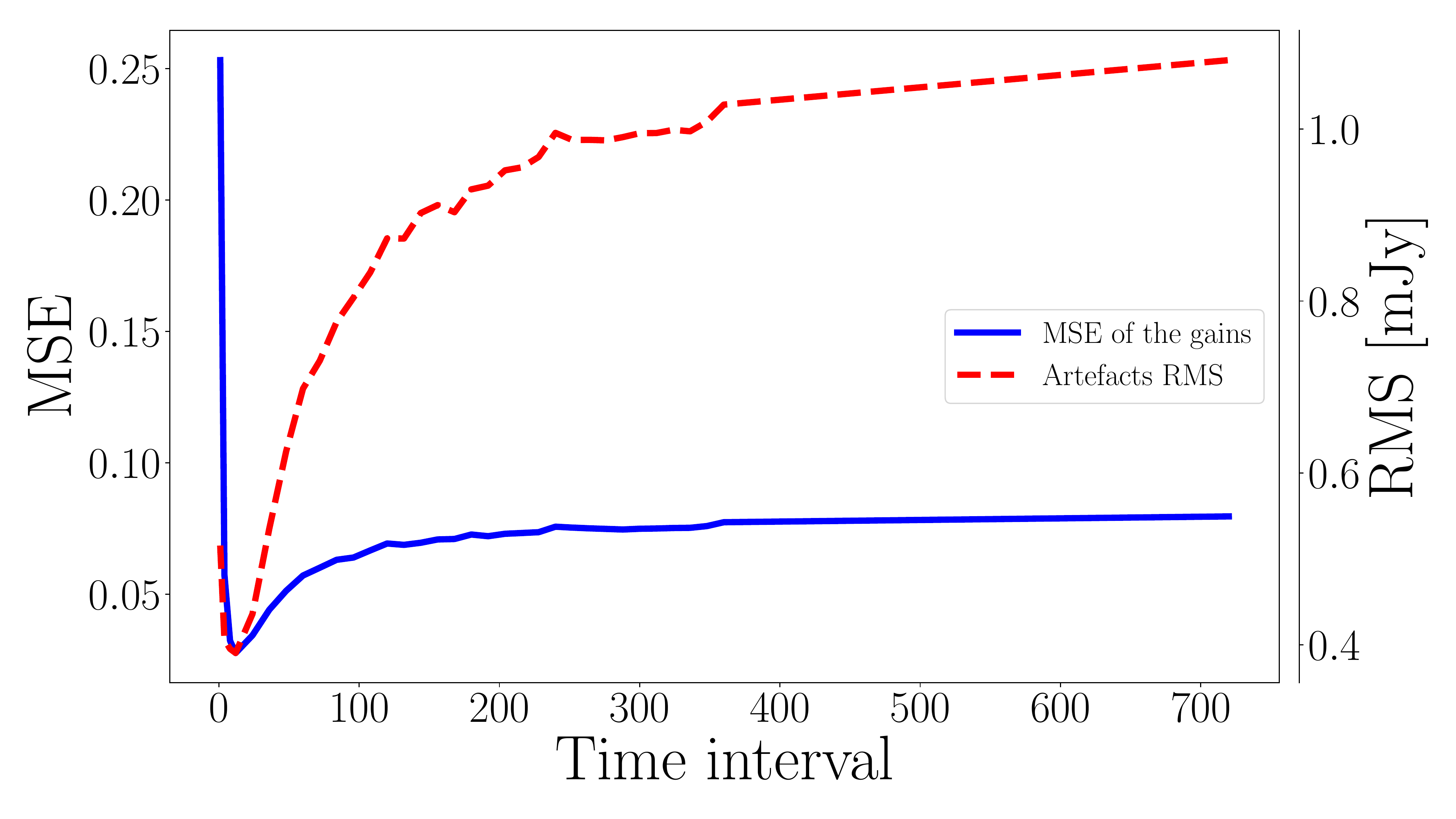} 
\caption{Complete sky model -- MSE}\label{complete_mse}
 \end{subfigure}
\begin{subfigure}{0.48\textwidth}
\centering
  \includegraphics[width=8cm,height=6cm,keepaspectratio]{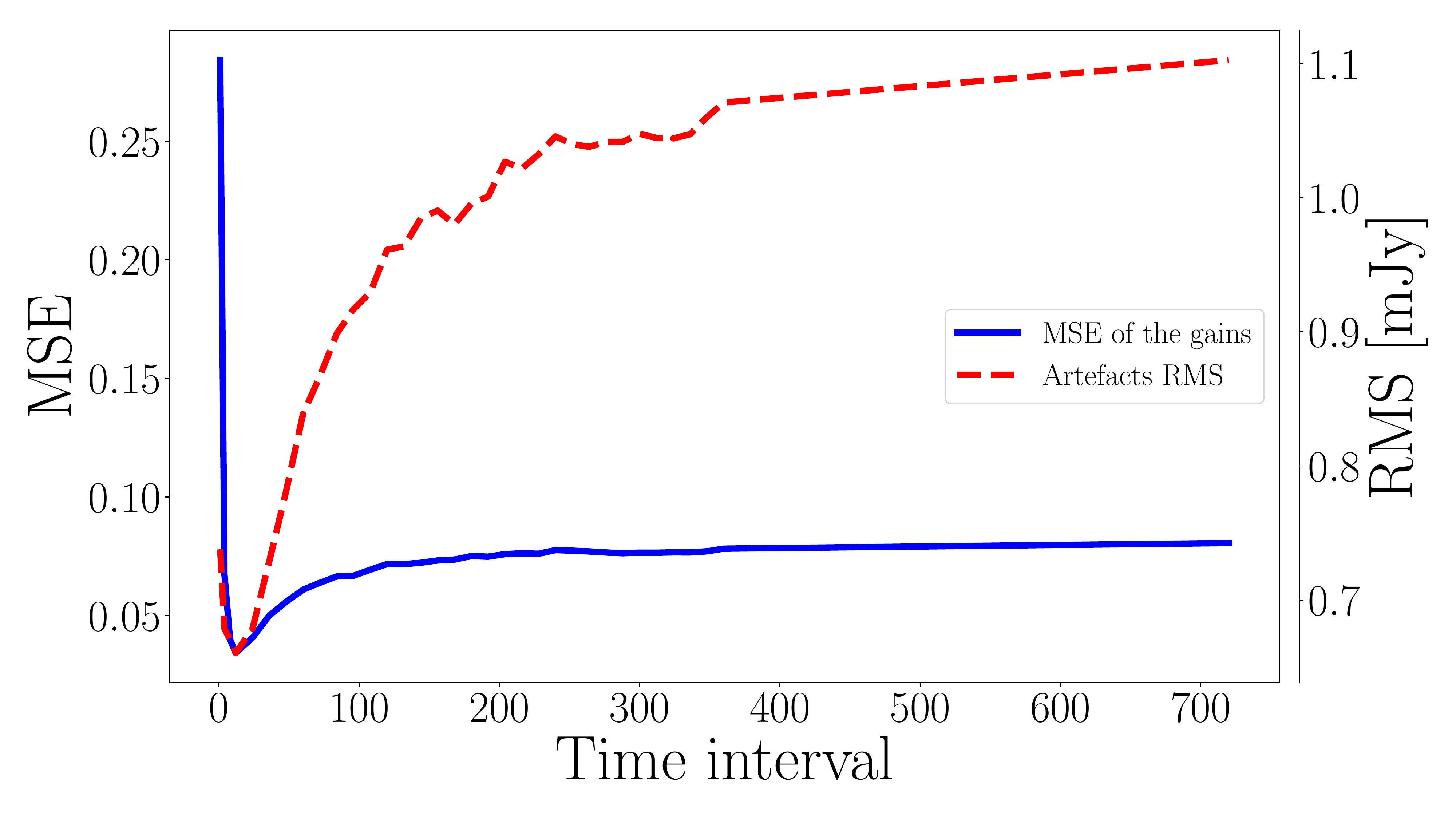}
 \caption{Incomplete sky model -- MSE}\label{incomplete}
 \end{subfigure}
 \begin{subfigure}{0.48\textwidth}
 \centering
  \includegraphics[width=8cm,height=6cm,keepaspectratio]{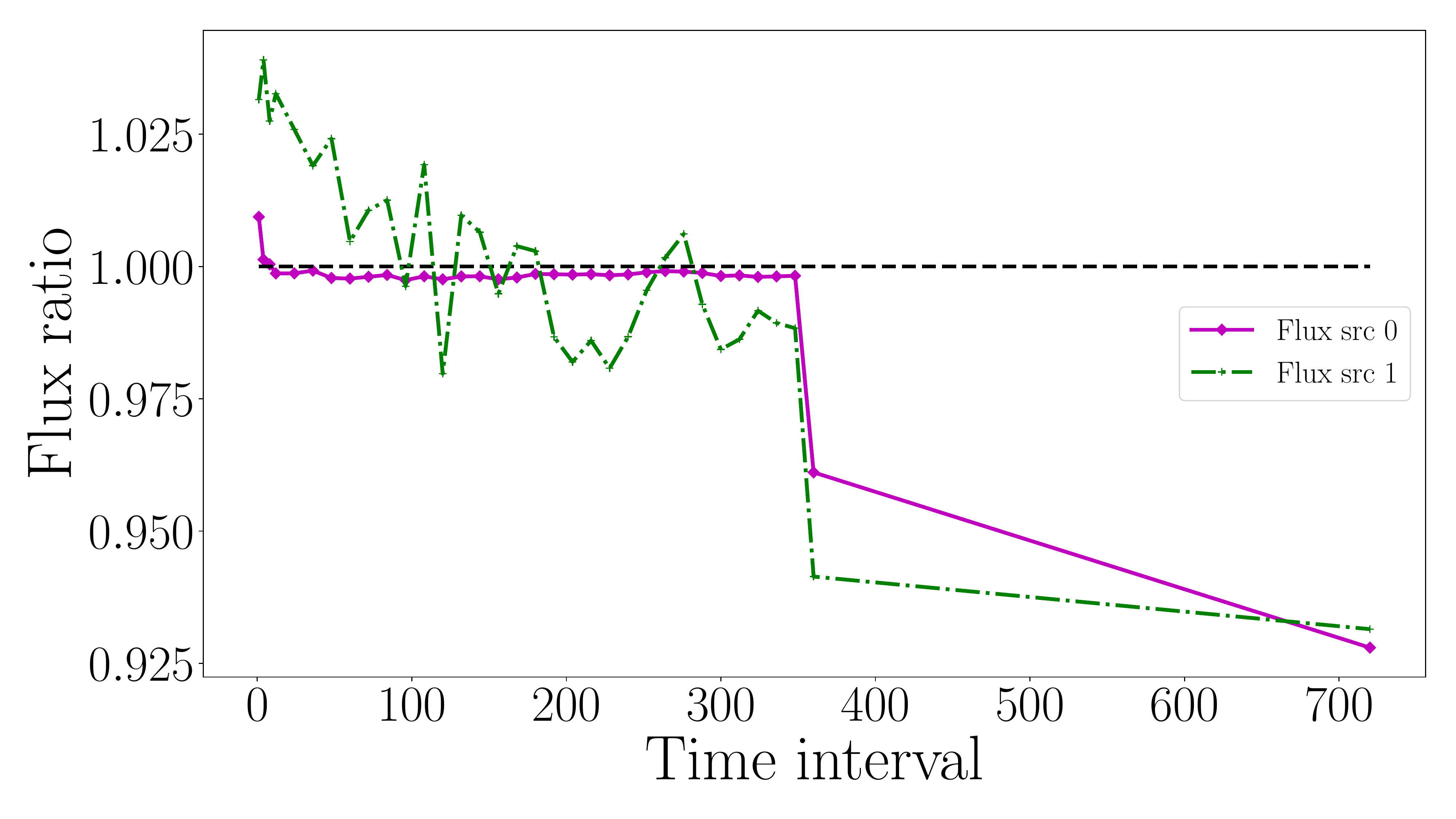} 
\caption{Complete sky model -- Flux}\label{complete_flux}
 \end{subfigure}
\begin{subfigure}{0.48\textwidth}
\centering
  \includegraphics[width=8cm,height=6cm,keepaspectratio]{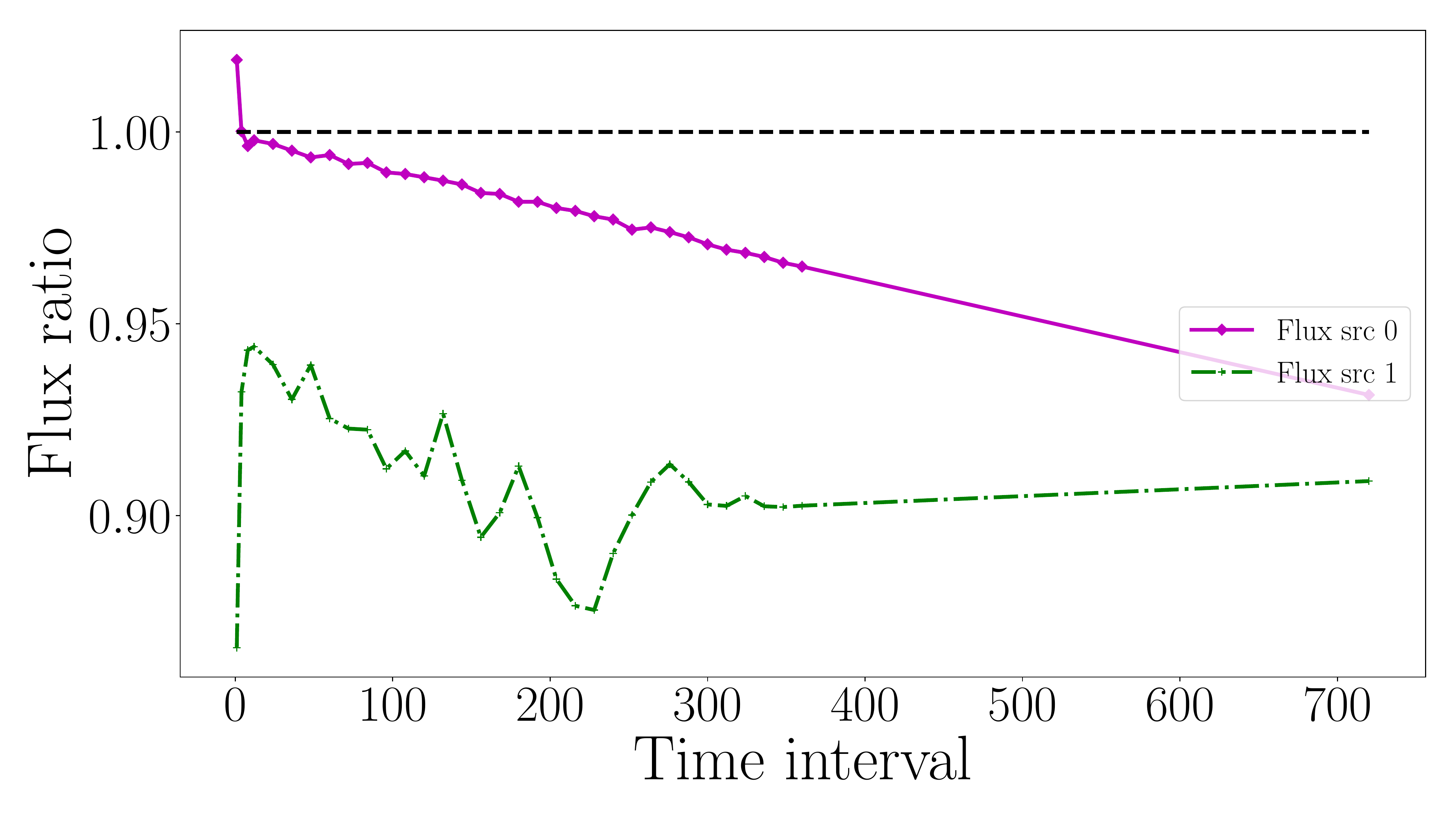}
 \caption{Incomplete sky model -- Flux}\label{incomplete_flux}
 \end{subfigure}
\caption[MSE of estimated gains and artefacts rms]{(a) and (b) are plots of the MSE of estimated gains (blue solid line) against the solution interval used on the left y-axis and the rms of the artefact images (red dashed line) on the right y-axis. These plots illustrate the trade-off between the error due to noise on the visibilities and the intrinsic variability of the gains as we increase the solution interval. There is a strong correlation between the MSE and the artefact rms. When the sky model is complete, the minimum rms occurs at approximately the same point as the minimum MSE, which is not guaranteed when the sky model is incomplete. (c) and (d) are plots of the ratio of recovered to the actual fluxes for the brightest source (green) and a faint source (magenta) respectively. The black dashed line indicates the flux ratio of $1$ for perfect reconstruction.}
\label{fig:MSE_example}
\end{figure*}

Fig. \ref{complete_mse} and Fig. \ref{incomplete} illustrate how the MSE of the estimated gains varies with solution intervals. Increasing the solution interval improves the SNR, by reducing the noise in the visibilities, but when the interval becomes too large, it becomes impossible to track the gain variations. The MSE then increases to an asymptotic value which depends on the gains' intrinsic variability. Another quantity we can look at is the noise contribution from the calibration errors. Hence, we also show in Fig. \ref{complete_mse} and Fig. \ref{incomplete} the rms of the artefact maps, $\bm{r}^{\Delta}_{pq}$, which we define as
\begin{equation}
\bm{r}^{\Delta}_{pq} = \hat{\Beps}_{pq}  - \bm{n}_{pq}, \qquad \hat{\Beps}_{pq} = \V_{pq} - \hat{\G}_p \C_{pq} \hat{\G}_q^H, \label{distill}
\end{equation}
where $\hat{\G}$ denotes the solution to Eq. \eqref{eq_rime_2}, and $\bm{n}_{pq}$ is the exact noise realisation added to the simulated visibilities, i.e. we subtract the exact input noise from the uncorrected residual visibilities to isolate the artefacts. The artefacts' rms follows a similar profile as the error in the gains but increases to larger values at longer intervals. We plot in Fig. \ref{complete_flux} and Fig. \ref{incomplete_flux} ratios of the fluxes recovered after calibration to the true fluxes for two sources namely:
\begin{itemize}
\item \textbf{src 0 --} the brightest source in the field (which has a flux of $0.5$ Jy),
\item \textbf{src 1 --} the brightest unmodelled source in the field (which has a flux of $0.05$ Jy) for the incomplete sky model case. 
\end{itemize}
For the complete sky model, we slightly overestimate the fluxes of the sources at the shortest interval because of the low SNR, but as we move to longer intervals, we start observing source suppression. For the incomplete sky model, the brightest source has a similar flux plot as in the case with a complete sky model, but here, source suppression in its flux is observed earlier and gets worse with increasing solution interval.
On the other hand, the unmodelled flux has a more complicated flux profile. Initially, the flux is largely suppressed, the suppression reduces with the solution interval, and then suppression starts increasing again. 

Isolating and understanding the effects of solution intervals is not a straightforward task, but Fig. \ref{fig:MSE_example} summarises most of what can happen as a result of solution intervals. From the simulations, long solution intervals which do not capture the variations in the gains could cause a flux suppression of up to $\approx$ 10 $\%$ even for modelled sources, while the variance in the gains due to the solution interval at low SNRs could increase the image rms by a factor of up to 3. \citet{marti2008spurious}, for example, describes the formation of spurious sources when phase self-calibration is done at low SNR and even derives an approximate expression for the fluxes of the spurious sources as a function of solution interval and number of antennas.

An appropriate definition for the optimal solution interval should then be one which minimises flux suppression or overestimation while maximising dynamic range. Thus, we define the optimal solution interval as the one which minimises the rms of the artefact image. Fig. \ref{fig:MSE_example} shows that the MSE accurately tracks the artefacts' rms. Hence, throughout the rest of the paper, for reasons of computational efficiency we will use the MSE as a reference metric during simulations. However, both the rms of the artefact image and the MSE are only available when the ground truth is known, i.e. only for simulations. We will, therefore, show in $\S$\ref{infering} that it is possible to derive a nearly equivalent metric which we can efficiently compute in practice.
\begin{table*}
    \small
    \captionsetup{font=footnotesize,labelfont=footnotesize}
    \centering
    \begin{tabular}{ | l | p{12cm} |}
    \hline
    \textbf{Identifier}& \centerline{\textbf{Setup}} \\ \hline
    \textbf{i}\label{t_sim_i} & Sky model: 100 sources (random positions and fluxes following a power law)\newline  
    Peak flux = 0.5 Jy \newline
    Input rms  = 1 Jy (per visibility) \newline
    Input gains from a GP with a squared exponential kernel ($\sigma_f \,= \,0.2,\, l\,=\,100$)\newline
    Array = MeerKAT\newline
    a) Calibrate with the complete sky model\newline
    b) Calibrate with an incomplete sky model (50\% of total flux). Peak unmodelled flux = 0.05 Jy \\ \hline
    \textbf{ii}\label{t_sim_ii} & Sky model: 1 Jy source at phase centre\newline
    Input rms = 2 Jy (per visibility)\newline
    No input gains, i.e. unity gains\newline
    Array = KAT-7, VLA, VLA-14 and MeerKAT\\\hline
    \textbf{iii}\label{t_sim_iii} & Sky model: two 0.05 Jy sources (at phase centre and at $12'$ off axis) \newline
    Input rms = 0.5 Jy (per visibility) \newline
    Array = MeerKAT \newline
    Single channel MS \newline
    No input gains, i.e. unity gains\newline
    a) Calibration with the phase centre source only\\\hline
  \textbf{iv}\label{t_sim_iv}& Sky model: two 0.05 Jy sources (at phase centre and at $12'$ off axis) \newline
    Input rms = 0.5 Jy (per visibility) \newline
    Array = MeerKAT \newline
    Single channel MS \newline
    No input gains, i.e. unity gains\newline
    a) Phase-only calibration\newline
    b) Amplitude-only calibration\newline
    c) Full-complex calibration\\\hline
    \textbf{v}\label{t_sim_v} & Sky model: 1 Jy source at phase centre\newline
    Input rms = 0 Jy (no noise added) \newline
    Input gains from a GP with a squared exponential kernel ($\sigma_f \,= \,0.5,\, l\,=\,200$) \newline 
    Array = MeerKAT\\\hline
    \end{tabular}
    \caption[Table with simulation setups]{Summary of the parameters for the different simulations. Note that certain simulations are repeated with calibration performed using complete and incomplete sky models. We use a square exponential covariance function to sample the gains. The length scale employed is in units of seconds, for example, $l = 100$ implies the gains vary on scales of 100 secs, i.e. ten units of integration time (10 secs). We use the same parameters for the amplitude and phase of the gains.} 
    \label{Table:sim_setups}
    \end{table*}

\subsection{Full-complex versus amplitude and phase-only solutions}

It is common practice, at least as far as DI (self-calibration) solutions are concerned, to induce additional regularisation by parametrising the gains in terms of separate amplitude and phase components (and, possibly delays, i.e. linear phase ramps in frequency). This is physically well-motivated by the fact that DI phases (in particular, those associated with the atmosphere) tend to fluctuate on much shorter time scales than amplitudes. Furthermore, most phase effects affect both polarization components equally (in RIME parlance, these are ``scalar'' effects). In linear feed systems, unmodelled Stokes $Q$ will tend to be absorbed into amplitude solutions. These factors tend to favour a calibration strategy where a number of scalar phase-only self-calibration cycles are performed using short solution intervals, followed by a round of amplitude self-calibration using longer solution intervals.

DD solutions, on the other hand, tend to employ longer solution intervals from the outset \citep{RIME3}; it is well-established that short DD solution intervals quickly lead to a proliferation of DoFs, and thence to undesirable effects such as severe flux suppression \citep{nunhokee2015link}. With longer intervals, there is no advantage to separating the amplitude and phase solutions, so a full-complex solver can be employed. 

Unfortunately, such parametrisations, being non-linear functions, are difficult to study analytically. In what follows, we therefore restrict our theoretical analysis to the more tractable case of the full complex gain solver. The phase-only and amplitude-only parametrisations \emph{can} be studied via simulations -- and even a simple simulation ($\S$~\ref{phase_amplitude}) shows that the story becomes considerably more ``complex'' in these cases. These scenarios clearly deserve an in-depth study that we must leave to a future work.

\section{Gain Errors}\label{sol_int}
The aim of this section is to use simulations to demonstrate how different factors affect calibration solutions and how these effects vary with the chosen solution intervals. In particular, the following factors will be investigated:
\begin{itemize}
\item the noise in the visibilities,
\item the intrinsic variability of the gains,
\item the degree of model incompleteness and RFI.
\end{itemize}

\subsection{Noise in the visibilities}\label{err_noise}
\subsubsection{Variance of estimated gains}\label{noise_variance_derivation}
In this section, we use the Fisher Information matrix (FIM; \citet{fisher1920012}) to find an expression for the minimum expected variance of the error in gains resulting from the noise in the visibilities during calibration. 
In simple terms, the FIM can be thought of as the amount of information from an unknown parameter that can be obtained from a measured data. Given an optimisation problem 
\begin{align}
    \min_{\vectb{\theta}}\sum_{t} \lvert \vectb{y}_{t} - \vectb{f}_{t}\rvert ^{2},  \nonumber
\end{align}
where $\vectb{y}_{t}$, $\vectb{f}_{t}$ and $\vectb{\theta}$ are the measured data, the model and the unknown parameter vector respectively, if we assume a Gaussian likelihood function, then the elements of its FIM, $\mathrm{F}$, are given by 
\begin{eqnarray}
\mathrm{F} \, &=&\, \sum_{t}{\frac{1}{\sigma^{2}}\frac{\partial \vectb{f}_{t}}{\partial \vectb{\theta}_{i}}\frac{\partial \vectb{f}_{t}}{\partial \vectb{\theta}_{j}}, }\label{fisher}
\end{eqnarray}
where $\sigma$ is the standard deviation of the errors in the measured data, and $t$ is a time index or data count index (note that the Gaussian noise approximation only holds for synthesis arrays like MeerKAT and VLA where the noise on each visibility is mutually independent, but this is generally not the case for aperture and phased arrays). From the Cramer-Rao bound (see \citet{Jansen2011}), the inverse of the FIM is a lower bound on the variance of the estimated parameters. 

\begin{figure}
	\hspace*{-1cm}
	\centering
	\includegraphics[width=0.9\linewidth]{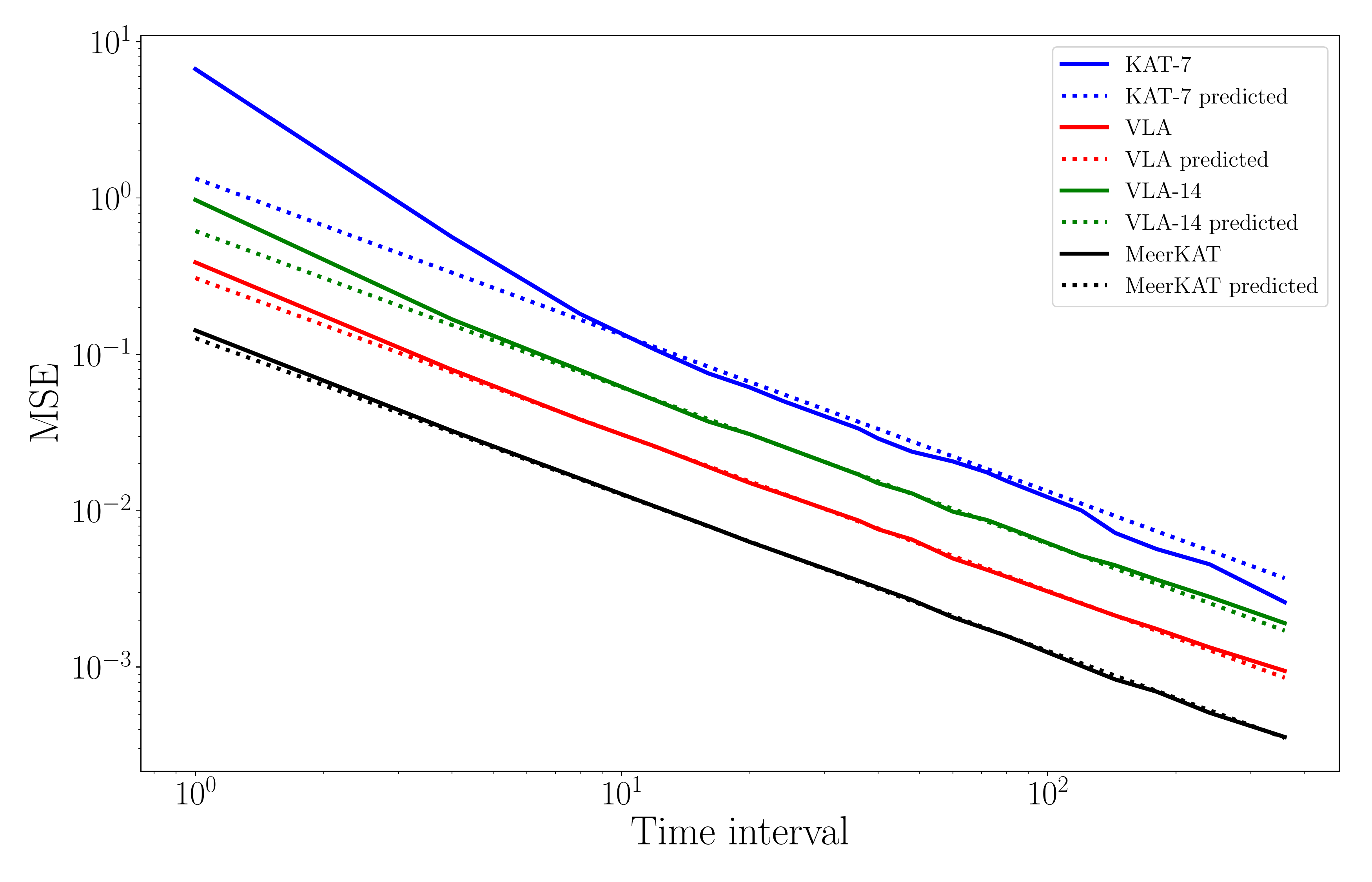}
    \caption[MSE of estimated gains against solution interval on a log scale]{MSE of estimated gains against solution interval on a log scale. In this simulation, the input gains are all set to $1$. The solid lines are the measured values while the dashed lines are the error values predicted using Eq. \eqref{vargp}. VLA-14 is a 14 antenna subset of the 27 VLA antennas \citep{perley2011expanded}. Eq. \eqref{vargp} is a close fit for MeerKAT because of its large number of antennas. For the other arrays, the predicted and measured values only start agreeing after a certain time interval.}
    \label{fig:rms only plots}
\end{figure}

In the simplest case of DI calibration of unpolarised single channel visibilities, the RIME at time $t$ is given by
\begin{eqnarray}
 \V_{pqt} \, &=&\,  \G_{pt}\C_{pqt}\G^H_{qt} \, + \, \epsilon, \label{dirime}
 \end{eqnarray}
 where 
\begin{equation}
\G_{pt} \, = \, \begin{pmatrix}
\vectb{g}^{(1)}_{pt}&0\\
0&\vectb{g}^{(2)}_{pt}
\end{pmatrix}, 
\end{equation}
where $\vectb{g}^{(1)}_{pt}$ and $\vectb{g}^{(2)}_{pt}$ are the diagonal terms of our gain matrix. Let us assume a classical dish such as MeerKAT, operating in the frequency range where the noise is dominated by the system contribution rather than the sky ($T_{\mathrm{sys}} \gg T_{\mathrm{sky}}$). The noise can then be seen as independent Gaussian: \begin{equation}
\epsilon \, \sim \, \mathrm{CN}(0,\mathrm{I}\sigma_{\mathrm{rms}}^2),
\end{equation}
where $\mathrm{I}$ denotes the identity matrix.
For a given solution time interval, $n_t$, using Eq. \eqref{fisher}, the diagonal elements of the Fisher matrix for this process are given by
 \begin{eqnarray}
 \mathrm{F}_{pp} \, &=&\, \frac{1}{\sigma_{\mathrm{rms}}^{2}}\sum_{t=1}^{n_{t}}{\sum_{p \neq q}^{N_{a}}{(\C_{pqt}\G^H_{qt})\circ(\C_{pqt}\G^H_{qt})}}, \label{fpp}
 \end{eqnarray} 
where $N_{a}$ is the number of antennas in our array and $\circ$ denotes the Hadamard (element-wise) product. Assuming that the Hessian matrix is diagonally dominant (which is a very plausible assumption as demonstrated by \citet{smirnov2015radio}), the inverse of the Fisher matrix becomes $\mathrm{F}_{ij}^{-1} \, = \, \frac{1}{\mathrm{F}_{ij}}$ (for all non-zero entries). Hence 
 \begin{align}
 \mathrm{F}^{-1}_{pp} \,= \, \mathrm{Var}(\hat{\vectb{g}}_{p}) = \frac{\sigma_{\mathrm{rms}}^{2}}{\sum_{t=1}^{n_{t}}{\sum_{p \neq q}^{N_{a}}{(\C_{pqt}\G^H_{qt})\circ(\C_{pqt}\G^H_{qt})}}}, \label{fppinv}
 \end{align}
where $\hat{\vectb{g}}_{p} \in \{\hat{\vectb{g}}^{(1)}_p, \hat{\vectb{g}}^{(2)}_p \}$.
If we assume constant gains $1+0\text{j}$ we can rewrite Eq. \eqref{fppinv} as
\begin{align}
 \mathrm{Var}(\hat{\vectb{g}_{p}}) \approx \frac{\sigma_{\mathrm{rms}}^{2}}{\sum_{t=1}^{n_{t}}{\sum_{p \neq q}^{N_{a}}{\C_{pqt}\circ\C_{pqt}}}}. \label{fppinv-1}
 \end{align}
Eq. \eqref{fppinv-1} can be further simplified if we consider a field with a single source having flux $S$ located at the phase centre. In this case, the error in the estimated gains resulting from the noise in the visibilities is given by
\begin{eqnarray}
\mathrm{Var}(\hat{\vectb{g}}_{p})\, & \approx & \frac{\sigma_{\mathrm{rms}}^{2}}{n_{t} (N_{a}-1) S^{2}}. \label{pre-vargp}
\end{eqnarray} 
Additionally, for a multichannel observation with solution frequency interval, $n_\nu$, Eq. \eqref{pre-vargp} becomes
\begin{eqnarray}
\mathrm{Var}(\hat{\vectb{g}}_{p})\, & \approx & \frac{\sigma_{\mathrm{rms}}^{2}}{n_{\nu}n_{t} (N_{a}-1) S^{2}} \; = \; \frac{\sigma_{\mathrm{rms}}^{2}}{n (N_{a}-1) S^{2}}, \label{vargp}
\end{eqnarray}
where $n=n_{\nu}n_{t}$. Eq. \eqref{vargp} is an already known result in the field of radio interferometry and is stated in \cite{synthesisImagingbook} in the following forms for phase-only and full-complex calibration respectively,
$$\mathrm{Var}(\hat{\vectb{g}}_{p})\, = \frac{\sigma_{\mathrm{rms}}^{2}}{n (N_{a}-2) S^{2}} \; ; \; \mathrm{Var}(\hat{\vectb{g}}_{p})\, = \frac{\sigma_{\mathrm{rms}}^{2}}{n (N_{a}-3) S^{2}}.$$
The factor $N_{a}-1$ is replaced by the factors $N_{a}-2$ and $N_{a}-3$ in order to account for the extra DoFs required for phase-only and full-complex calibration respectively. For consistency throughout the paper, we will use Eq. \eqref{vargp} with the factor $N_{a}-1$ as an estimate of the expected variance on gains solutions resulting from the noise in the visibilities. \\[0.2cm]
\indent We check the validity of Eq. \eqref{vargp} using \textbf{Simulation-ii}. This simulates a 1 Jy source at phase centre using a variety of array layouts (see Table \ref{Table:sim_setups}), and corrupts each visibility with 2 Jy rms noise. Fig. \ref{fig:rms only plots} is a plot of the MSE against solution interval for the different datasets. The plot shows that the estimated MSE are in close agreement with Eq. \eqref{vargp}, particularly for larger arrays. The predicted and measured MSE for MeerKAT are very close even at the shortest time interval of 1. For smaller arrays (VLA-14 and KAT-7), the predicted MSE matches better towards longer time intervals.

For a complex field with multiple sources, we can use the absolute value of the model visibilities to define the SNR, thus rewriting Eq. \eqref{fppinv-1} as 
\begin{align}
 \mathrm{Var}(\hat{\vectb{g}}_{p}) \approx \frac{\sigma_{\mathrm{rms}}^{2}}{\sum_{t=1}^{n_{t}}{\sum_{p \neq q}^{N_{a}}{|\C_{pqt}|^2}}}. \label{fppinv-2}
\end{align}
 Eq. \eqref{vargp} is simply a special case of Eq. \eqref{fppinv-2} for a single source. For more complex fields, we can still use Eq. \eqref{vargp} by substituting the flux $S$ with the mean of $|\C_{pqt}|$, i.e.
 \begin{align}
    S = \mathrm{mean}(|\C_{pqt}|) .  \label{fppinv-3}
 \end{align}

\subsubsection{Effects of noisy gains on corrected visibilities}\label{noise_corr_data}
In this section, we demonstrate the effects of applying noisy gains obtained from calibration at low SNR on corrected images. Furthermore, we discuss the presence of unmodelled sources and show that their effects only become noticeable during calibration when their flux is above the noise level.

From Eq. \eqref{eq_rime_simple}, we can write down and decompose the corrected visibilities into several components. Splitting the coherency into $\C_{pq} = \C^{\mathrm{m}}_{pq} + \C^{\mathrm{um}}_{pq}$ gives
\begin{align}
\V_{pq}^{c} &= \hat{\G}_p^{-1} \V_{pq} \hat{\G}_q^{-H}, \label{corr_c1a}\\
            &= \hat{\G}_p^{-1} \G_p \C_{pq} \G_q^H\hat{\G}_q^{-H} + \hat{\G}_p^{-1}\bm{\epsilon}_{pq}\hat{\G}_q^{-H},\label{corr_c2a}\\
&= \hat{\G}_p^{-1} \left(\G_p (\C^{\mathrm{m}}_{pq} +  \C^{\mathrm{um}}_{pq})\G_q^H + \bm{\epsilon}_{pq}\right)  \hat{\G}_q^{-H}\label{corr_c3a},\\
&= \hat{\G}_p^{-1}\G_p \C^{\mathrm{m}}_{pq} \G_q^H\hat{\G}_q^{-H} + \hat{\G}_p^{-1}\G_p \C^{\mathrm{um}}_{pq}\G_q^H\hat{\G}_q^{-H} \nonumber \\
&\quad + \hat{\G}_p^{-1}\bm{\epsilon}_{pq}\hat{\G}_q^{-H}, \label{corr_c3}
\end{align}
where $\C^{\mathrm{m}}_{pq}$ and $\C^{\mathrm{um}}_{pq}$ denote the  modelled and unmodelled components, respectively. 
{\color{black}Thus, we refer to the three terms in Eq. \eqref{corr_c3} as the \textit{corrected modelled} ($\mathrm{CM} \; =\; \hat{\G}_p^{-1}\G_p \C^{\mathrm{m}}_{pq} \G_q^H\hat{\G}_q^{-H}\nonumber$), \textit{corrected unmodelled} ($\mathrm{\CU} \; =\; \hat{\G}_p^{-1}\G_p \C^{\mathrm{um}}_{pq} \G_q^H\hat{\G}_q^{-H}\nonumber$) and the \textit{corrected noise} ($\mathrm{C}\sigma \;=\;\hat{\G}_p^{-1}\bm{\epsilon}_{pq}\hat{\G}_q^{-H} \nonumber$) terms respectively.} 

If the calibration is perfect, $\hat{\G}_p$ = $\G_p$ and $\hat{\G}_p^{-1} \G_p = \mathbf{I}$. The \CM\ and \CU\ terms then correspond to the true sky visibilities, and the \Cnoise\ term is the noise in the visibilities slightly modified by the inverse of the gains. However, at low SNR, the solver fits noise rather than the model, thereby transferring the model flux into the gains. When such gains are applied to the noise, the resultant \Cnoise\ term will contain ghost sources at the position of the modelled sources that leads to an amplification of these sources in the corrected image. This problem is not only related to solution intervals but forces us to re-assess our definition of corrected visibilities. How exactly should we apply the gains to the data without modifying the underlying structure of the noise in the data? 

\begin{figure*}
\begin{subfigure}{\linewidth}
    \centering
    \vspace*{-3.5cm}
    \includegraphics[width=0.8\linewidth,height=6.5cm, keepaspectratio=True]{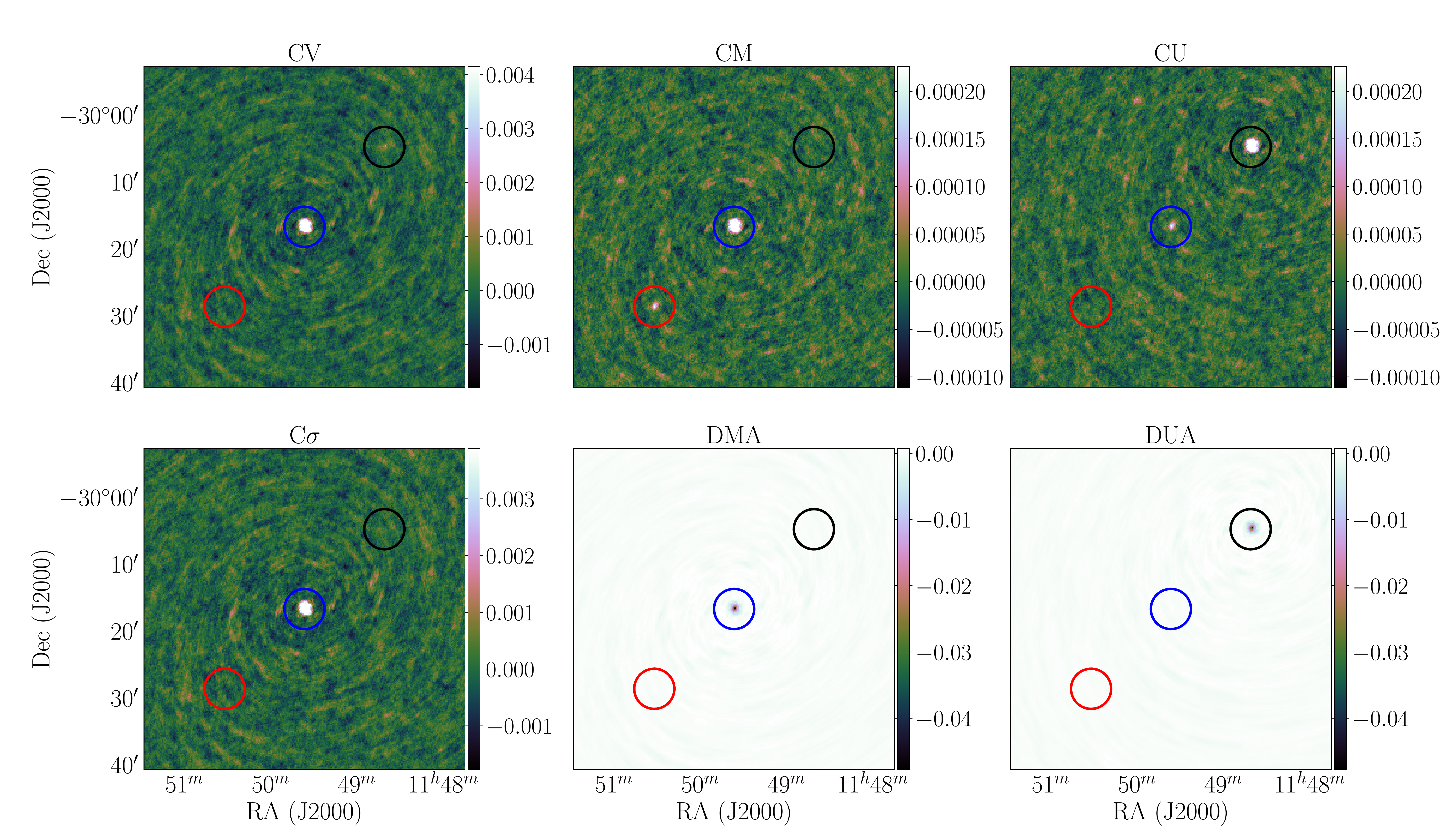}
    \caption{Time interval = 1 (10 secs)\\Because of the low SNR in this case, we fit noise instead of modelled visibilities. The modelled source is absorbed into the gains and its signature appears in the blue circle of the \Cnoise\ term. In the \CM\ and \CU\ terms, we can see ghost sources appearing in the red and blue circles respectively. These ghost sources result from the unmodelled source and causes further suppression in the fluxes of the sources. Due to the large suppression caused by the noise in the \DMA\  and \DUA\  terms, we can only see negative peaks in the blue and black circles at the position of the sources, confirming that their fluxes have been significantly suppressed.} 
    \label{fig:corr_noise_1}
 \end{subfigure}
\begin{subfigure}{\linewidth}
    \centering
    \includegraphics[width=0.8\linewidth,height=6.5cm,keepaspectratio=True]{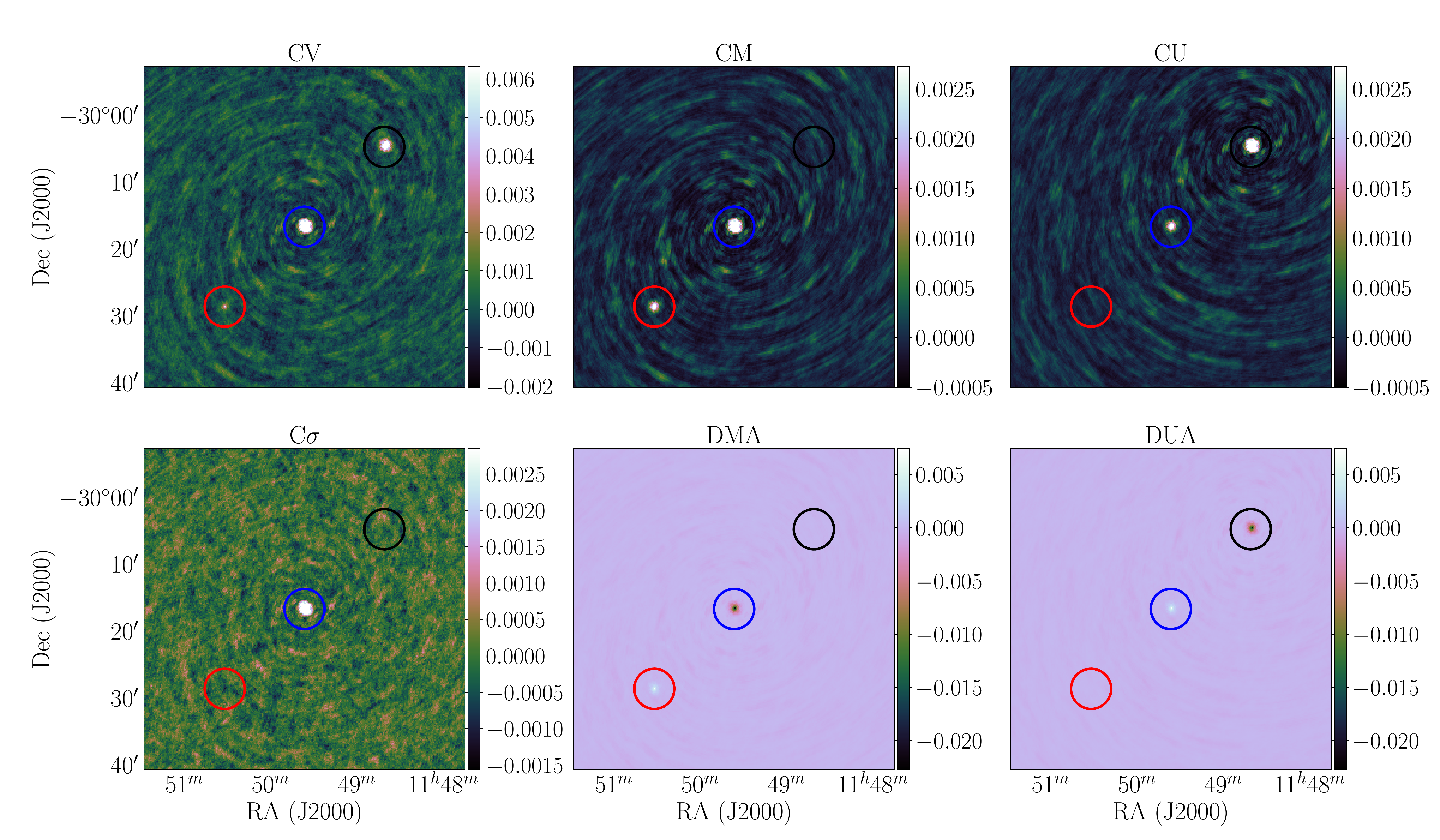}
    \caption{Time interval = 8 (80 secs)\\Here, the SNR is higher, hence the unmodelled source has a stronger influence on the calibration. This results in brighter ghost sources which we can see in the corrected visibilities (red circle), the \CM\ term (red circle) and the \CU\ term (blue circle). Also, we can see positive peaks in the red and blue circles of the \DMA\ and \DUA\ terms, confirming the stronger influence of the unmodelled source in this case.}
    \label{fig:corr_noise_8}
 \end{subfigure}
\begin{subfigure}{\linewidth}
    \centering
    \includegraphics[width=0.8\linewidth,height=6.5cm,keepaspectratio=True]{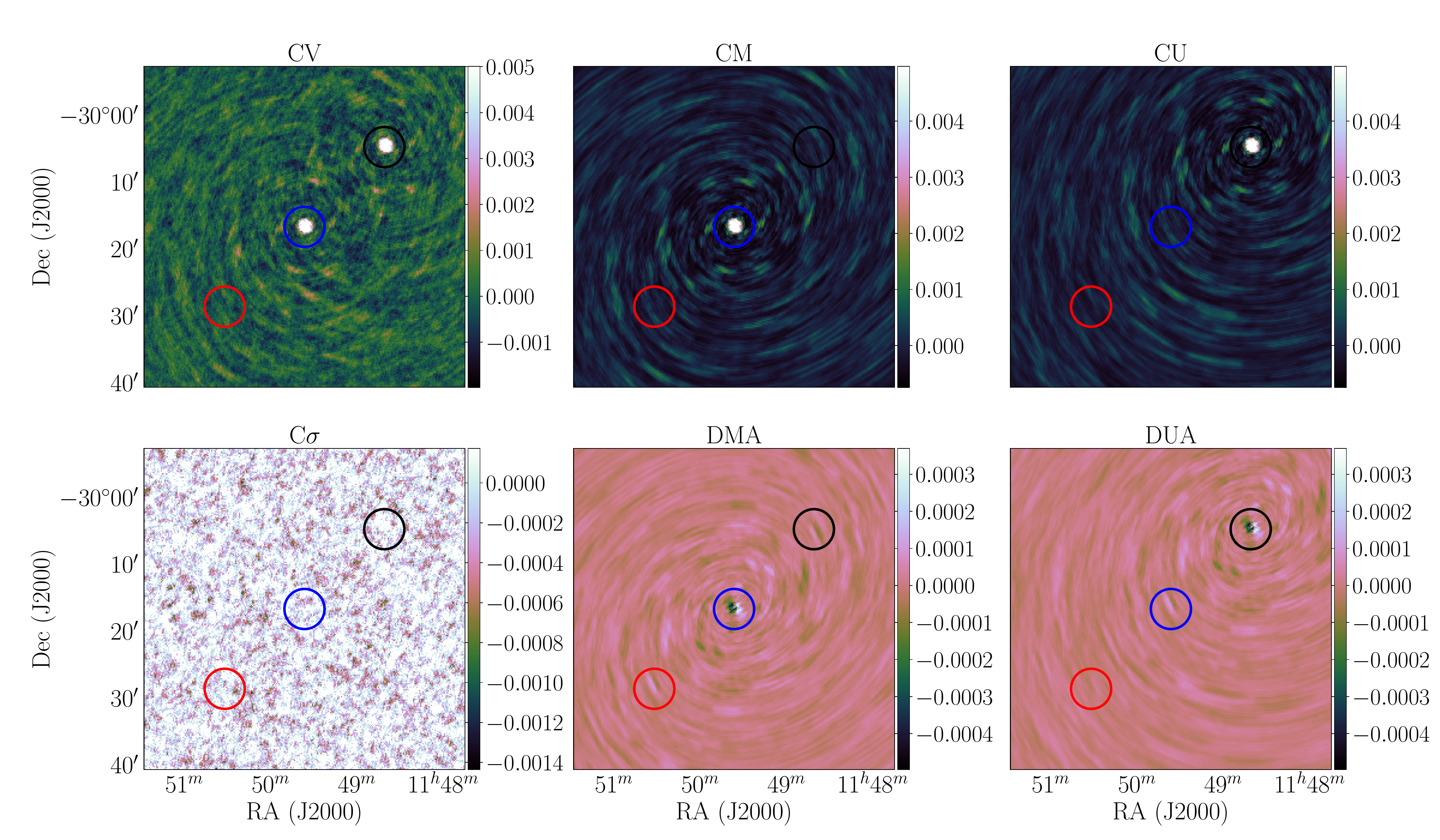}
    \caption{Time interval = 720 (2 hours)\\ In this case, the effects of the noise and the unmodelled source are ultimately averaged out because of the very long solution interval. Hence, in the \Csky\, \CM\ and \CU\ terms, we see that the modelled source and unmodelled source in the blue and black circles have not been suppressed. The \Cnoise\ term is entirely noise-like as we expect, while the \DMA\ and \DUA\ images exhibit only very low-level PSF-like structures.}
    \label{fig:corr_noise_720}
 \end{subfigure}
 \caption[Corrected images at different time intervals]{Images of the different terms in Eq.~\eqref{corr_c3} as well as the \DMA\ and \DUA\ maps. Each $2\times 3$ block corresponds to a specific time solution interval (a: 1 i.e. 10 secs, b: 8 i.e. 80 secs,  c: 720 secs i.e. 2 hours). On the top row, from left to right, we have the images of the \Csky\, the \CM\ and the \CU\ terms, respectively. On the bottom row, from left to right, we have the \Cnoise, \DMA\, and \DUA\ terms, respectively. The blue and black circles indicate the position of the modelled and unmodelled sources. The red circle indicates a position at which we expect a ghost source to form due to the unmodelled source.} \label{fig:noise_corr_vis}
 \end{figure*}

We perform a two-source simulation (\textbf{Simulation-iii}) to investigate the contribution of the different terms in Eq. \eqref{corr_c3} to the corrected visibilities at low SNRs. The setup consists of a 0.05 Jy source at the phase centre and a 0.05 Jy source at an offset position, with only the former included in the calibration sky model (see Table \ref{Table:sim_setups}).
\thispagestyle{plain}
Fig. \ref{fig:noise_corr_vis} shows images of the various components of the corrected visibilities (\Csky) after calibration. Alongside these, we have plotted images corresponding to the \textit{distilled modelled} (\DMA) and \textit{distilled unmodelled artefacts} (\DUA):
\begin{align}
  \mathrm{\DMA} &= \mathrm{\CM} - \C^{\mathrm{m}}_{pq}, \nonumber \\
  \mathrm{\DUA} &= \mathrm{\CU} - \C^{\mathrm{um}}_{pq}, \nonumber \\
  \mathrm{\Csky} &= \mathrm{\CM} + \mathrm{\CU} + \mathrm{C}\sigma.
 \end{align}

Intuitively, the \DMA\ and \DUA\ maps correspond to the distortions induced by gain corrections in the modelled and unmodelled components of the sky (compare with Eq.~\eqref{corr_c3}).

Each $2\times 3$ block of images corresponds to a specific time interval. On the top row, from left to right, we have the images of the \Csky\ and the \CM\ and \CU\ terms, respectively. On the bottom row, from left to right, we have the \Cnoise, \DMA, and \DUA\ maps, respectively. The blue and black circles indicate the position of the modelled and the unmodelled sources. The red circle indicates a position at which we expect a ghost source to form because of the unmodelled source. It is worth paying attention to the colour bars when looking at these images. These images are shown on the best possible colour scales to illustrate the subtle differences between the various components at different solution intervals. 

When the time interval is 1 (see Fig.~\ref{fig:corr_noise_1}), the SNR is very low. Hence, we tend to fit noise rather than the modelled visibilities. In this case, the flux of the modelled source is absorbed into the gains. Applying these gains to the data suppresses the model source in the \CM\ term, and its signature instead appears in the \Cnoise\ term. Because of the strong influence of the noise, in this case, the \Csky\ look very similar to the \Cnoise\ term. Similarly to the \CM\ term, the unmodelled source appears to be largely suppressed in the \CU\ term. In images of both the \CM\ and the \CU\ terms, we can see artefacts caused by the presence of the unmodelled source. As shown by \citet{grobler2014calibration} and \citet{grobler2016calibration}, in the two-source scenario, an unmodelled source will lead to the formation of a string of ghost sources along the line between the modelled source and the unmodelled source. To paraphrase, in this simulation we are observing flux suppression due to two factors: firstly because of the noise in the data and the short solution interval, and secondly, because we have an unmodelled source in our field.  Because of the short time interval used here, the effect of noise dominates the calibration, and the ghost sources caused by the unmodelled sources are too faint to detect in the distilled maps. 

When the time interval is 8 (see Fig.~\ref{fig:corr_noise_8}), the SNR is improved. Because of the drop in noise level on the calibration solutions, the ghost sources from the unmodelled source become more visible and can be seen in the corrected image. Likewise, the amount of flux transferred to the \Cnoise\ term is significantly reduced, and some of the ghost sources are visible in the distilled maps. Using an interval of 720 (see Fig.~\ref{fig:corr_noise_720}), both the effects of the noise and the unmodelled source are entirely averaged out. The corresponding images look exactly like what we expect from a perfect calibration. Note that using such a long interval is only possible here because we did not include any gain variations in the simulation.  
 \subsubsection{Phase-only vs Amplitude-only vs Full-complex calibration flux suppression}\label{phase_amplitude}
 In this section, we investigate what causes more flux suppression between phase-only, amplitude-only and full-complex (amplitude and phase) calibration. We use the setup of \textbf{Simulation-iv} for this purpose. We simulate a 0.05 Jy source at the phase centre and a 0.05 Jy source at an offset position using a single frequency MeerKAT array layout. We add noise with an rms of 0.5 Jy to the data but we do not include any gain corruptions. Next, we perform phase-only, amplitude-only and full-complex calibration to compare how well each calibration scenario recovers the fluxes of the sources.
 
\begin{figure}
	\centering
	\includegraphics[width=0.9\linewidth]{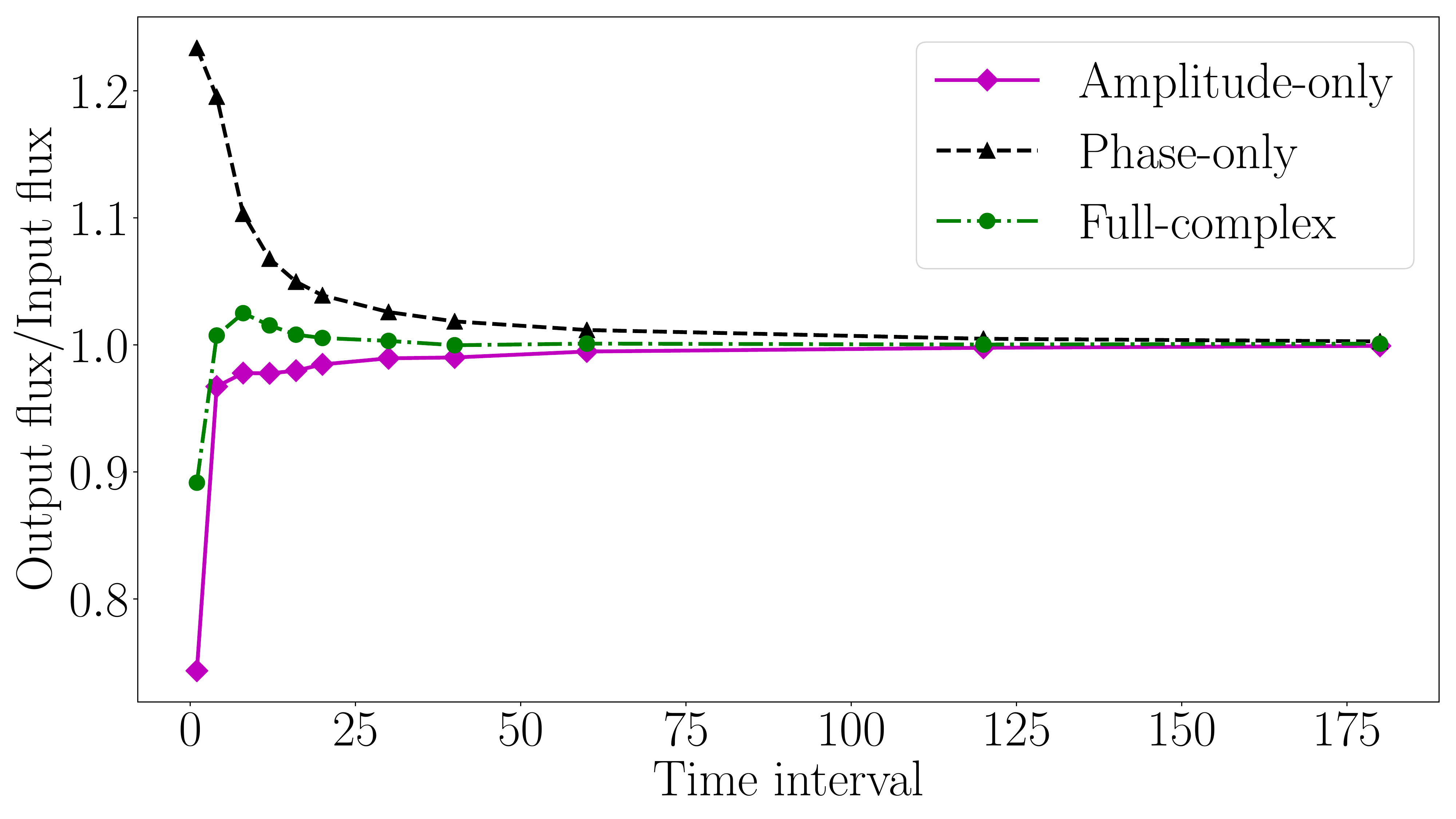}
    \caption[Phase-only vs full-complex flux suppression]{This figure shows the average output to input flux ratio against solution intervals for phase-only calibration in black, amplitude-only calibration in magenta and full-complex (amplitude and phase) calibration in green.}
    \label{fig:phase_vs_amplitude}
\end{figure}

\begin{figure*}
\begin{subfigure}{\textwidth}
\includegraphics[width=1\linewidth]{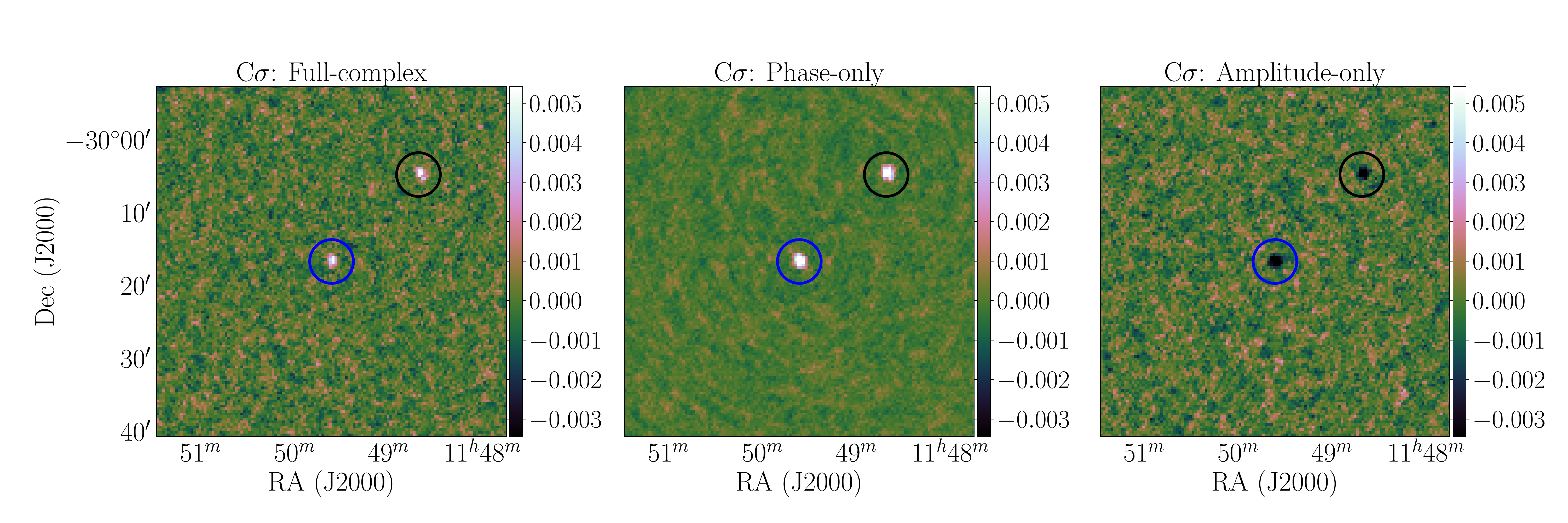}
\end{subfigure}
\caption[Corrected noise maps]{Images of the \Cnoise\ noise term for full-complex calibration (left), phase-only calibration (middle) and amplitude-only calibration (right) with a solution time interval of 4. All images are shown on the same colour scale.}
\label{fig:ampvsphase_corr_noise}
\end{figure*}

 We show in Fig. \ref{fig:phase_vs_amplitude}, the ratio (average for the two sources) of the output flux to input flux against solution intervals for the three calibration scenarios. From Fig. \ref{fig:phase_vs_amplitude}, we observe that at low SNR (short solution intervals), phase-only calibration leads to flux amplification while both amplitude-only and full-complex calibration predominantly causes flux suppression. This result can be explained by the fact that both flux suppression or amplification results from the solver trying to compensate for missing signal (i.e., unmodelled sources, RFI or noise) in the measured visibilities. In the cases of amplitude-only and full-complex calibration, the missing signal is easily transferred to the amplitude of the gain solutions leading to flux suppression. On the other hand for phase-only calibration, because there is no gain amplitude to absorb the missing flux, the solver tends to amplify the fluxes of the model signal. This result here is in agreement with \citet{grobler2016calibration}, where the ``anti-ghost'' responsible for flux amplification in phase-only calibration is $N_a$ times (where $N_a$ is the number of antennas) brighter than that in full-complex calibration. We further illustrate this by looking at maps of the \Cnoise\ term for full-complex, phase-only and amplitude-only calibration using a time interval of 4 in Fig. \ref{fig:ampvsphase_corr_noise}. In these maps we have bright positive ghost sources (at the position of the real sources) in the phase-only map and bright negative ghost sources in the amplitude-only map.  The ghost sources in the full-complex map are positive but are faint compared to those in the phase-only and amplitude-only maps. Thus, the positive ghost sources at the position of real sources will lead to flux amplification while the negative ghost sources will lead to flux suppression.
 
 Hence, a good indicator for extremely short solution intervals during phase-only calibration or significantly incompletely sky models is the amplification of the model fluxes after calibration. 
 
 \subsection{Intrinsic variability of the gains}\label{err_gains}
The expressions derived in $\S$\ref{err_noise} rely on the assumption that we have an unbiased estimator, the residual visibilities or noise is normally distributed, and  that the Hessian is diagonally dominant. This is almost never true in practice, because the gains are not constant in either time or frequency. This section focusses on the bias term of Eq.~\eqref{mse2}. This term is the error in the estimated gains due to the chosen solution interval. This term is complicated to understand and to model analytically because, in real life, we do not have any independent information about the intrinsic variability of the gains. 

\begin{figure}
	\centering
	\includegraphics[width=0.8\linewidth]{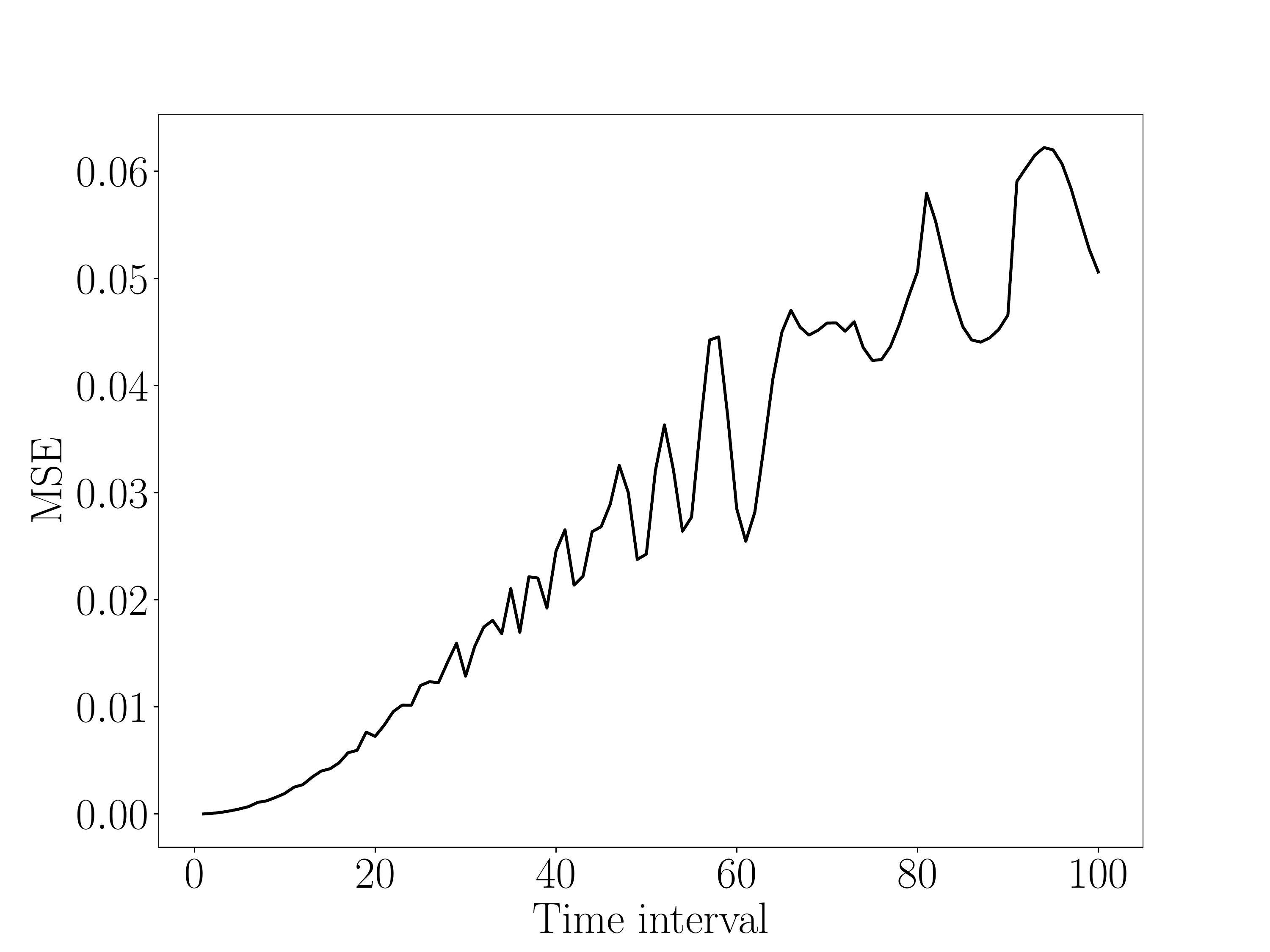}
    \caption[Gain's bias against solution interval]{This plot shows the error in the gains caused by the averaging as a function of the solution interval. This error term, contrary to that from the noise, increases with increasing solution intervals.}
    \label{fig:mse-bias}
\end{figure}

Fig. \ref{fig:mse-bias} is an illustration of how this term will vary with solution intervals. We make this plot by computing the MSE when a highly variable gain (from a GP) is reconstructed using the solution interval boxcar formulation by assuming the gain values to be equal to their mean values in every solution block for various solution intervals.
As expected, the error increases with the solution interval. The oscillations in the curve is a result of the fact that not all intervals divide the data equally, i.e. all the solution interval blocks do not have the same number of visibilities. This effect increases with solution intervals as we get more and more unequal divisions with longer intervals. This is an additional overhead which makes it difficult to formulate the solution interval problem as an optimisation problem. In practice, this might even be more complicated, since observations are usually done in scans (switching between different fields), with each scan potentially having a different number of visibilities and timestamps.
 
\begin{figure*}
\begin{subfigure}{\textwidth}
\includegraphics[width=1\linewidth]{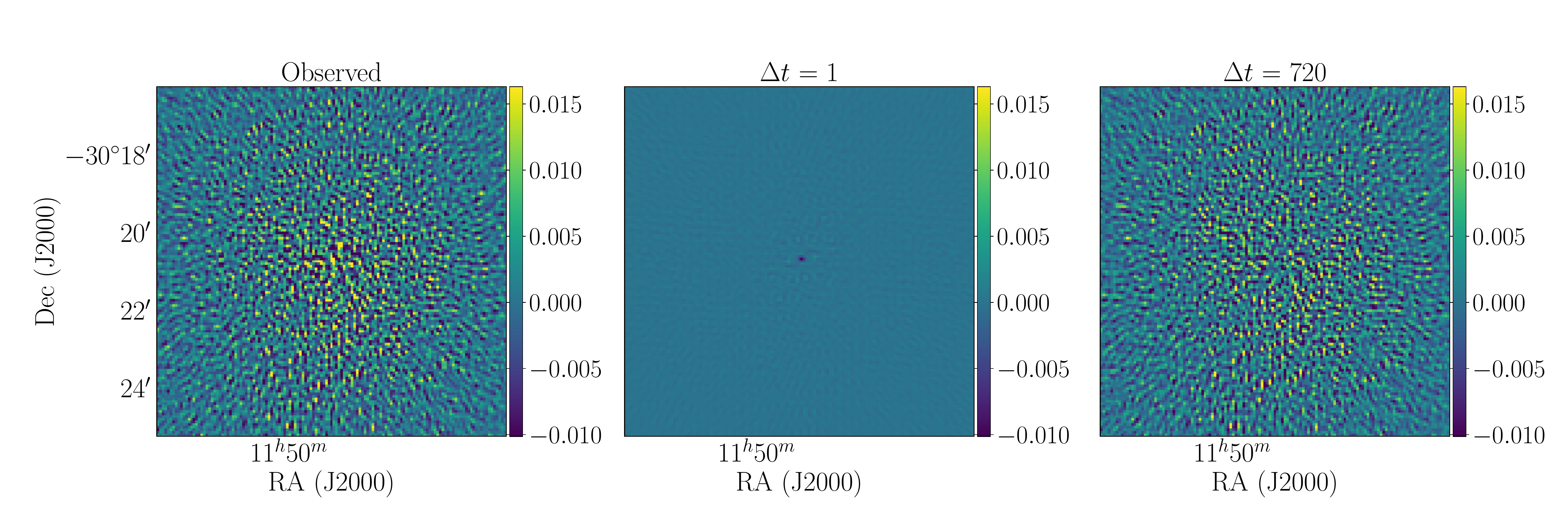}
\end{subfigure}
\caption[Artefact maps for intrinsic gains variability]{Left is an image of the artefacts introduced by the gains (corrupted visibilities minus model visibilities). The middle and right images are the artefacts' maps after calibration (corrected visibilities minus model visibilities) with time intervals of 1 and 720 respectively.}
\label{fig:gains_corr_vis}
\end{figure*}

We use \textbf{Simulation-v} to visualise the effects of the intrinsic variability of the gains on corrected visibilities. We simulate a 1 Jy source at phase centre as observed by MeerKAT, and corrupt it with rapidly varying gains. We add no noise to the data to ensure a high SNR, and to avoid obscuring the effects of the gain variations by the effects of noise. The results are plotted in Fig. \ref{fig:gains_corr_vis}. 
\noindent The leftmost image shows the artefacts added to the data by the gains. This image is computed by subtracting the model visibilities from the corrupted visibilities. For the middle and the rightmost images, we subtract the model visibilities from the corrected visibilities. These images show how well we have removed the artefacts introduced by the gains. With a time interval of 1, almost all the artefacts introduced by the gains are removed because we have the highest time resolution possible tracking the gain variations.
On the other hand, if a time interval of 720 is used, we have no time resolution at all, and cannot track the gain variations. As seen earlier in Fig. \ref{fig:MSE_example}, this leads to artefacts that increase the noise in the image, particularly around sources. The noise contributed by such artefacts is not necessarily Gaussian or following a symmetric i.i.d distribution, and thus does not decrease with averaging. Hence, even if we perform multi-frequency observations over a long period, if the solver does not accurately capture the variations in the gains, the noise in the averaged image will be higher than expected. 
\subsection{Model Completeness \& Radio Frequency Interference}\label{sec:model completeness}
\citet{sob2019robust} showed that the effects of unmodelled sources depend on their brightness relative to the noise in the visibilities. An unmodelled source with flux $\ll$ noise rms will have minimal impact on calibration, especially in the case of DI calibration. For DD calibration, given the low number of constraints per DoF, the faint sources may have a stronger impact. On the other hand, a source that is relatively bright compared to the noise and the calibration model will have a substantial contribution. This is precisely what we observe in $\S$\ref{err_noise} for \textbf{Simulation-iii}. Furthermore, \citet{sob2019robust} showed that the main effect of numerous unmodelled point sources is to make the data seem noisier, i.e. it increases the variance of the residual visibilities, and that these effects could be mitigated using a weighted maximum likelihood approach.

Another parameter which complicates the task of selecting an optimal solution interval is the presence of RFI in the data\footnote{By which we mean low-level RFI that has gotten past the usual flagging techniques.}. If the RFI is localized to a few time chunks and frequency channels, then using long solution intervals can significantly reduce the effects of the RFI. As discussed by \citet{sob2019robust}, the preferred solution to tackle unflagged RFI is to use an iterative weighting approach which attempts to down-weigh RFI-corrupted visibilities during calibration. Hence, for the real datasets in the upcoming sections, we will employ the robust-2x2 solver in CubiCal \citep{kenyon2018cubical} to tackle any low-level unflagged RFI.  

Additionally, the presence of strong RFI in the raw data means the data has to be carefully flagged before any calibration. Unfortunately, the fraction of visibilities flagged per antenna, timestamp and frequency channel is not uniform. This leads to a situation whereby, once we have split the dataset into regular chunks, different chunks can end up with substantially different SNRs due to a varying flagged fraction. Hence, within the same dataset, it may be ``optimal'' to calibrate certain data chunks with shorter intervals, while other data chunks may require longer intervals. As we highlight throughout the following discussions, using variable solution intervals across a dataset may be a solution to this problem (despite the obvious difficulties of implementation). In $\S$\ref{meerkatdata}, we describe a detailed example of this in a real MeerKAT observation.

\section{Searching for optimal solution intervals}
\label{sec:interval_search}
After discussing the concept of solution intervals, and most of its effects in different calibration scenarios, we now present a brute force optimal solution interval search algorithm. Before describing the algorithm in $\S$\ref{infering}, we first discuss the idea of a \emph{calibration minor cycle} in $\S$\ref{ssec:minor_cycle}.
\subsection{Calibration minor cycle}\label{ssec:minor_cycle}
The process of finding an optimal solution interval for calibration requires repeating calibration numerous times and identifying the interval which produces the best residuals. Given the large size of RI datasets, this is not a practical approach, and any inference method which requires repeating the calibration and computing residual visibilities will be extremely inefficient. 

{\color{black}We suggest a framework termed the \emph{calibration minor cycle} in analogy to the major/minor cycles used by deconvolution algorithms such as \citet{clarkefficient1980}}. The minor cycle here refers to projecting the problem from the high dimensional visibility space to the low dimensional gain space for computational efficiency. If we solve for the gains using the full time and frequency resolutions, then, because of the high number of DoFs, the gain solutions will be very noisy. However, once we have a maximum likelihood estimate of $\g$, $\hat{\g}$, the equivalence
\begin{equation}
\P(\g|\V) = \frac{\P(\V|\g, \Lambda) \P(\g)}{\P(\V)} \approx \P(\g|\hat{\g}) = \frac{\P(\hat{\g}|\g, \Lambda_g) \P(\g)}{\P(\hat{\g})} \label{eq:v_g_proj}
\end{equation}
is approximately true. The reason for this is that $\hat{\g}$ is a stationary point of Eq. \eqref{eq_rime_2} and $\Lambda_g$ is the Cramer-Rao bound (i.e. defined by the inverse of Eq. \eqref{fisher}) and plays a similar role as the PSF during imaging. 
Hence, following Eq.~\eqref{eq:v_g_proj}, given an initial noisy estimate of the gains, the search for an optimal interval can be done directly in gain space without reincurring the computational cost of manipulating visibilities. Using the solution interval parametrisation below
\begin{equation}
\g_{p}(\thetab_p) = X \thetab_p, \label{sol_int_param}
\end{equation}
where $X$ is a suitable design matrix\footnote{See the Appendix \ref{sec:appendixC} for the explicit form of $X$.}, we seek to find
\begin{equation}
\underset{\thetab}{\mbox{min}}\;\chi^2, \quad \mbox{where} \quad \chi^2 = \left(\hat{\g} - X \thetab\right)^H \Lambda_{g}^{-1} \left(\hat{\g} - X \thetab \right),
\label{sol_int_param_chi2}
\end{equation}
i.e. we assume that the gain solutions are normally distributed around their true value (that we wish to approximate with a boxcar model) with noise drawn from $\mathrm{CN}\left(0, \Lambda_{g}\right)$. This formulation is per antenna and correlation, allowing one to parallelise over these dimensions. This is important because we have to solve the problem many times in order to find the optimal solution interval.

\subsection{Optimal solution interval search algorithm}\label{infering}
Following up from $\S$\ref{ssec:minor_cycle}, our goal is to find solution intervals for which the gains parametrised by Eq.~\eqref{sol_int_param} minimise the $\chi^2$ in Eq.~\eqref{sol_int_param_chi2}. This could be done by a full Bayesian evidence approximation,
but given how prohibitively expensive this is computationally, we suggest rather the following simplistic procedure to find adequate solution intervals for calibration.
 
The first step in the process is identifying a minimum solution interval for which the maximum likelihood problem gives a certain target SNR, $\mathrm{T}$, per antenna\footnote{This value being arbitrary and entirely up to the user to specify. The value of 3 is a traditional rule of thumb, and is used by calibration packages such as CASA \citep{mcmullin2007casa} as a threshold SNR value for identifying invalid solutions. See \citet{brogan2018advanced} for more discussions on how to choose this threshold.}. Thus, for a field with a peak flux of $P$, observed with an interferometer consisting of $N_{a}$ antennas, having visibilities with noise $\sigma_{\mathrm{rms}}$ per baseline, we want, from Eq. \eqref{vargp}, an interval satisfying
\begin{equation}
\sigma_{a} \leq \frac{P}{\mathrm{T}} \quad \mbox{where} \quad \sigma_{a} = \frac{\sigma_{\mathrm{rms}}}{P\sqrt{N_{a}-1}}.
\end{equation}
For a given interval containing $n = n_t n_\nu$ combined time and frequency grid points, we have,
\begin{equation}
\sigma_{a} = \frac{\sigma_{\mathrm{rms}}}{P\sqrt{n(N_{a}-1)}} \quad \Rightarrow \quad n \geq \frac{\mathrm{T}^2\sigma^2_{\mathrm{rms}}}{P^2(N_{a}-1)}. 
\label{selfcal_tip}
\end{equation}
Once the minimum solution interval has been obtained, the next step is to perform calibration with this minimum interval to obtain a maximum likelihood solution, $\hat{\thetab}$, and an updated noise covariance, $\hat{\Lambda}_g$. By using a relatively short interval, the maximum likelihood solution is unlikely to be significantly biased and can be used to map the problem into gain space. 

To avoid having to specify informative priors, we attempt to identify an easily computable statistical inference criterion for our model selection problem of finding the optimal interval given an initial maximum likelihood estimate. We choose the Akaike Information Criterion (AIC; see \citet{akaike1973information}) defined as \begin{equation} 
 \qquad \qquad \text{AIC}  = 2 N_{p} - \ln\left(\sum_i\frac{\vectb{r}_i^2}{\sigma_i^2}\right) + \frac{2N_{p}^{2} + 2N_{p}}{N_{g}-N_{p}-1}, \label{AIC}
 \end{equation}where $\vectb{r}$ is the difference between the estimated gains and reconstructed boxcar gain vector separated into its real and imaginary parts, $N_{g}$ and $N_{p}$ are the number of real terms in the gains at full resolution and the boxcar parameter vector respectively, and $\sigma_i$ is the uncertainty in the gain with index $i$.
 
The AIC is a model selection criterion that minimises the information loss by connecting the Kullback-Leibler measure (see \citet{kullback1951information}) and the maximum likelihood estimation method. The AIC is a robust statistic for this specific model selection problem because it corrects for the low number of DoFs and sample sizes. Precisely, the last term of Eq.~\eqref{AIC} is a penalty term which corrects for the low number of DoFs The best model is taken to be the one with the minimum AIC. We summarise the proposed search algorithm as follows:\\[0.05cm]
 \begin{enumerate}
    \vspace{-0.65cm}
     \item Read the observation parameters from a measurement set, i.e. integration time, channel width, number of channels, number of antennas, flags, and scans information.
     \item Estimate the noise in the visibilities from the data and model columns, or by using the System Equivalent Flux Density.
     \item Estimate the number of grid points per solution interval, $n$, necessary to reach a certain minimum SNR (for example, 3 or 5).
     \item Define the minimum solution interval by first choosing the largest frequency interval matching $n$ (i.e. set $n_\nu = n$ and $n_t = 1$). Ideally, anyone performing self-calibration would have already done bandpass calibration. Hence, the variations in frequency are expected to be slow compared to time. We can also do the contrary (i.e. set $n_\nu = 1$ and $n_t = n$) if we expect the specific gain to vary faster in frequency compared to time. A conservative approach will be to use $\sqrt{n}$ for both time and frequency (i.e. set $n_\nu = \sqrt{n}$ and $n_t = \sqrt{n}$). In the case $n>$ the total number of channels possible, $\mathrm{max}(n_\nu)$, then we set $n_\nu = \mathrm{max}(n_\nu)$ and $n_t = \ceil*{\frac{n}{\mathrm{max}(n_\nu)}}$, and likewise we will use $n_\nu = \ceil*{\frac{n}{\mathrm{max}(n_t)}}$ and $n_t = \mathrm{max}(n_t)$ when we expect more rapid variations in frequency compared to time.
     \item Perform calibration with the selected minimum interval and using the estimated gains, search for the optimal interval using the AIC.
 \end{enumerate}

\subsection{Verification}\label{multi_chan}
Before moving on to apply the proposed algorithm on a real dataset we first verify that it works on a simulated dataset. We replicate a VLA observation of the ``VIDEO'' field (J2000, RA=$2^{\mathrm{h}}11^{\mathrm{m}}21.09^{\mathrm{s}}$, Dec=$-04^\circ 11' 13.5''$) via simulations and use the proposed algorithm to find the optimal solution interval for DI amplitude and phase calibration of this dataset. After validating our approach on this simulated dataset, we will use it to find the optimal solution interval for the real observation. 

The observation in question covers a frequency range of 1--2 GHz split into 16 spectral windows with 64 channels each. The integration time was $\approx$ 9 seconds on average. 28 antennas were used for the observation, with a maximum baseline of 36.4 km. The target field was observed in 9 different scans with three groups of consecutive scans which can be calibrated as single chunks. The dataset has had initial flagging and (1GC) calibration applied to it in the standard way using CASA (see \citet{IanVIDEO} for more details).

We image the 1GC-corrected data and extract a component-based sky model using PyBDSF \citep{mohan2015pybdsf}. The sky model consists of over 200 Gaussian sources with the brightest source of $\approx$ 0.02 Jy using an island and pixel threshold of 5 and 10 $\sigma$ for the source extraction using PyBDSF. Note that this is an apparent sky model since no primary beam correction has been done. We then take this sky model as the ground truth for the simulation.

We use the MeqTrees software package \citep{noordam2010meqtrees} to replicate the observation and simulate visibilities corresponding to our apparent sky model. We corrupt the simulated visibilities with gains drawn from a GP with a Mat\'ern covariance function\footnote{\color{black}The following parameters are used. Amplitude: $(\sigma_f, \mu, l)$ = $(0.5, 3/2, 0.2)$ for time and $(\sigma_f, \mu, l)$ = $(0.1, 7/2, 0.5)$ for frequency. Phase: $(\sigma_f, \mu, l)$ = $(0.5, 3/2, 0.08)$ for time and $(\sigma_f, \mu, l)$ = $(0.1, 7/2, 0.5)$ for frequency. Here, the units of the length, $l$, have been normalised to the range $[0,1]$. See Appendix \ref{sec:appendixD} for the definition of the Mat\'ern covariance function.}. The gains are constructed to have rapidly varying phases and slowly varying amplitudes with a shorter length scale in time than frequency. We add Gaussian noise with an rms value of 0.16 Jy to the corrupted visibilities. The added rms corresponds to the estimated rms from the real dataset. The real data also contains RFI, but we do not include RFI in the simulation. Note that the effect of RFI in this dataset is discussed in a previous work \citep{sob2019robust}.

Following the steps described in $\S$\ref{infering}, we first compute the solution interval that provides a minimum SNR of at least 3. For this computation, we use the simulated noise rms of 0.16 Jy and estimate the peak flux using Eq. \eqref{fppinv-3}. The latter is used to estimate the peak flux since the sky model consists of numerous sources. The estimated peak flux is 0.029 Jy. Using Eq. \eqref{selfcal_tip}, we get the minimum combined frequency and time interval, $n$, to be 64 using a per antenna SNR threshold of 3. Because the gains are slowly varying in frequency compared to time, we set the minimum frequency and time intervals to 64 channels (64 MHz) and 1 timeslot (9 secs) respectively. 

\begin{figure*}
 \begin{subfigure}{0.45\textwidth}
	\includegraphics[width=\linewidth]{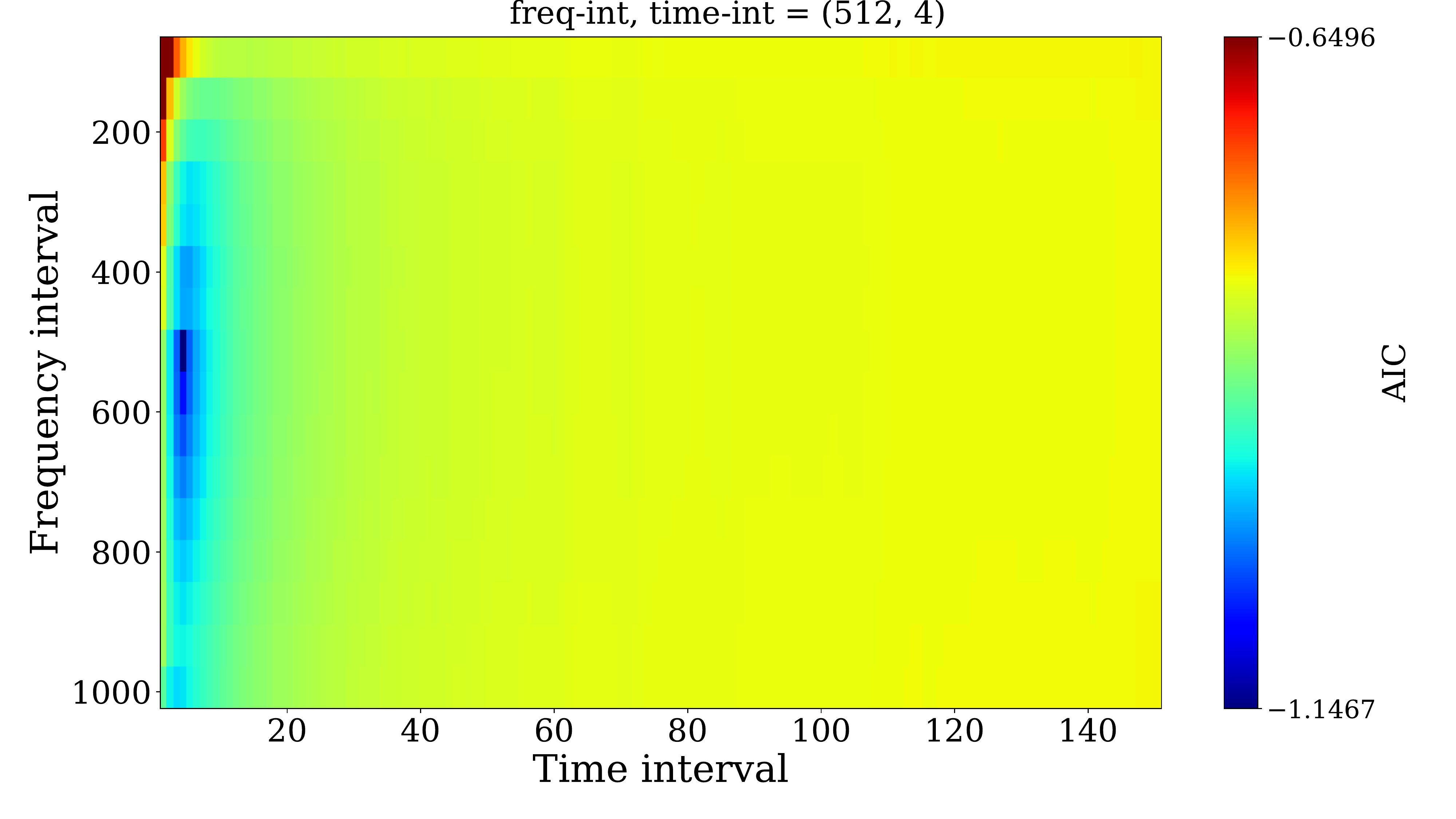}
    \caption{AIC}
    \label{video_sim_chidof}
 \end{subfigure}
 \begin{subfigure}{0.45\textwidth}
	\includegraphics[width=\linewidth]{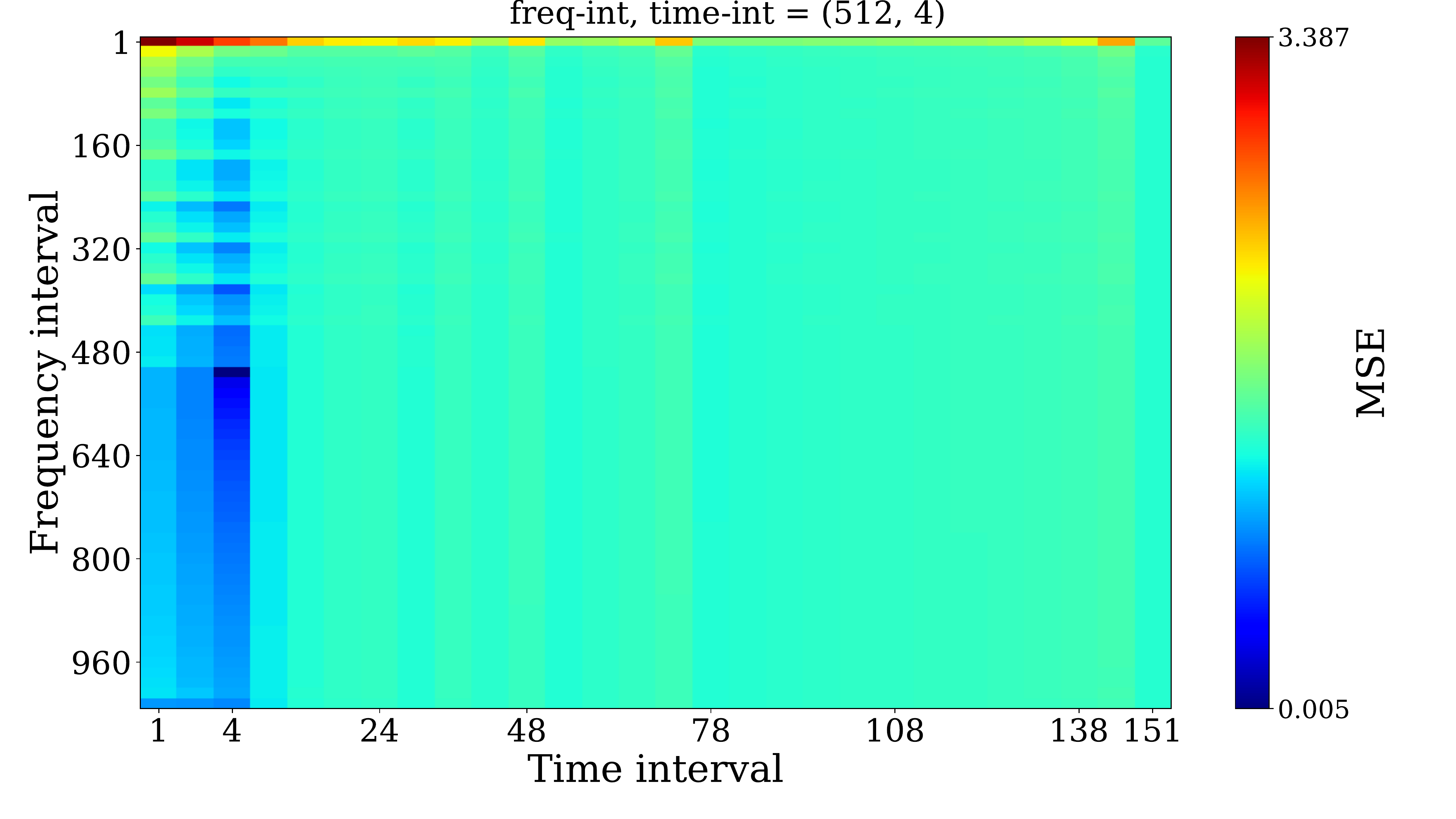}
    \caption{MSE}
    \label{video_chidof}
 \end{subfigure}
 \caption[AIC of ``VIDEO'' field]{Images of the AIC and the MSE of the gains for the simulated VLA ``VIDEO'' datasets on a two dimensional grid of time and frequency intervals. The optimal solution interval suggested by the AIC and the MSE of the gains is 4 (36 secs) and 512 MHz for time and frequency respectively.}
 \label{2d_chidof}
\end{figure*}

We perform calibration with this interval and use the estimated gains to search for the optimal time and frequency interval with the AIC. Fig. \ref{2d_chidof} shows images of the AIC and the MSE of the gains computed on a two dimensional grid of time and frequency intervals. For computational reasons, we can not evaluate the MSE of the gains on a full grid since this requires repeatedly calibrating the data. Instead, we use a grid with a frequency interval step size of 12 and time interval step size of 6.  Because the AIC suggests an optimal solution interval of 512 and 4 for frequency and time, respectively, we add the time intervals 2 and 4 to our grid. The MSE plot agrees with the AIC, and both metrics suggest the optimal interval to be 512 (512 MHz) in frequency and 4 (36 secs) in time.

\begin{figure*}
\begin{subfigure}{.45\textwidth}
\centering
  \includegraphics[width=0.8\linewidth]{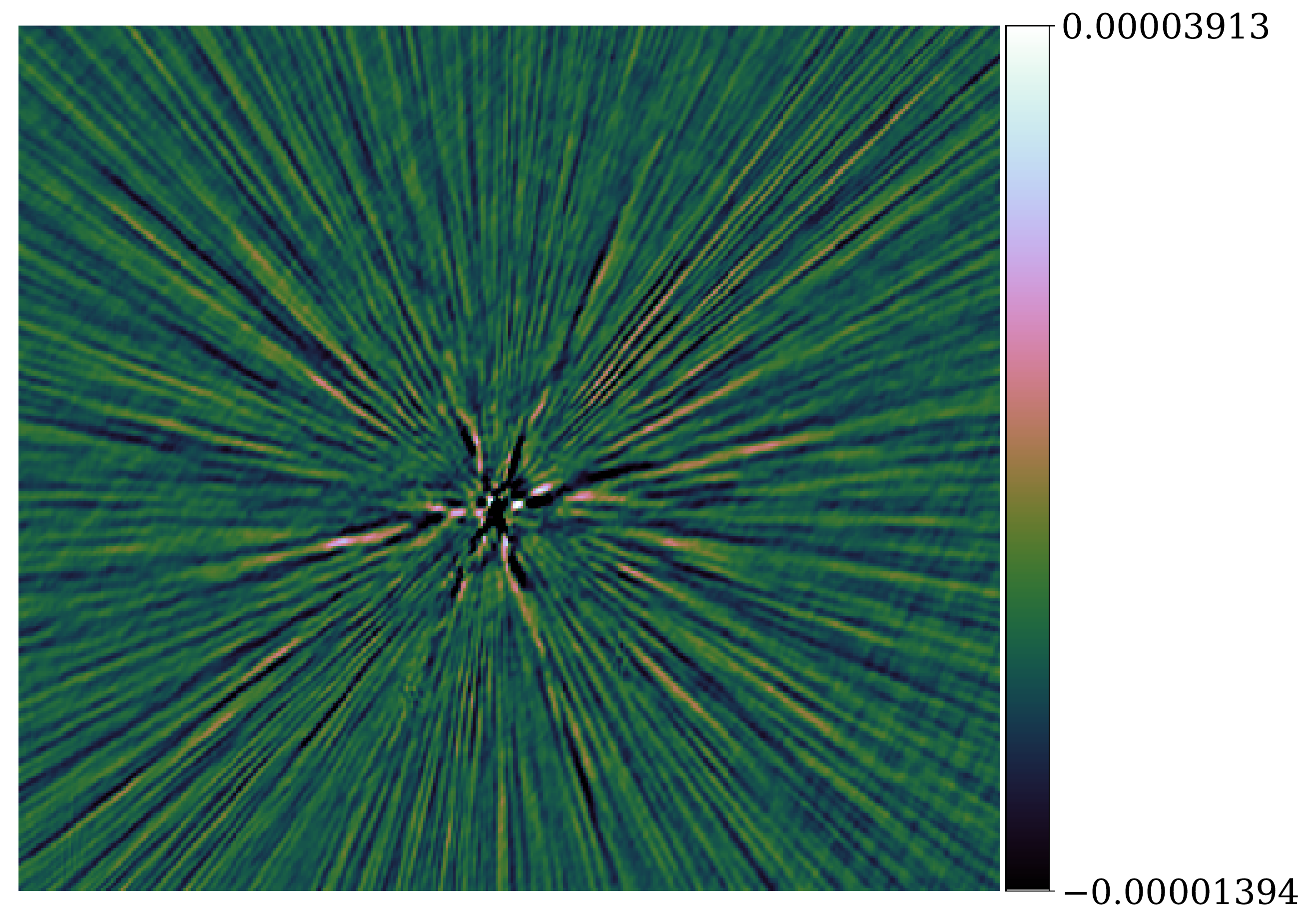} 
\caption{before calibration}\label{video_sim}
 \end{subfigure}
\begin{subfigure}{.45\textwidth}
\centering
  \includegraphics[width=0.8\linewidth]{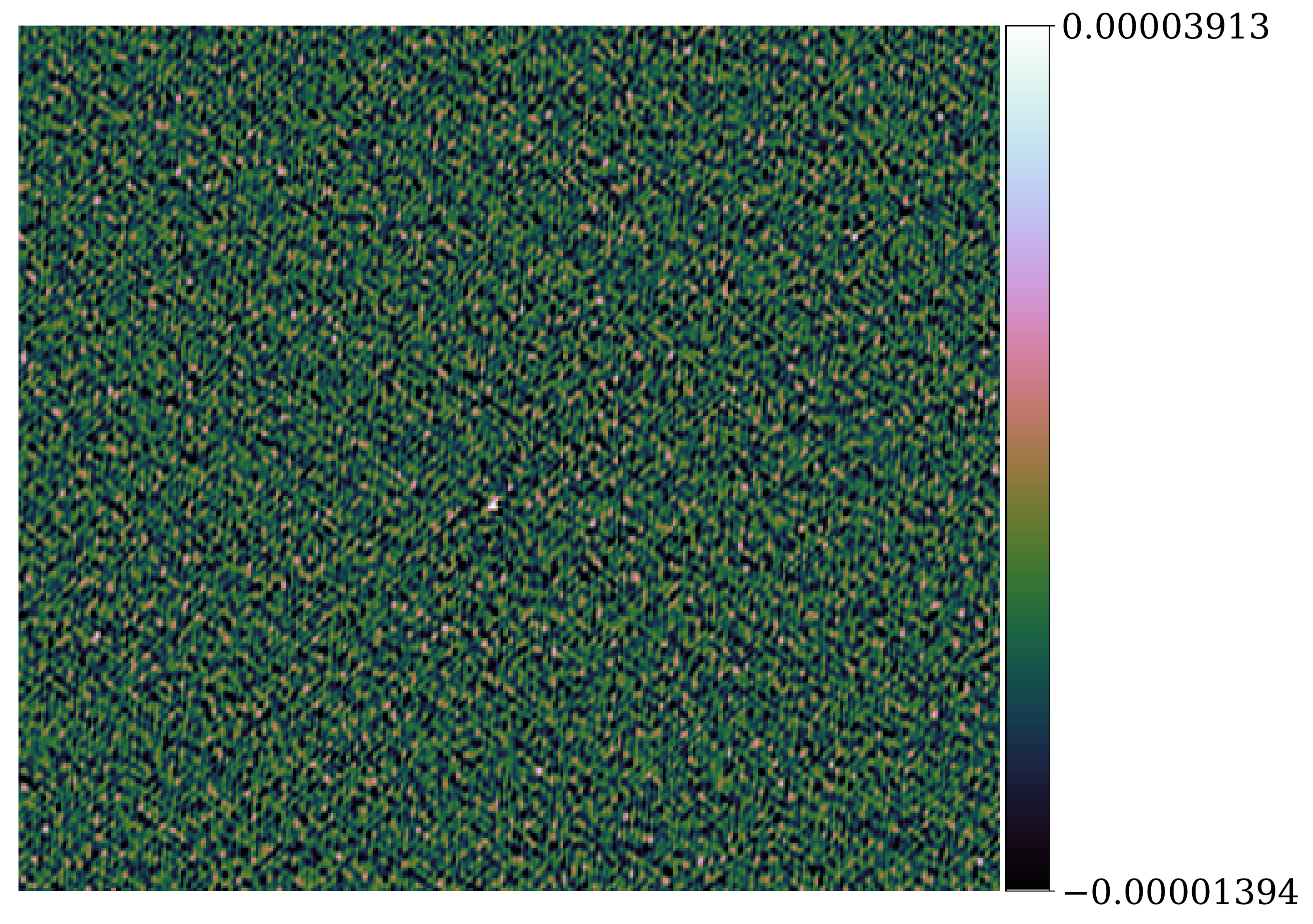}
    \caption{freq-int = 64 MHz, time-int = 9 secs} \label{complex_tint1a}
 \end{subfigure}
\begin{subfigure}{.45\textwidth}
\centering
  \includegraphics[width=0.8\linewidth]{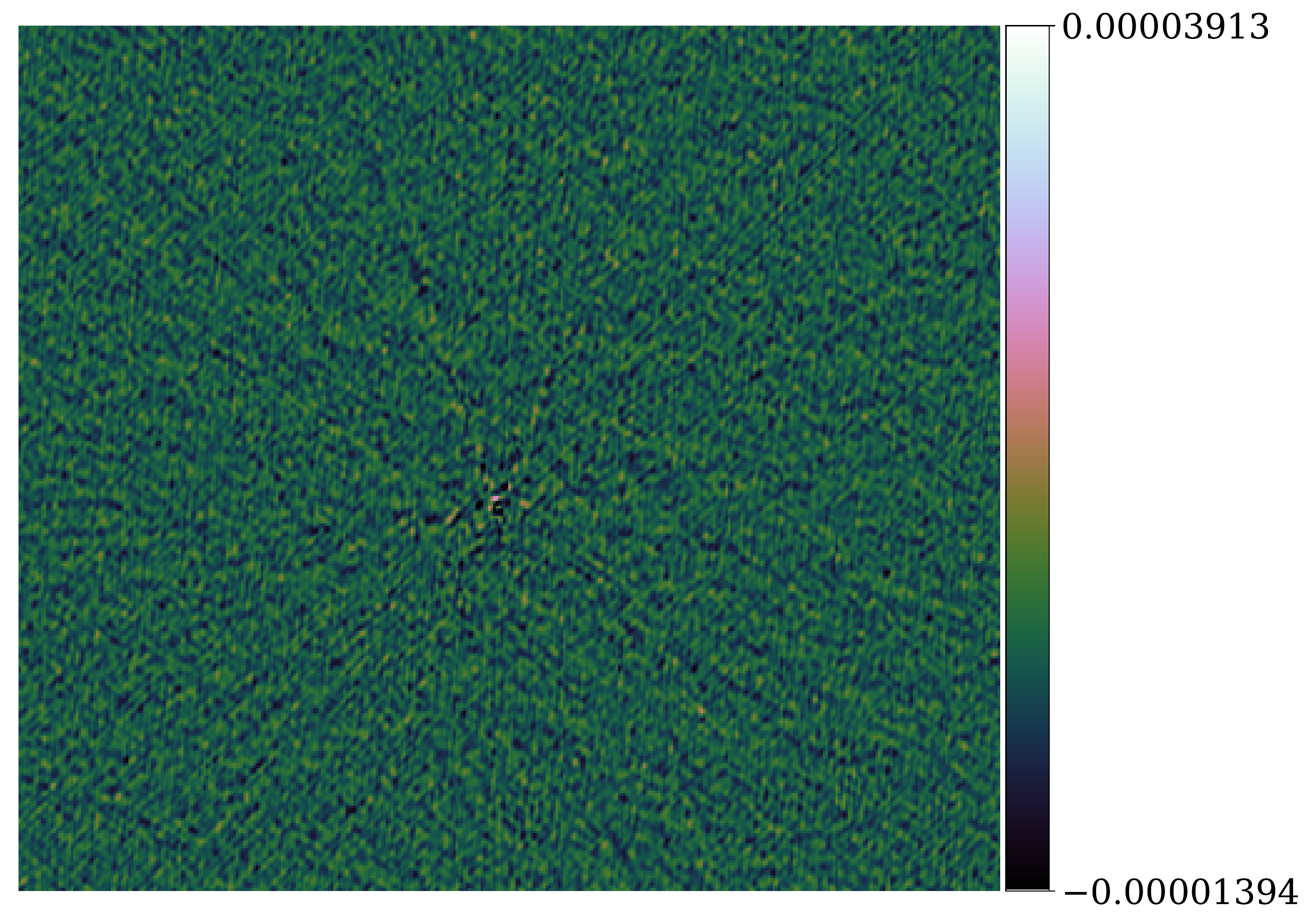}
\caption{freq-int= 512 MHz, time-int = 36 secs}\label{complex_tint24}
 \end{subfigure}
\begin{subfigure}{.45\textwidth}
\centering
 \includegraphics[width=0.8\linewidth]{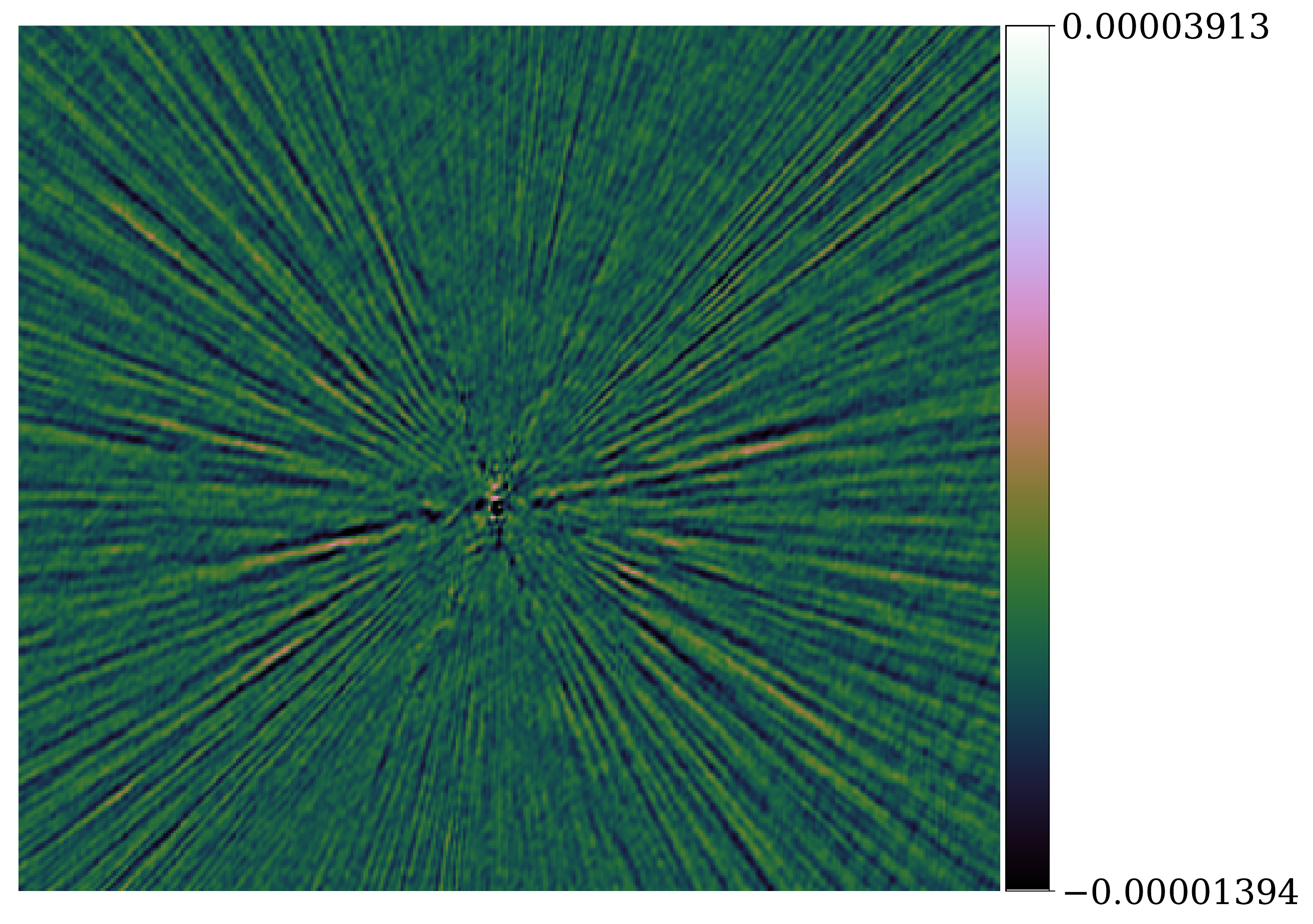}
\caption{freq-int = 512 MHz, time-int = 23 mins} \label{complex_tint222}
 \end{subfigure}
 \caption[Residual images of the simulated ``VIDEO'' field]{Residuals images for a patch at the field centre of the simulated ``VIDEO'' field. (a) corrupted visibilities minus model visibilities, showing the artefacts introduced by the gains. (b), (c) and (d) are images of corrected visibilities minus model visibilities, at different time and frequency solution intervals.}
\label{video_sim images}
\end{figure*}

We show residual images after calibration (similar to Fig. \ref{fig:gains_corr_vis}) for a patch of sky at the field centre in Fig. \ref{video_sim images}. Fig. \ref{video_sim} shows the artefacts introduced by the gains (i.e. an image of the difference between the corrupted and the model visibilities). The remaining images neatly demonstrate the different calibration regimes induced by different choices of solution intervals.  
Fig. \ref{complex_tint1a} is the residual image obtained with the shortest possible interval (frequency interval = 64, time interval = 1); this appears noisier compared to all the other images because here the solutions are in a low SNR regime. 
At the suggested optimal interval (frequency interval = 512, time interval = 4) (see Fig. \ref{complex_tint24}), we see that the artefacts have been considerably attenuated and the noise in the image appears much lower. Fig. \ref{complex_tint222} shows the result with a longer time interval (frequency interval = 512, time interval = 151). Since this cannot track the actual variations in the gain, the image retains most of the gain-induced artefacts.

These images confirm that the solution interval of 512 MHz in frequency and 36 secs in time are near optimal in this context for this particular simulated dataset. It is important to note that such a brute force technique only suggests appropriate solution intervals for a given dataset. Depending on the variability of gains and noise in the data, different combinations of time and frequency intervals may be optimal for a given dataset. For example, if we have fairly constant gains, a high SNR, and no significant unmodelled flux, then using long or short solution intervals will not produce significantly different results. 

Our experiment confirms that it is possible to determine the optimal solution intervals using the AIC computed from the gain solutions [at the shortest possible intervals] and their boxcar approximations. Appendix~\ref{results} presents supplementary simulations further emphasizing this result. 

\section{Application to Real Data}\label{realdata}
Following the successful application of the proposed method to synthetic data, in this section we test its performance on real datasets. The first dataset we consider is a VLA observation of the same ``VIDEO'' field that we simulated above. The second dataset is a MeerKAT observation for which we discuss how the algorithm can be applied for DD calibration. Furthermore, this section highlights prominent practical difficulties we may encounter with real data.

\subsection{VLA VIDEO observation}\label{real_video_data}
This section focusses on the calibration of the real ``VIDEO'' dataset. As noted earlier, the dataset has already had 1GC calibration applied to it. This section concerns itself with the choice of optimum solution intervals for the subsequent self-calibration steps. 
Firstly, like any real observation, this dataset contains flagged visibilities due to RFI, as well as unflagged low-level RFI. The RFI occupancy is not necessarily constant in time. For these (and perhaps other) reasons, the noise level is not entirely constant over time and frequency. Hence, ideally for a real dataset the rms needs to be computed separately for every antenna and frequency channel to account for the percentage of visibilities flagged in the individual solution bins. 

However, such an approach is not always feasible in practice. Hence, we suggest using a single data chunk for this computation. For this purpose, we split the data in chunks of 128 channels and 64 timestamps. We then select the data chunk whose flag percentage is closest to the mean flag percentage of the chunks to compute the minimum required solution interval.

\begin{figure*}
 \begin{subfigure}{0.45\textwidth}
 \centering
	\includegraphics[width=\columnwidth]{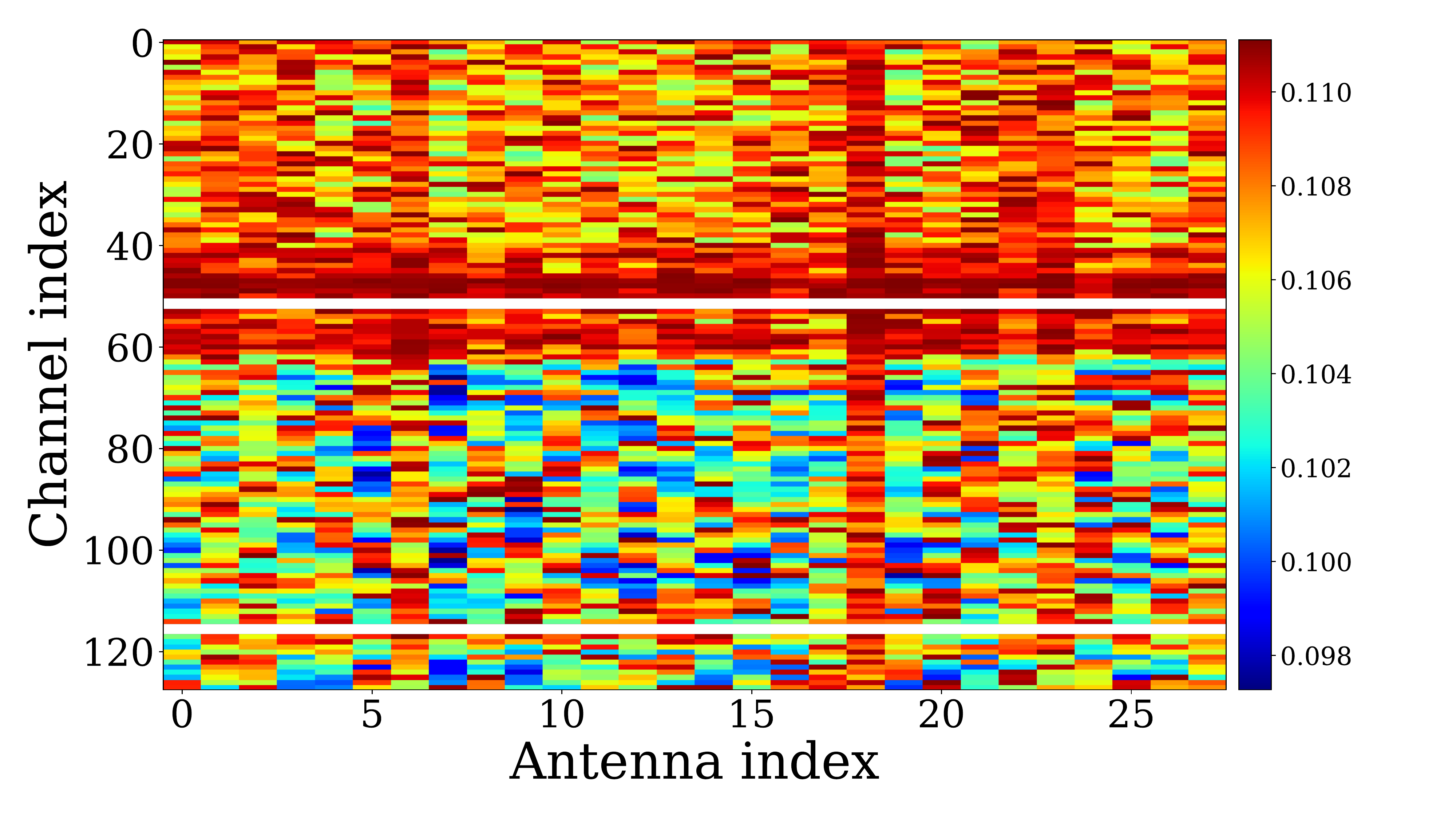}
    \caption{Channel rms per antenna}
    \label{rms_scan_freq}
 \end{subfigure}
 \begin{subfigure}{0.45\textwidth}
 \centering
	\includegraphics[width=\columnwidth]{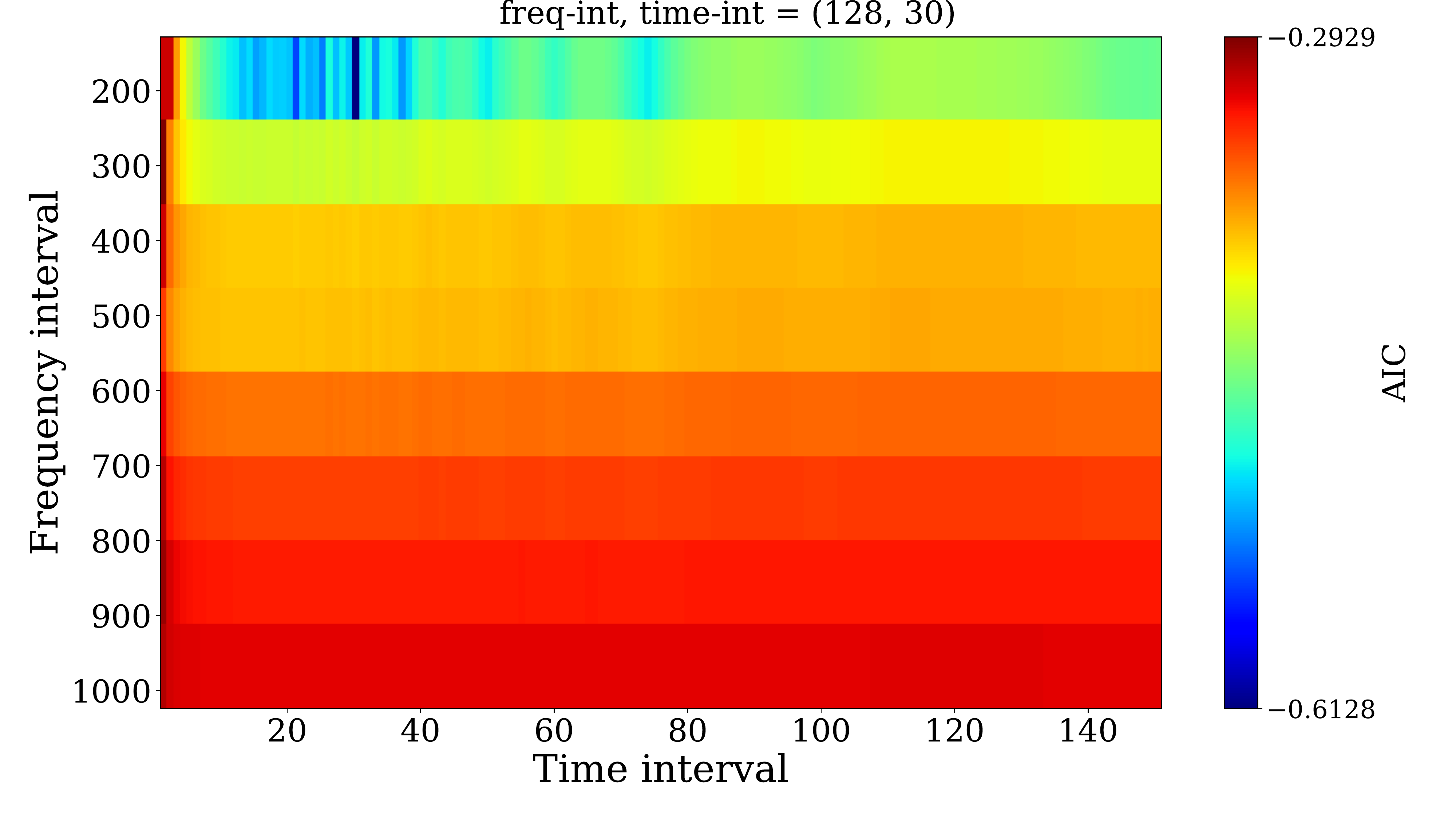} 
    \caption{AIC for the real dataset}
    \label{real_aic}
 \end{subfigure}
 \caption[``VIDEO'' field rms per channel]{(a) is the estimated rms per channel for each antenna. The white stripes represent completely flagged antennas. The peak estimated rms is $\approx$ 0.16 Jy. (b) is a plot of the AIC for the real ``VIDEO'' dataset at different frequency and time intervals suggesting an optimal solution interval of 128 MHz in frequency and 30 (4.5 mins) in time.}
 \label{2d_rms}
\end{figure*}

Fig.~\ref{rms_scan_freq} shows the estimated rms per channel and antenna for the selected data chunk. For this specific dataset the flagging is fairly uniform across the antennas and channels, hence the rms does not vary much. Thus we can use a single rms value to estimate the minimum required interval. We use an SNR of 5 because of the flagged data (this reduces the SNR because the valid number of visibilities at each time is now less than what it should be) and the peak rms from Fig.~\ref{rms_scan_freq} to compute the minimum combined interval. We find the minimum combined interval, $n$, to be 128. Similarly, as in $\S$\ref{multi_chan}, because the data is already bandpass calibrated, we set the frequency interval equal to the combined interval, i.e. we set the frequency interval to 128 MHz and the initial time interval to 1 timeslot (9 secs). Using this solution interval, we perform amplitude and phase calibration on the data and use the AIC of the estimated gains to search for the optimal solution interval.  We show the AIC obtained from the estimated gains with the minimum possible interval in Fig. \ref{real_aic}. The suggested optimal interval is 128 MHz in frequency and 30 (4.5 mins) in time.

\begin{figure*}
\begin{subfigure}{.45\textwidth}
\centering
  \includegraphics[width=0.8\linewidth]{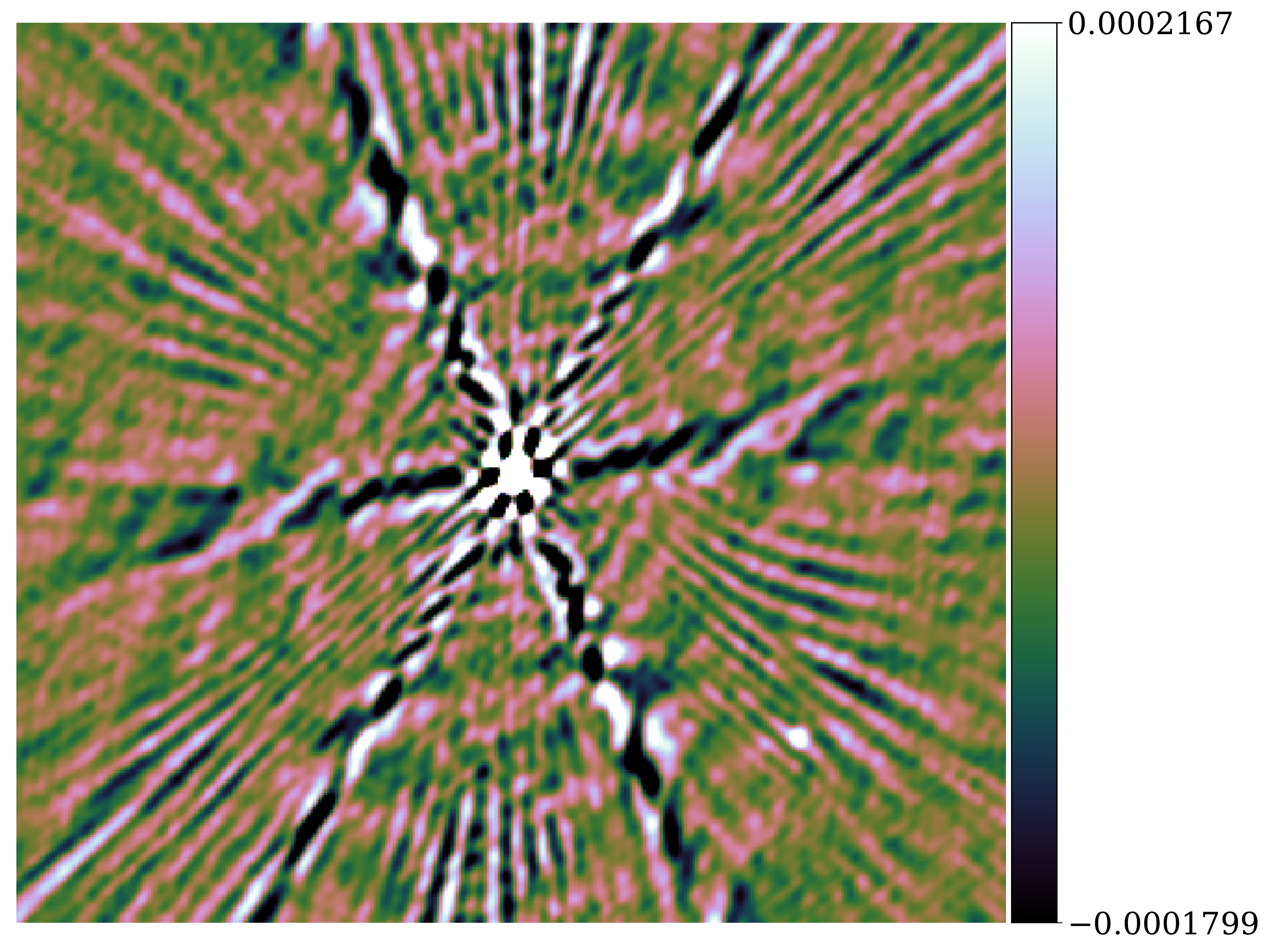} 
\caption{before calibration}\label{video_real}
 \end{subfigure}
\begin{subfigure}{.45\textwidth}
\centering
  \includegraphics[width=0.8\linewidth]{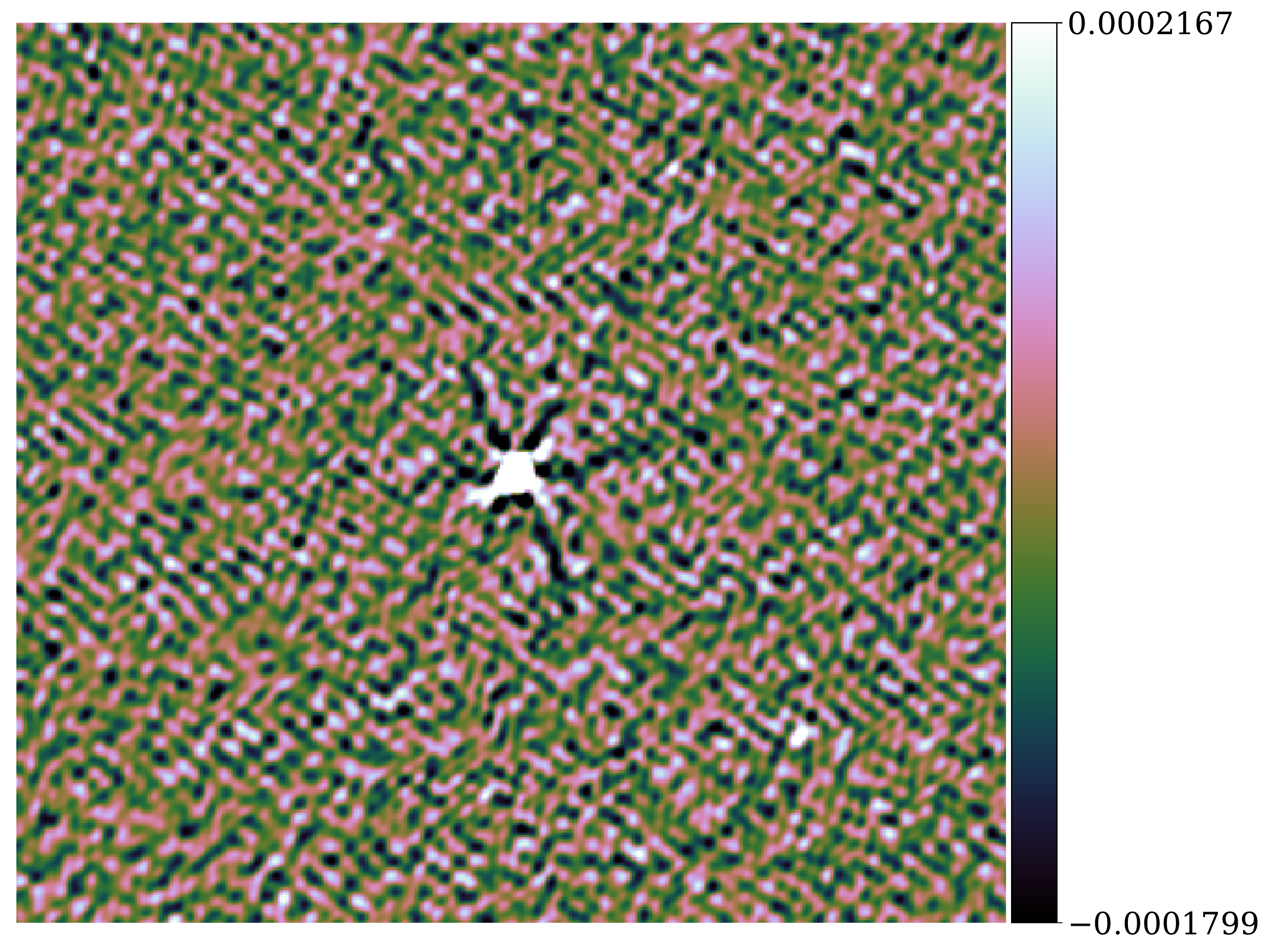} 
    \caption{freq-int = 128 MHz, time-int = 9 secs} \label{real_tint1}
 \end{subfigure}
\begin{subfigure}{.45\textwidth}
\centering
  \includegraphics[width=0.8\linewidth]{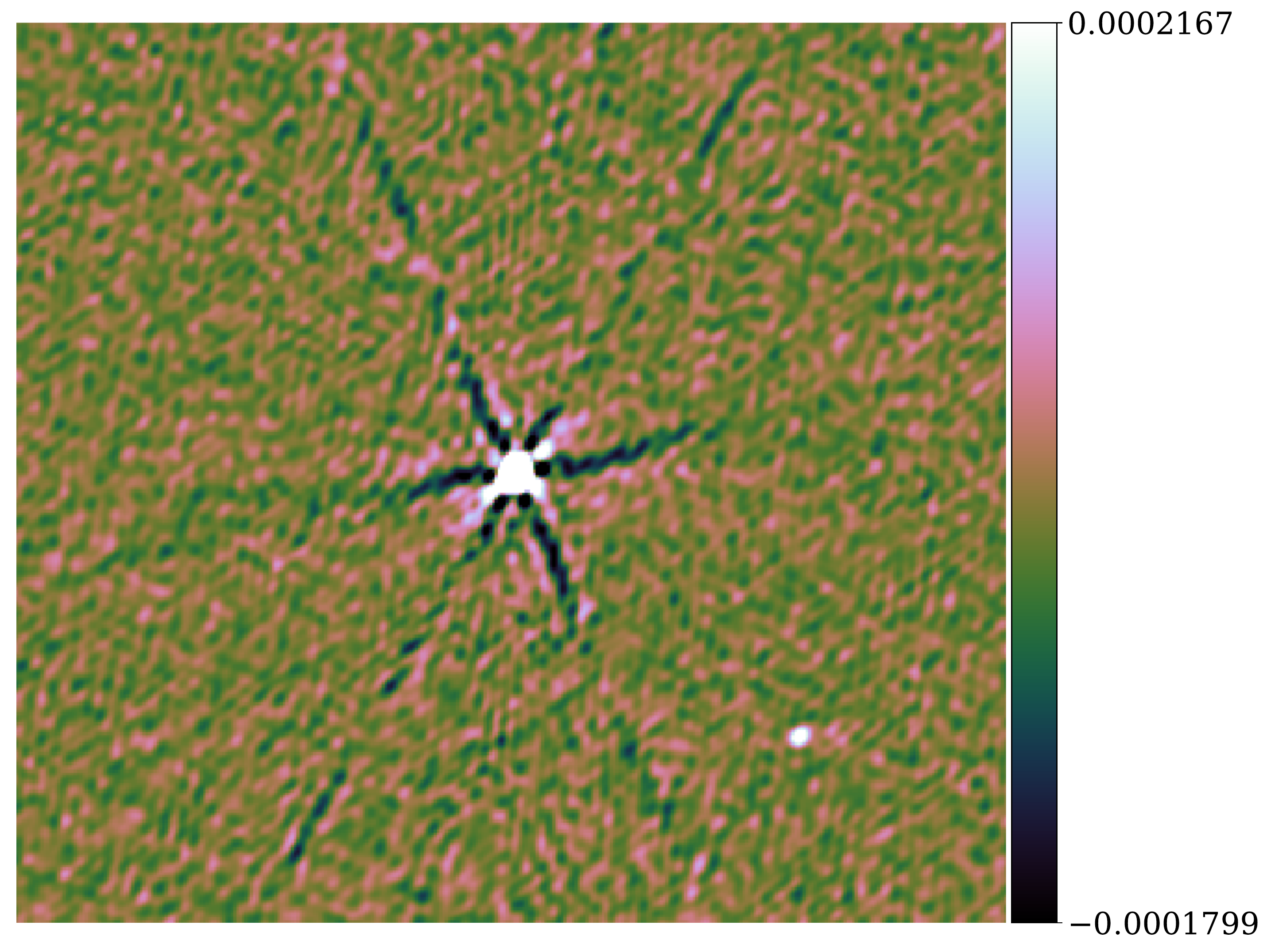} 
\caption{freq-int = 128 MHz, time-int = 4.5 mins}\label{real_tint11}
 \end{subfigure}
\begin{subfigure}{.45\textwidth}
\centering
 \includegraphics[width=0.8\linewidth]{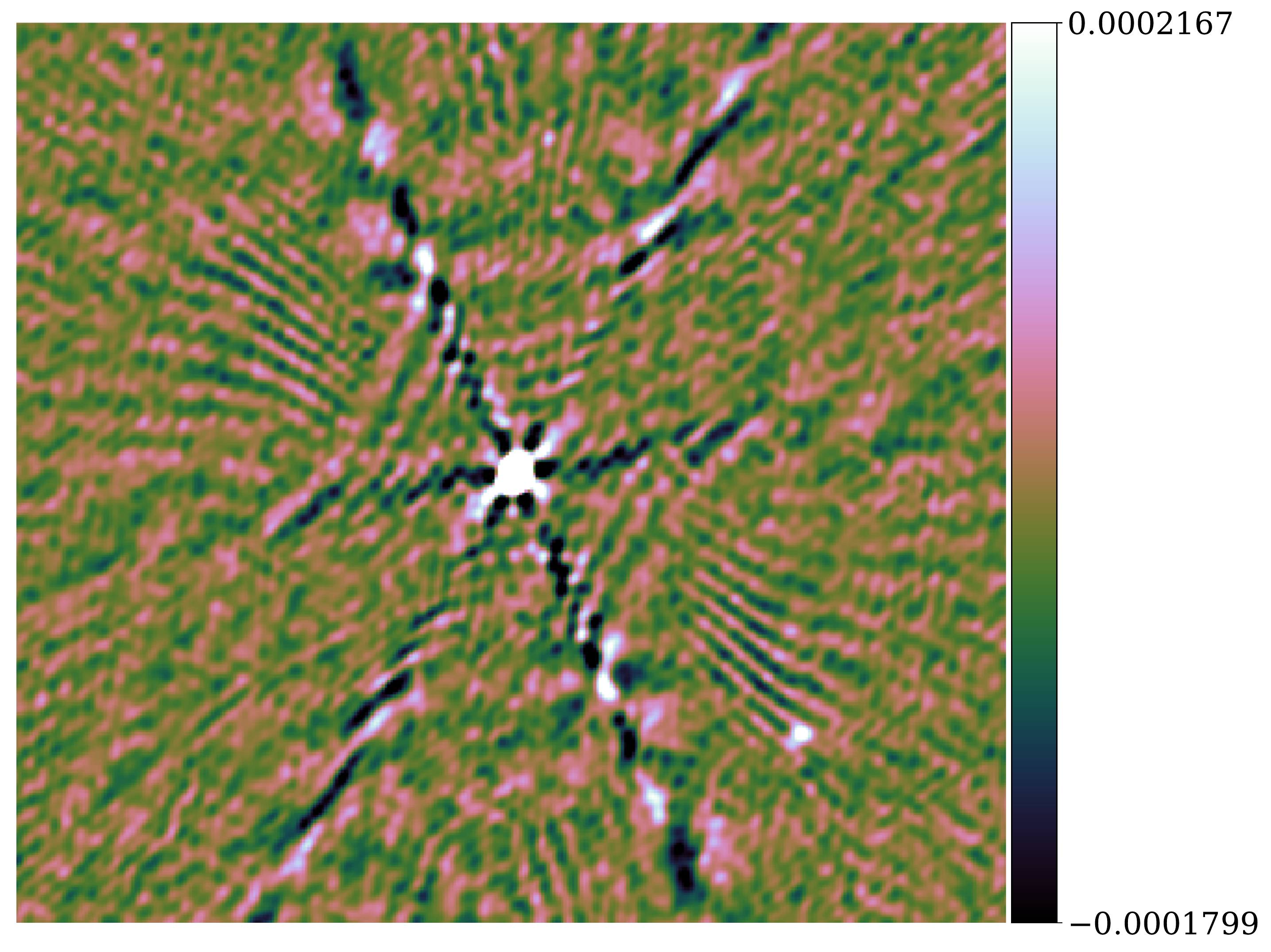}
\caption{freq-int = 512 MHz, time-int = 23 mins} \label{real_tint222}
 \end{subfigure}
 \caption[Residual images of the real``VIDEO'' observation]{Images for a small
 patch at the centre of the field for the real ``VIDEO'' observation. (a) before self-calibration. (b), (c) and (d) after self-calibration with a frequency interval of 128 MHz and a time interval of 9 secs, a frequency interval of 128 MHz and a time interval of 4.5 mins, and a frequency interval of 512 MHz and a time interval of 23 mins respectively.
}
\label{video_real images}
\end{figure*}

Fig.~\ref{video_real images} shows images of the same patch of sky as in $\S$\ref{multi_chan} for the initial (1GC-calibrated) data, and the corrected data after self-calibration. Note that every calibration in this section is performed using the robust solver in CubiCal \citep{sob2019robust}, in order to mitigate the unflagged RFI still present in the data. These images show the same trends that we observed in the simulations. Here, the artefacts are more pronounced compared to the simulated data (see Fig. \ref{video_sim images}), which we attribute to DD effects (DDEs) due to the primary beam (note that these effects were not included in the simulated data). 

The results demonstrate the ``optimality'' of the chosen solution intervals, in terms of trading off artefact suppression versus increased noise. At the shortest possible interval (128 MHz, 9 secs), even though the artefacts appear to have been removed, the noise is higher. Also, some effects of low-level RFI are visible in the map (see Fig.~\ref{real_tint1}). At the optimal interval (128 MHz, 4.5 mins), the artefacts are attenuated, and the noise is lower (see Fig.~\ref{real_tint11}). Finally, at the longest solution interval (512 MHz, 23 mins), the artefacts dominate (see Fig.~\ref{real_tint222}).
\subsection{MeerKAT observations}\label{meerkatdata}
We now attempt to search for optimal DD solution intervals on a MeerKAT dataset with a substantial amount of unflagged RFI.

The dataset in question is a 3 hour observation of the field centred on the source 4C12.02 (J2000, RA=$0^{\mathrm{h}}4^{\mathrm{m}}50.2^{\mathrm{s}}$, Dec=$12^\circ48'40''$). As in the previous example, the dataset has already had standard 1GC and flagging applied to it, but, as we shall see, some substantial RFI remains unflagged. This could probably be improved by redoing the 1GC and flagging more carefully, but this is outside the scope of this work. Rather, here, we treat the data as it is, and proceed to do self-cal and DD calibration. The field contains a bright source to the North-West. Due to DDEs, this source induces artefacts that dominate the background of resulting image (see Fig. \ref{before_peeling}). We proceed to solve for a DD gain term in the direction of the troublesome source and subtract (peel) it from the field. Hence, we consider the following RIME model (for $N_d$ directions),
\begin{equation}
\V_{pq} =  \G_{p} \left(\sum_{d=1}^{N_d} \E_{dp}(\C_{pqd})\EH_{dq}\right) \G_{q}^{H},
\label{selfcal_9}
\end{equation}
where $\E$ denotes DD gains and all other terms are defined as before.
\subsubsection{Direction Dependent Models}\label{DD_model_peeling}
All the examples and simulations we considered in the preceding sections were for DI calibration. For DD calibration, longer solution intervals are generally employed, not only to improve the SNR, but, more importantly, to make the calibration problem well-posed since the number of unknown parameters increases rapidly with the number of directions. While DDEs (for example, primary beam effects) may still vary on relatively short frequency and time scales, short-interval DD gain solutions are only practical in the direction of very bright sources. This approach was originally pioneered as ``peeling'' by \citet{noordam2004lofar}. 

In the DD case, we have a set of separate model visibilities for every ``direction'' (i.e. every independent DD gain solution). Eq. \eqref{vargp} can still be used to compute the minimum solution interval required to attain a certain SNR. To achieve the required SNR in all directions, we need to base the computation on the directional model with the lowest SNR. Our model  for the 4C12.02 field consists of two directions: one for the troublesome source, and one for the rest of the field\footnote{The measurement set has a MODEL_DATA column which contains visibilities for all the sources in our current sky model for the field. Additionally, visibilities for the troublesome source have been computed separately and stored in a new measurement set column called PEELED_MODEL_DATA. We compute the model visibilities for the rest of the field by subtracting the PEEL_MODEL_DATA column from the MODEL_DATA column.}. The latter is the weakest model and is used in this section to compute the optimal solution intervals.

\subsubsection{Effects of RFI and data flags}\label{RFI_flags_peeling}
In contrast to the ``VIDEO'' dataset, this dataset still contains some noticeable RFI.
Potentially, this can be fixed by more careful flagging in the preprocessing. However, in the age of high data rates and, consequently, more and more automated pipelines, such careful preprocessing of each and every observation is no longer feasible. We therefore try to treat such residual RFI as a given, and attempt to design a calibration procedure that is robust against it.

\begin{figure*} 
\begin{subfigure}{0.32\textwidth}
\centering
  \includegraphics[width=\linewidth]{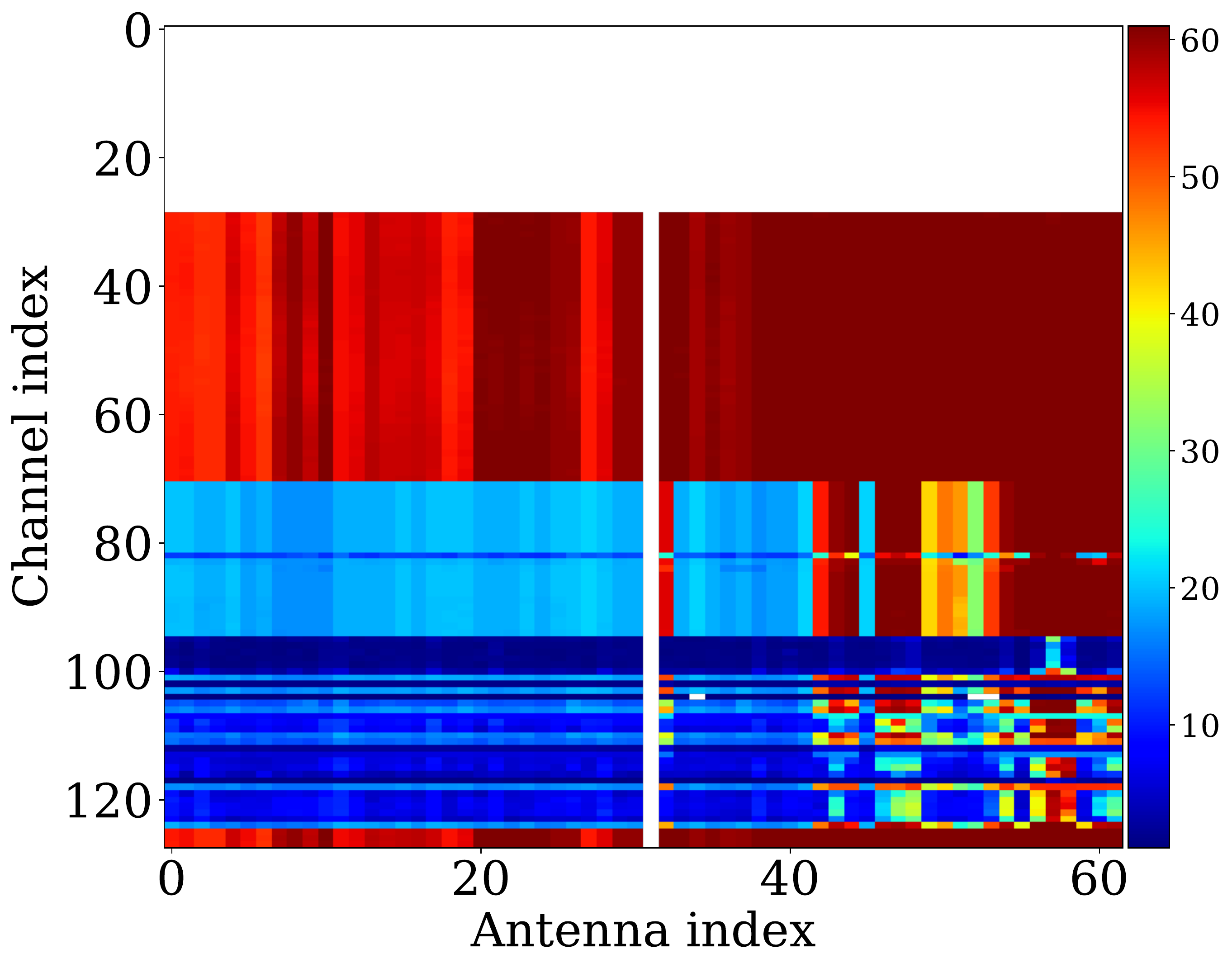}
  \caption{Antennas count}\label{fig:ant_c}
 \end{subfigure}\hspace{\fill}
 \begin{subfigure}{0.32\textwidth}
 \centering
  \includegraphics[width=\linewidth]{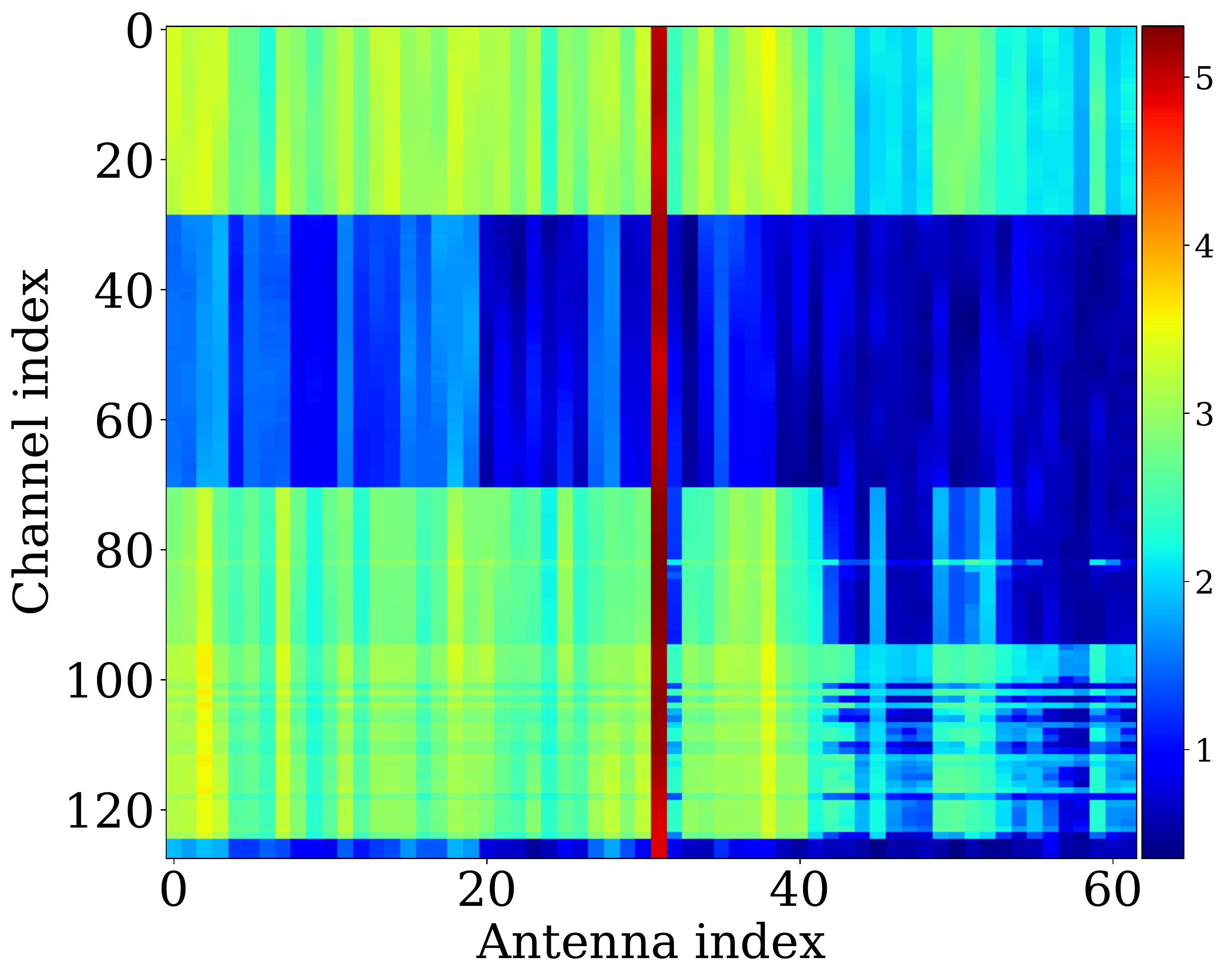}
  \caption{Channel rms per antenna}\label{fig:rms}
 \end{subfigure}\hspace{\fill}
 \begin{subfigure}{0.32\textwidth}
 \centering
  \includegraphics[width=\linewidth]{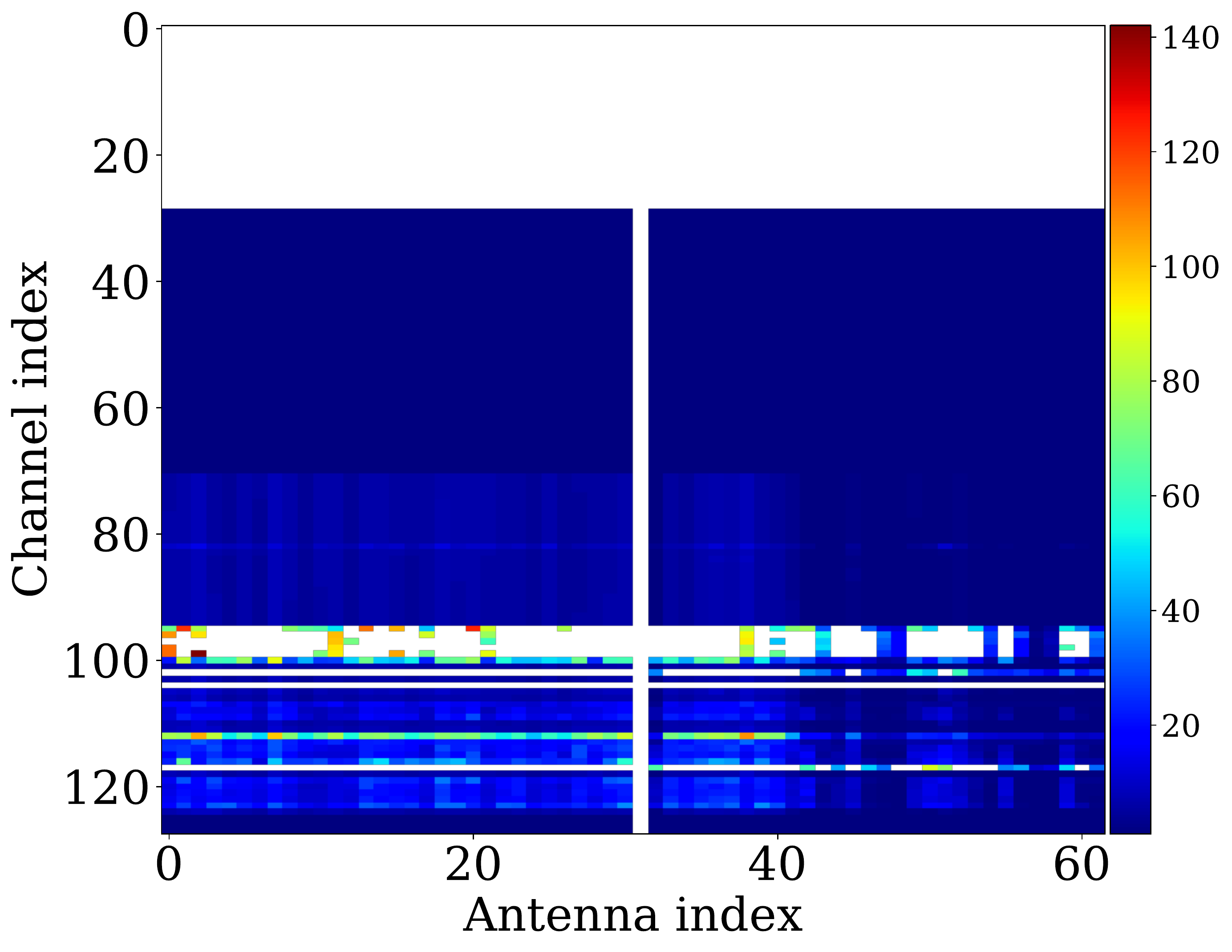}
  \caption{Solution intervals}\label{fig:nv_nt}
 \end{subfigure}\hspace{\fill}
\caption{Plots of the estimated antenna counts per channel for each antenna (a), rms per channel for each antenna (b) and the minimum required solution for each antenna. The white stripes represent flagged antennas. Every antenna having a count less than 2 in solution block is flagged. Here, we can observe the discrepancies in the size of solution intervals required for different solution bins for the same data chunk.}
\label{fig:intervals-stats}
\end{figure*}

Furthermore, the fraction of visibilities flagged (and thus SNR) per antenna, timestamp and frequency channel is not uniform. In contrast to the ``VIDEO'' dataset in $\S$\ref{real_video_data}, we can not use a single rms value to estimate the minimum required solution interval for the entire dataset. As in $\S$\ref{real_video_data}, we split the data in chunks of 128 channels and 64 timestamps and select a specific chunk for which we compute the minimum required solution intervals separately for each antenna and frequency channel. We show the results in Fig. \ref{fig:intervals-stats}. 

Fig. \ref{fig:nv_nt} shows the minimum required interval per antenna and channel. We can observe from this figure that most of the antennas only require a combined time and frequency interval of $N<20$. Some even have enough SNR for $N=1$, while some need $N>140$. To understand why we would need such long intervals for certain antennas, we look at Fig. \ref{fig:ant_c} and Fig. \ref{fig:rms}. Fig. \ref{fig:ant_c} is a plot of antenna count, i.e. the minimum number of valid baselines containing every antenna per channels (this is an average for all the timestamps). This plot shows the discrepancies in the antenna counts due to flagging. While some antennas have counts $>60$, others have counts of $<10$. Similarly, Fig. \ref{fig:rms} shows how the estimated rms varies for the different antennas and channels. In contrast to Fig. \ref{rms_scan_freq}, this plot shows large variations across the different channels and antennas. Hence, the suggested approach is to compute the intervals separately per antenna and channel, and select the largest one for calibration. If we estimate a single solution interval for the whole data chunk by computing a single rms value and number of antennas for the whole data chunk, we will infer a short solution interval. This short solution interval will be enough for most solution bins, but certain solution bins will not have sufficient SNR. 

\begin{figure*}
\begin{subfigure}{.45\textwidth}
 \centering
  \includegraphics[width=0.8\linewidth]{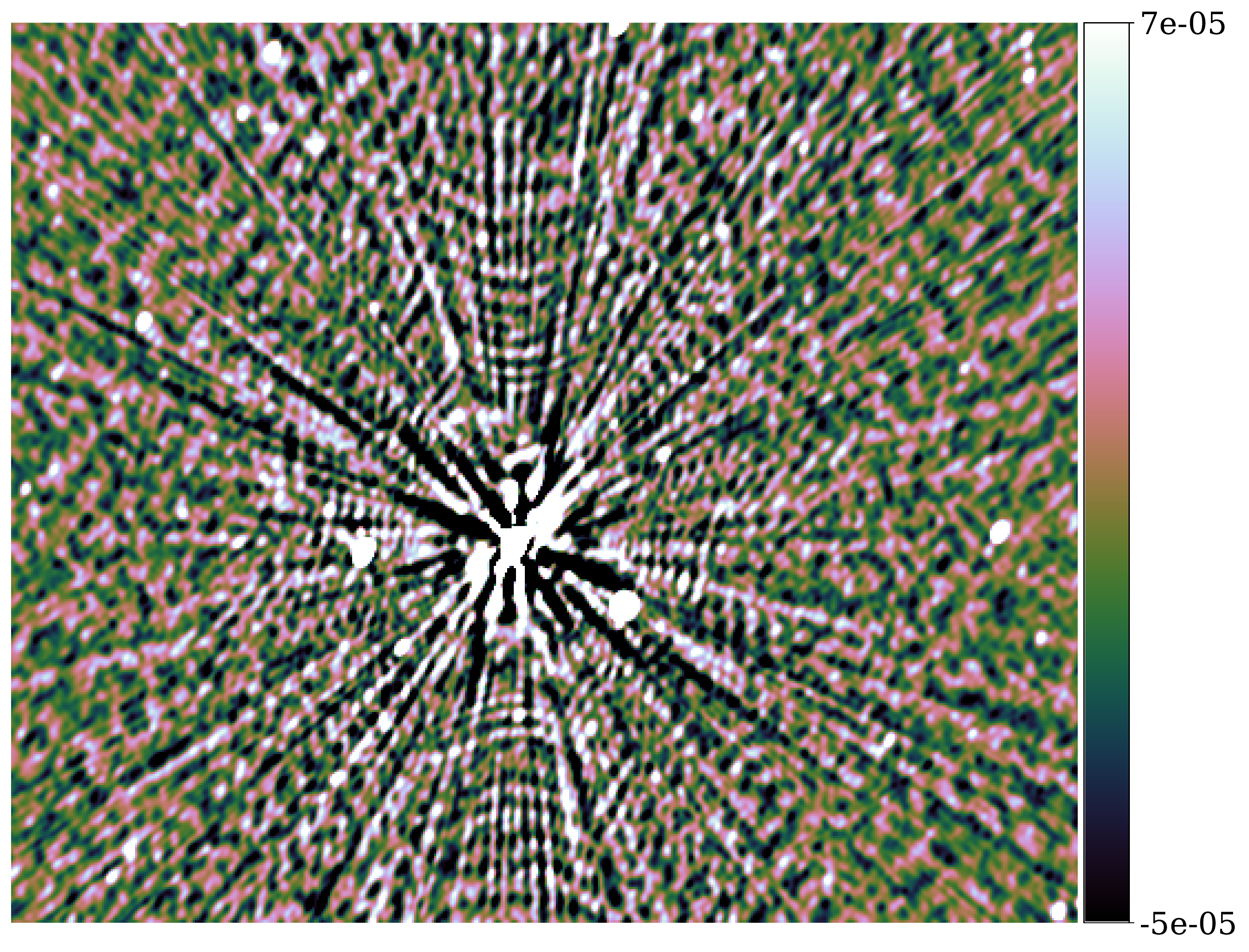} 
\caption{before peeling}\label{before_peeling}
 \end{subfigure}
\begin{subfigure}{.45\textwidth}
 \centering
  \includegraphics[width=0.8\linewidth]{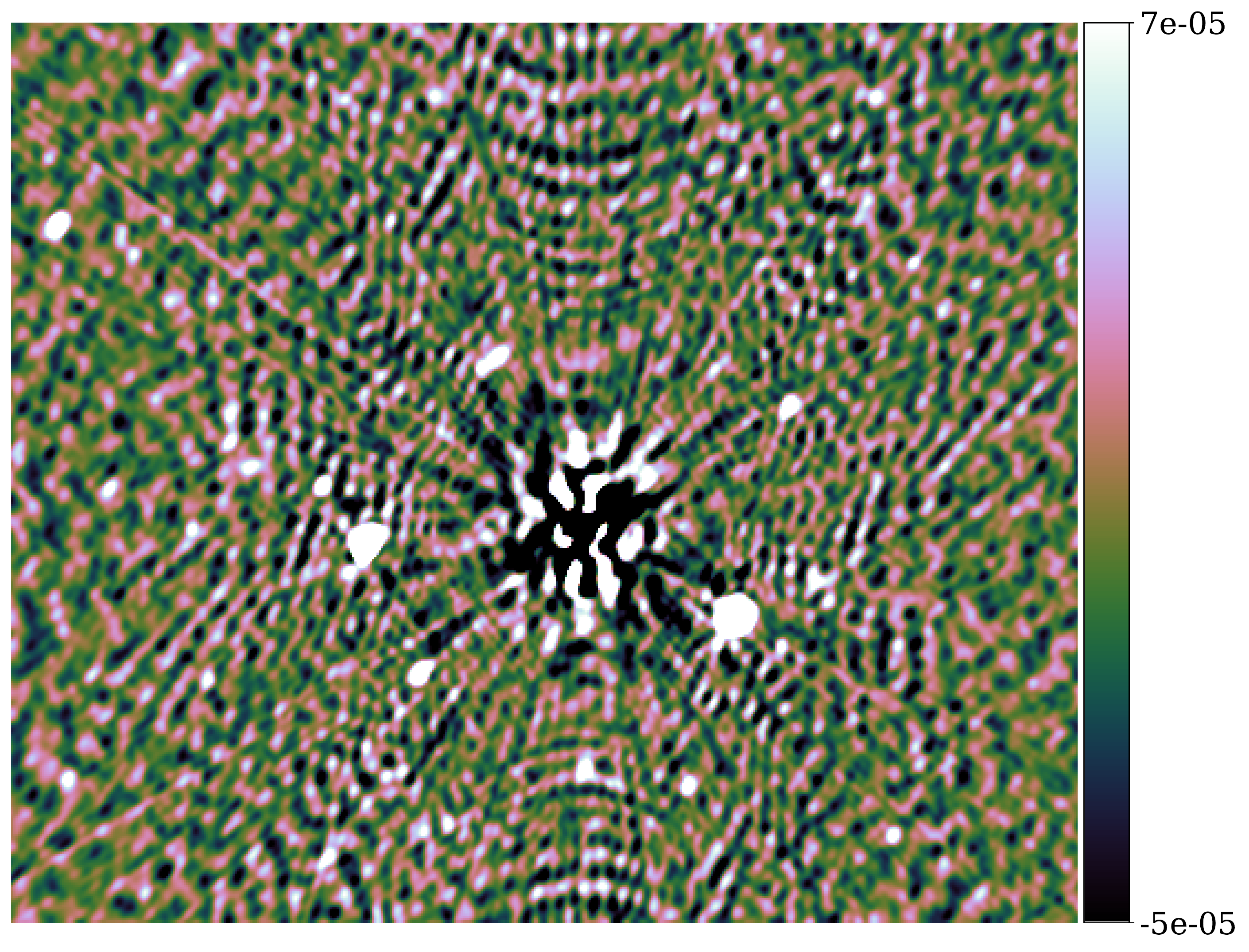} 
    \caption{$n_t$ = 1, $n_\nu$ = 1} \label{peel_fint1}
 \end{subfigure}
\begin{subfigure}{.45\textwidth}
 \centering
  \includegraphics[width=0.8\linewidth]{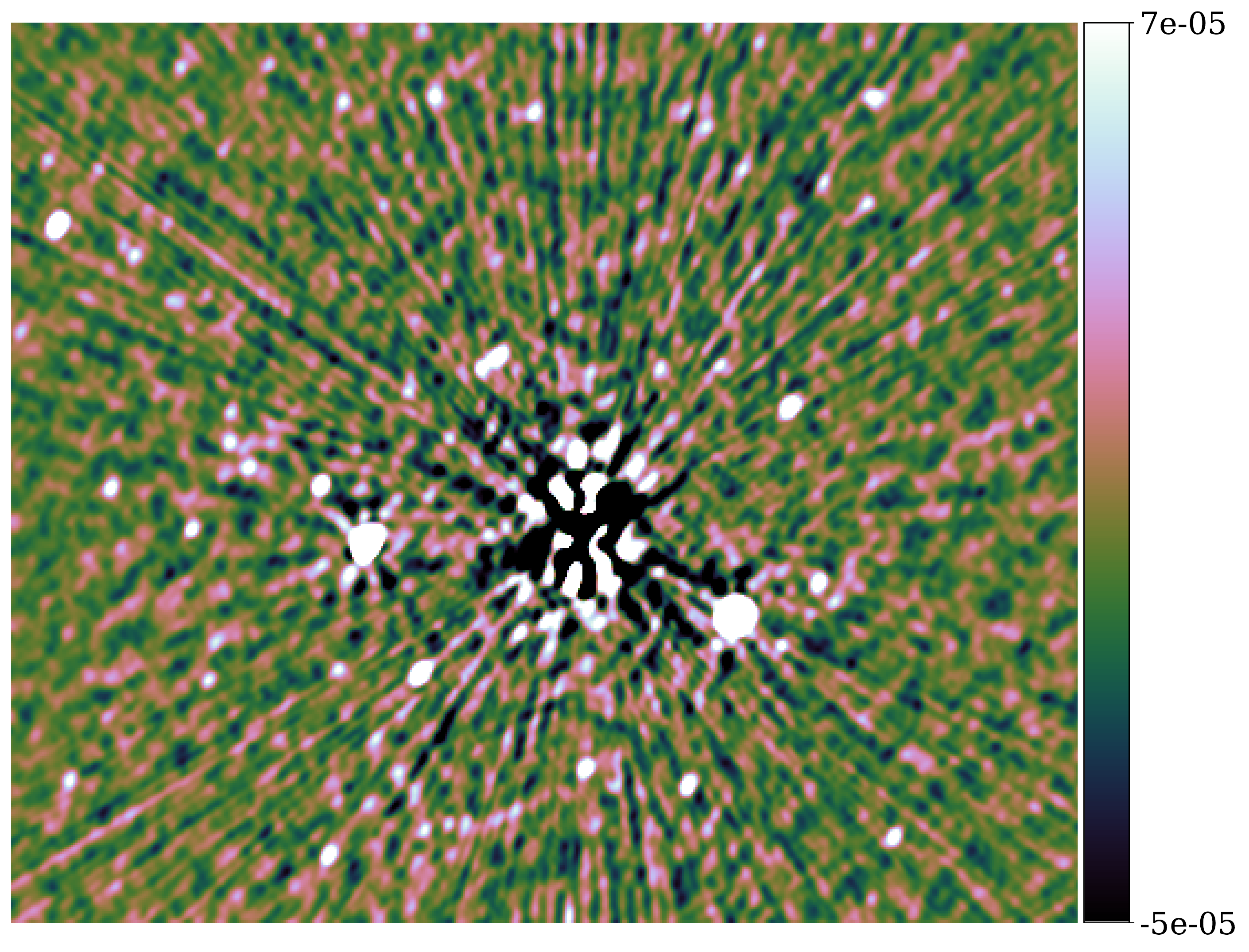} 
\caption{$n_t$ = 2, $n_\nu$ = 128}\label{peel_fint128}
 \end{subfigure}
\begin{subfigure}{.45\textwidth}
 \centering
 \includegraphics[width=0.8\linewidth]{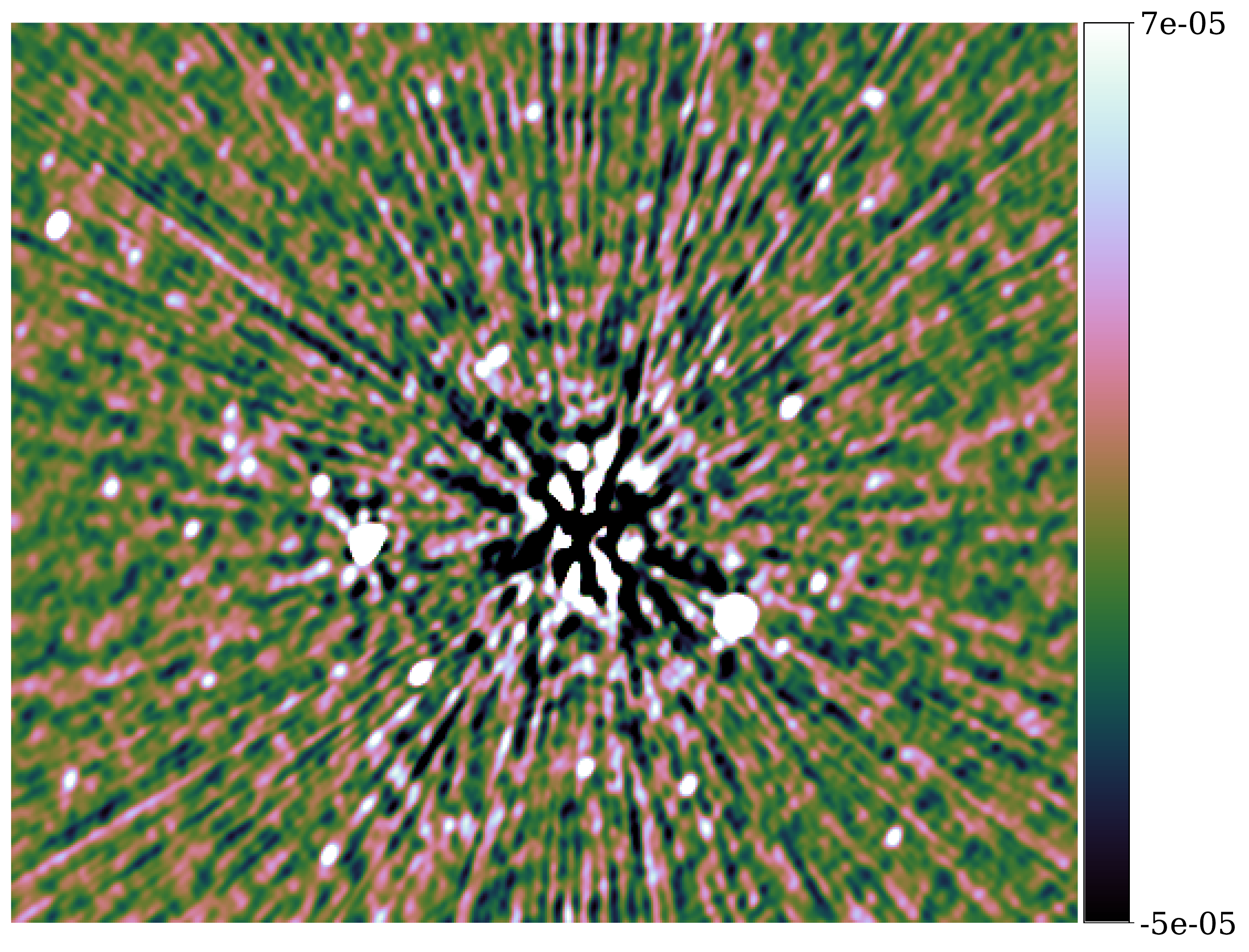}
\caption{$n_t$ = 64, $n_\nu$ = 1024} \label{peel_fint1024}
 \end{subfigure}
 \caption[4C12.02 images]{Cross-section images of the field centred around the troublesome source. (a) before peeling. (b), (c) and (d) are corrected images after peeling with solution intervals ($n_t$ = 1, $n_\nu$ = 1), ($n_t$ = 2, $n_\nu$ = 128) and ($n_t$ = 64, $n_\nu$ = 1024) respectively. 
}
\label{mk_real_images}
\end{figure*}

We elucidate the above point by performing  peeling using different solution intervals. In the first case, we use the shortest solution interval suggested by Fig. \ref{fig:nv_nt}, i.e. $N=1$, hence $n_t$ = 1 and $n_\nu$ = 1. Next, we use the minimum suggested interval required for all antennas, i.e. $N$ = 140. Thus, using our frequency chunking of 128 channels, we set the solutions intervals to $n_t$ = 2 and $n_\nu$ = 128. Lastly, we use the conservative approach of employing very long solution intervals and set the solution intervals to $n_t$ = 64 and $n_\nu$ = 1024. We show cutout images around the troublesome source before and after peeling with different solution intervals in Fig. \ref{mk_real_images}. Fig. \ref{before_peeling} shows the source before peeling. This is dominated by DDE-related artefacts. Fig. \ref{peel_fint1} show the image after peeling the source using  the shortest solution interval ($n_t$ = 1, $n_\nu$ = 1). Here, artefacts are still prominent. This is due to some antennas not having enough SNR on short solution intervals, and thus, are being flagged by the solver -- the visibilities corresponding to these antennas are then peeled without any correction. In Fig. \ref{peel_fint128}, when we use the minimum required solution interval for all the antennas, we obtain the best result. The troublesome source and its associated artefacts are mostly subtracted. In this case, we apply valid corrections to all subtracted visibilities, as no antennas are flagged on low SNR. Finally, when we employ very long solutions intervals, we are less able to track variations in the gains. Hence, in Fig. \ref{peel_fint1024}, we retain more artefacts.

\section{CONCLUSIONS}\label{conclusions}
We presented an analysis of the various mechanisms by which the choice of gain solution intervals impacts the  calibration quality (and therefore, ultimately, the science goals). Furthermore, we proposed a practical statistical approach for choosing adequate solution intervals during calibration, and demonstrated this on simulated and real data. Although most of the analysis focussed on DI calibration, we showed in $\S$\ref{meerkatdata} that the proposed approach can also be applied to DD calibration. 

In the real data application, when searching for the optimal solution interval, we used the robust solver in the CubiCal package in order to mitigate the low-level RFI still present in the data. As discussed in $\S$\ref{meerkatdata}, the presence of RFI and data flags makes the selection of solution intervals during calibration complicated. Generally, RFI occupancy (and the flagged data fraction) varies with time and frequency. As a result, parts of the data can be calibrated with shorter solution intervals, while other parts require longer intervals. In that sense, the truly optimal solution interval will actually vary across a dataset. This may be quite complex to implement in practice.

A factor we did not consider in our study is spectral variation. For a field dominated by steep-spectrum sources, the reduction in SNR at higher frequencies may warrant longer solution intervals at the top end of the band. Furthermore, for completeness, we need to consider polarimetric calibration in future works. This brings in the additional complication of the SNR being substantially different for the diagonal and off-diagonal visibility components.

In summary, solution intervals are a very convenient tool for regularising the calibration problem, because treating each interval independently makes the solution both simple to implement and   embarrassingly parallel. This alone probably explains their enduring popularity in current packages. They do, however, have significant limitations, which this work has highlighted. Firstly, because of intrinsic gain variability, the presence of RFI, unmodelled sources and inhomogeneously flagged data, it is not trivial to determine an optimal solution interval. Indeed, a single solution interval is not guaranteed to be optimal across the full domain of the problem. Secondly, the enormous data rates of modern radio telescopes necessitate more data parallelism while, simultaneously, faint science targets require increased SNR for successful calibration. This is clearly problematic if the only tool we have at our disposal for increasing SNR is making the solution intervals larger. 

Forunately, a number of alternative ways to approach the problem have already been signposted in the literature, for example non-linear filtering \citep{tasse2014nonlinear}, consensus and stochastic optimisation \citep{10.1093/mnras/stv596, yatawatta2019stochastic}, as well as approaches utilising Gaussian random field priors \citep{Arras_2019, Albert_2020}. This work documents some of the subtleties involved in the traditional solution intervals approach, and therefore argues for exploring these techniques further. We hope that future radio astronomers will consider solution intervals along the same vein that modern programmers consider the \emph{goto} statement: a brutal, inelegant and primitive relic.

\section*{Acknowledgements}
This research is supported by the South African Research Chairs Initiative of the Department of Science and Technology and National Research Foundation, the Royal Society and the Newton Fund (grant NA150184), and the South African Radio Astronomy Observatory (SARAO).

\section*{Data Availability}
The data underlying this article will be shared upon reasonable request to the corresponding author.
\bibliographystyle{mnras}
\bibliography{Bibliography} 
\appendix
\section{Gaussian Processes}
\label{sec:appendixD}
Gaussian processes (GP) are a family of stochastic processes defined such that every finite subset of random variables drawn from it follows a multivariate Normal distribution (see \citet{rasmussen2006gaussian} for more details). Mathematically, a GP is specified by two functions, namely  a \textit{mean} and a \textit{covariance} or \textit{kernel} function. For a real process, $f(x)$, on an input field, $x$, the mean function, $m(x)$, and the covariance function, $k(x,x')$, are given by
\begin{align}
m(x) & =  \mathop{\mathbb{E}}[f(x)],  \label{eq:meanfunc} \\
k(x,x') &=  \mathop{\mathbb{E}}[(f(x)-m(x))(f(x')-m(x')] .\label{eq:covfunc} 
\end{align}
The covariance function determines the nature of the process and encodes all prior information about the specific process. In order to sample a GP with a specific prior covariance, $k(x,x')$, and a mean function, $m(x')$, we first construct its Gram matrix, $K$, as
\begin{align}
 K_{ij} = k(x_i^*, x_j^*), \label{eq:Gram_matrix}  
\end{align}
where $x_i^*$ and $x_j^*$ are the $i^{th}$ and $j^{th}$ desired input field locations. Using the constructed Gram matrix, samples at input field points $x_*$ are simply drawn from the following Normal distribution,
 \begin{align}
f(x^*) \sim  \mathrm{N}\left(m^*, K\right), \label{eq:sampling}
\end{align}
where $m^* = m(x^*)$ is the mean function evaluated at the desired field points. Throughout this work, we used the following two covariance functions to sample gains for the various simulations.
\subsection{Squared Exponential Covariance function}
\label{chap3:ssec:SE_cov}
The Squared Exponential (SE) covariance function is defined as 
\begin{align}
k(\tau) &= k(x-x') =  \sigma_{f}^2 \exp\left(\frac{{(x-x')^2}}{2l^2}\right) = \; \sigma_{f}^2 \exp\left(\frac{{\tau^2}}{2l^2}\right). \label{eq:SE_eqn}
\end{align}
The two hyper-parameters $\sigma_f$ and $l$ define the standard deviation of the GP and its characteristic length scale (i.e. the input separation required for the GP to vary significantly) respectively. The SE covariance function is infinitely differentiable, and hence samples drawn from processes with SE kernels are extremely smooth. Its power spectrum exists and is given by,
\begin{align}
S(s) &= \sigma_{f}^2\left(2 \pi l^2\right)^{D/2}\exp(-2\pi^2l^2s^2),  \label{eq:SE_ps}
\end{align}
where $D$ represents the dimension of the input field vector.
\subsection{The Mat\'ern Class of Covariance functions}
\label{chap3:ssec:matern_cov}
The extreme smoothness of the SE covariance function makes it an unrealistic assumption for certain physical processes. The Mat\'ern  class of covariance functions is less smooth and is defined as follows \citep{rasmussen2006gaussian}
\begin{align}
k_{\mathrm{matern}} \label{eq:matern_eqn}(\tau) &= \sigma_{f}^2 \frac{2^{1-\mu}}{\Gamma(\mu)}\left(\frac{\sqrt{2\mu}\tau}{l}\right)^\mu K_\mu\left(\frac{\sqrt{2}\mu\tau}{l}\right),
\end{align}
where $K_\mu$ is the modified Bessel function, and $\Gamma$ is the Gamma function. $\sigma_{f}$ and $l$ play the same role as in the SE covariance function. The additional parameter $\mu$ controls the smoothness of the GP. A GP with Mat\'ern covariance is $\mu-1$ times differentiable. Extremely smooth processes are generated by using high values for $\mu$, while rough processes are generated using small values for $\mu$. Note that, as $\mu$ approaches $\infty$, the Mat\'ern covariance turns into the SE covariance. Its generalised power spectrum is given by
\begin{align}
S(s) =  \sigma_{f}^2 \frac{2^D\pi^{D/2}\Gamma(\mu+D/2)(2\mu)^2}{\Gamma(\mu)l^{2\mu}}\left(\frac{2\mu}{l^2} + 4\pi^2s^2\right)^{-(\mu+D/2)}.
\end{align}

\section{Boxcars parametrisation of the gains} 
\label{sec:appendixC}
Using a suitable choice of design matrix, Eq.~\eqref{sol_int_param} can be written as a linear matrix vector equation. For example, for each frequency, we may define
\begin{equation}
\underset{N_\nu \times M_\nu }{R_\nu} = \begin{bmatrix} 
 \mathrm{ones}(n_\nu, 1) & 0 &  \cdots \\
 0 & \mathrm{ones}(n_\nu, 1) &  \cdots \\
 \vdots & \vdots & \ddots
\end{bmatrix},
\end{equation}
where $\mathrm{ones}(n, m)$ represents an $n \times m$ matrix of ones. Next, we stack $n_t$ copies of $R_\nu$ on top of each other to define the design matrix for a single time and frequency interval i.e.
\begin{equation}
\underset{n_t N_\nu \times M_\nu}{R} = \begin{bmatrix}
R_\nu \\ R_\nu \\ \vdots
\end{bmatrix}.
\end{equation}
Finally, for the full design matrix $X$, we simply stack $M_t$ copies of $R$ into the ``diagonal'' as follows
\begin{equation}
\underset{N_t N_\nu \times M_t M_\nu}{X} = \begin{bmatrix}
R & 0 & \cdots \\
0 & R & \cdots \\
\vdots & \vdots & \ddots
\end{bmatrix}.
\end{equation} 
Thus, assuming we have a $M_t \times M_\nu$ matrix of parameters, $\Theta_p$ say, the gain can be written as
\begin{equation}
\g_p = X \theta_p,
\end{equation} 
where $\theta_p = \mbox{vec}(\Theta_p)$ is the vector obtained when stacking rows of $\Theta_p$ on top of each other.

\section{Supplementary simulations}\label{results}
\begin{figure}
	\centering
	\includegraphics[width=\linewidth]{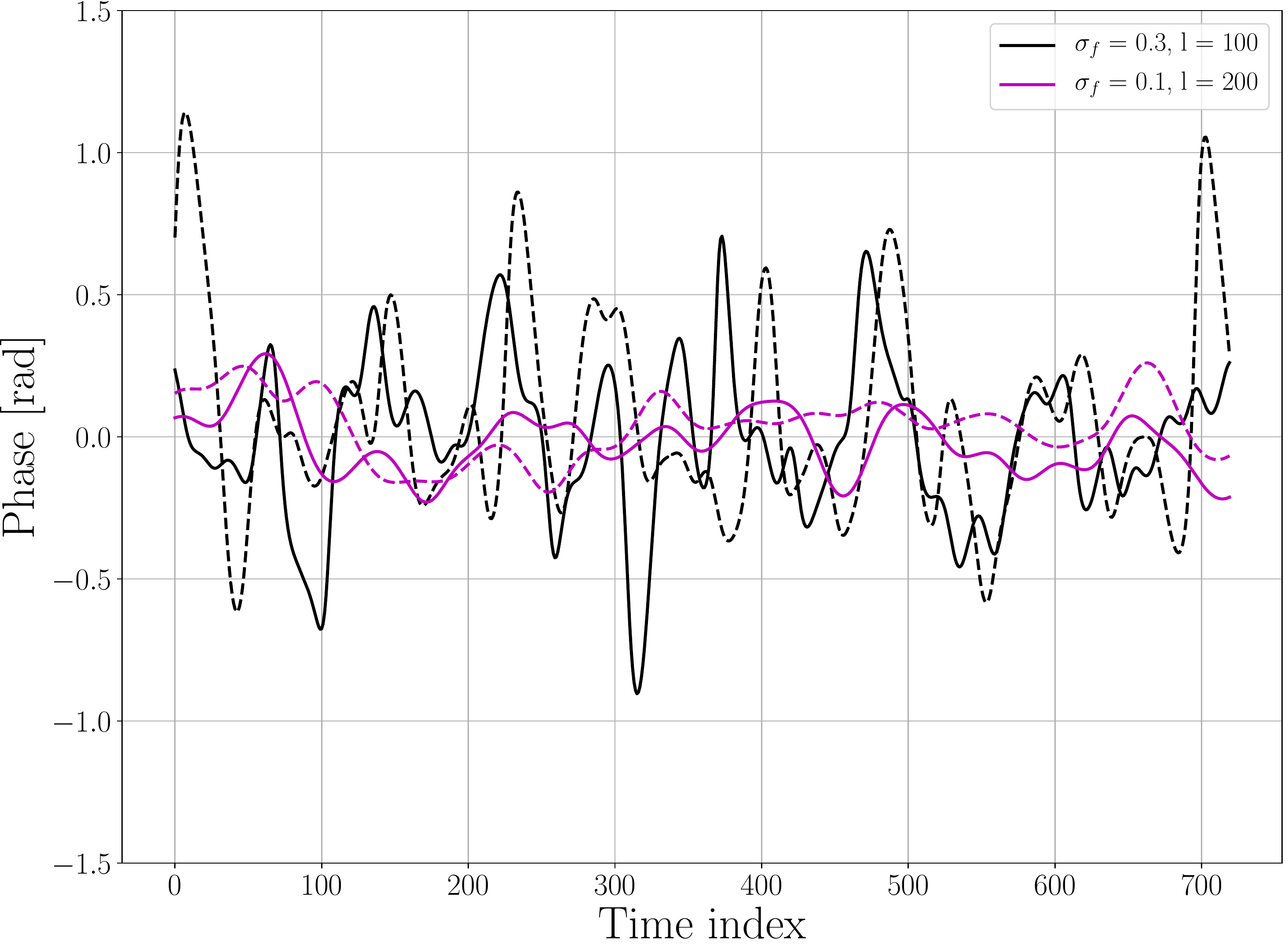}
    \caption[GP simulated gains]{Plots of a few gain realisations (phase against time index) with different GP parameters. All the plots are shown on the same scale to illustrate the differences in variability across the different experiments clearly. The dashed and solid line curves correspond to different gain realisations using the same parameters.}
    \label{fig:1d_input_gains}
\end{figure}

This appendix contains supplementary single channel simulations used to test how well the AIC tracks the MSE of the gains. For these simulations, we used a MeerKAT array layout with a bandwidth of 1 MHz, a start frequency of 0.9 GHz with 2 hours synthesis and 10 seconds integration.
\begin{table}
\centering
	\begin{tabular}{|l|c|c|r|} 
		\hline
		Simulation & $\sigma_f$ & $l$ & $\sigma_{\mathrm{rms}}$\\
		\hline
		a & 0.3 & 100 &0.2 Jy\\
        b & 0.1 & 200 &0.8 Jy\\
        c & 0.1 & 100 & 2 Jy\\
        d & 0.3 & 100 & 0.6 Jy\\
		\hline
	\end{tabular}
    \caption[Gaussian Process parameters and input rms (simulation \textbf{v})]{GP parameters for the gains and the noise rms used in the different simulations.}
\label{tab:1d_GP_parameters}
\end{table}
We used the same sky model (100 sources with random positions and fluxes drawn from a power law), but we change the gains and the noise rms across the simulations in order to vary from low to high SNR regimes as well as from slowly to rapidly varying gains. 
Because of the entanglement between the gain variability and the SNR of the data, it is difficult to cleanly separate the presentation here into low vs high SNR and slowly vs rapidly varying gains, but we will mention the specific regimes when describing the results. Table \ref{tab:1d_GP_parameters} shows the parameters of the squared exponential covariance function used for the gain realisations and the rms of the noise added to the visibilities ($\sigma_{\mathrm{rms}}$). The phases of the input gains for a few antennas are shown in Fig. \ref{fig:1d_input_gains}. Here, we made the amplitudes and phases to have similar variations, but in practice the phases vary on shorter time scales compared to the amplitudes.
\begin{figure*}
\begin{subfigure}{.45\textwidth}
  \includegraphics[width=\linewidth]{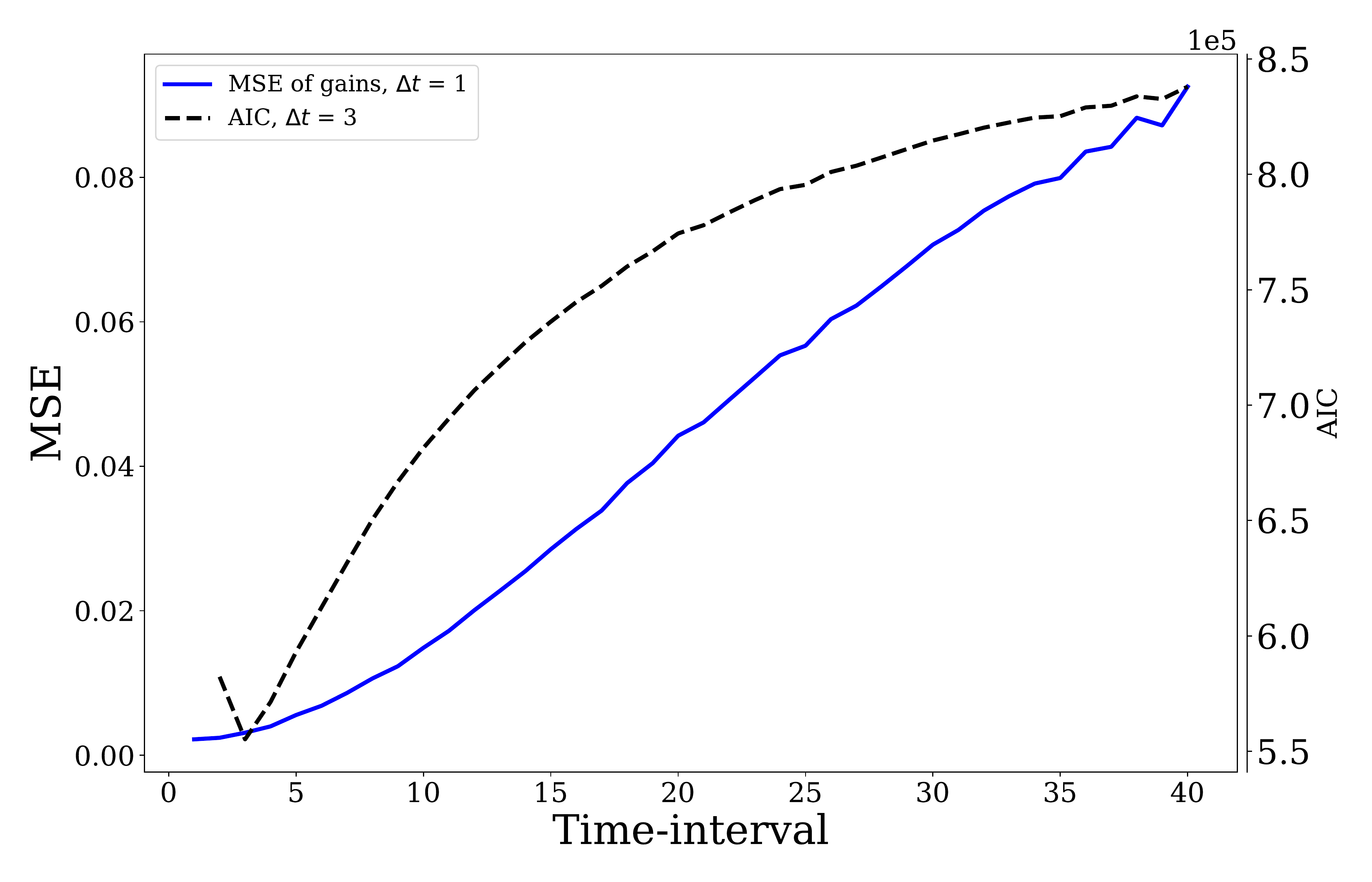} 
 \caption{$\sigma_f \,=\, 0.3, l \,= \,100, \,\sigma_{\mathrm{rms}} \,=\, 0.2$ Jy}\label{fig:1dchi1}
 \end{subfigure}
\begin{subfigure}{.45\textwidth}
  \includegraphics[width=\linewidth]{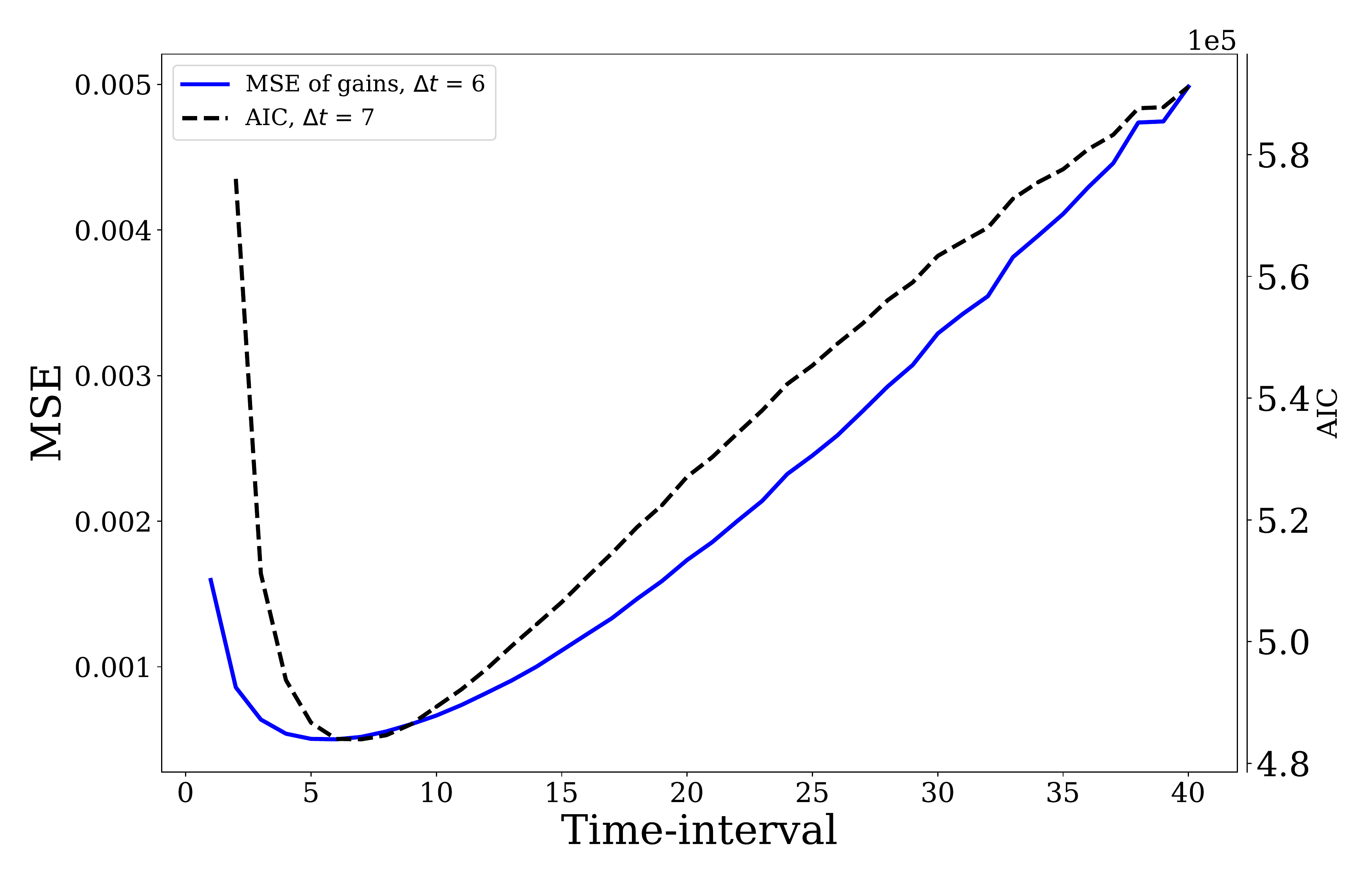} 
     \caption{$\sigma_f \,=\, 0.1, l \,= \,200, \,\sigma_{\mathrm{rms}} \,=\, 0.8$ Jy}\label{fig:1dchi2}
 \end{subfigure}
\begin{subfigure}{.45\textwidth}
  \includegraphics[width=\linewidth]{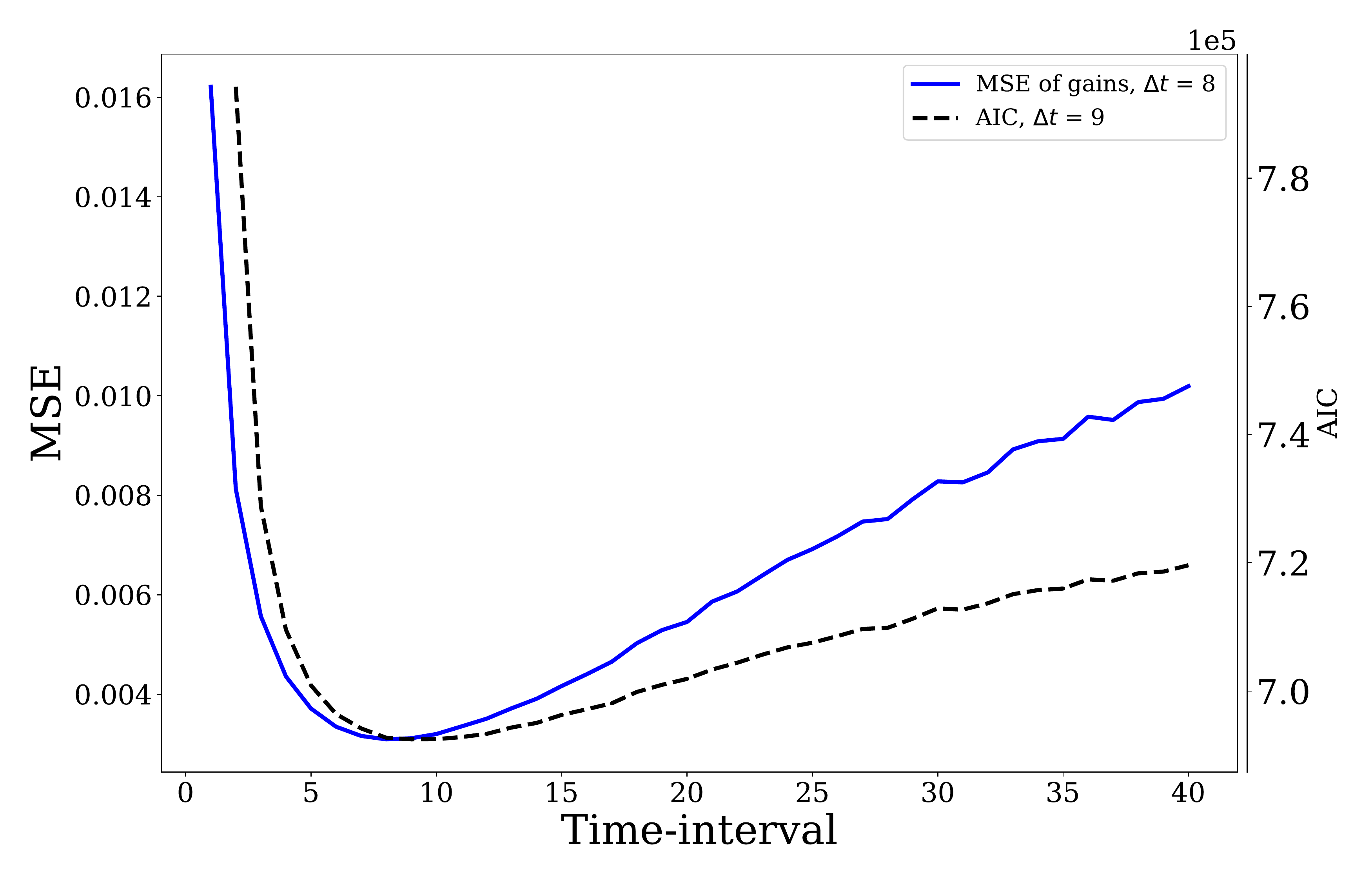} 
\caption{$\sigma_f \,=\, 0.1, l \,= \,100, \,\sigma_{\mathrm{rms}} \,=\, 2$ Jy}\label{fig:1dchi3}
 \end{subfigure}
\begin{subfigure}{.45\textwidth}
 \includegraphics[width=\linewidth]{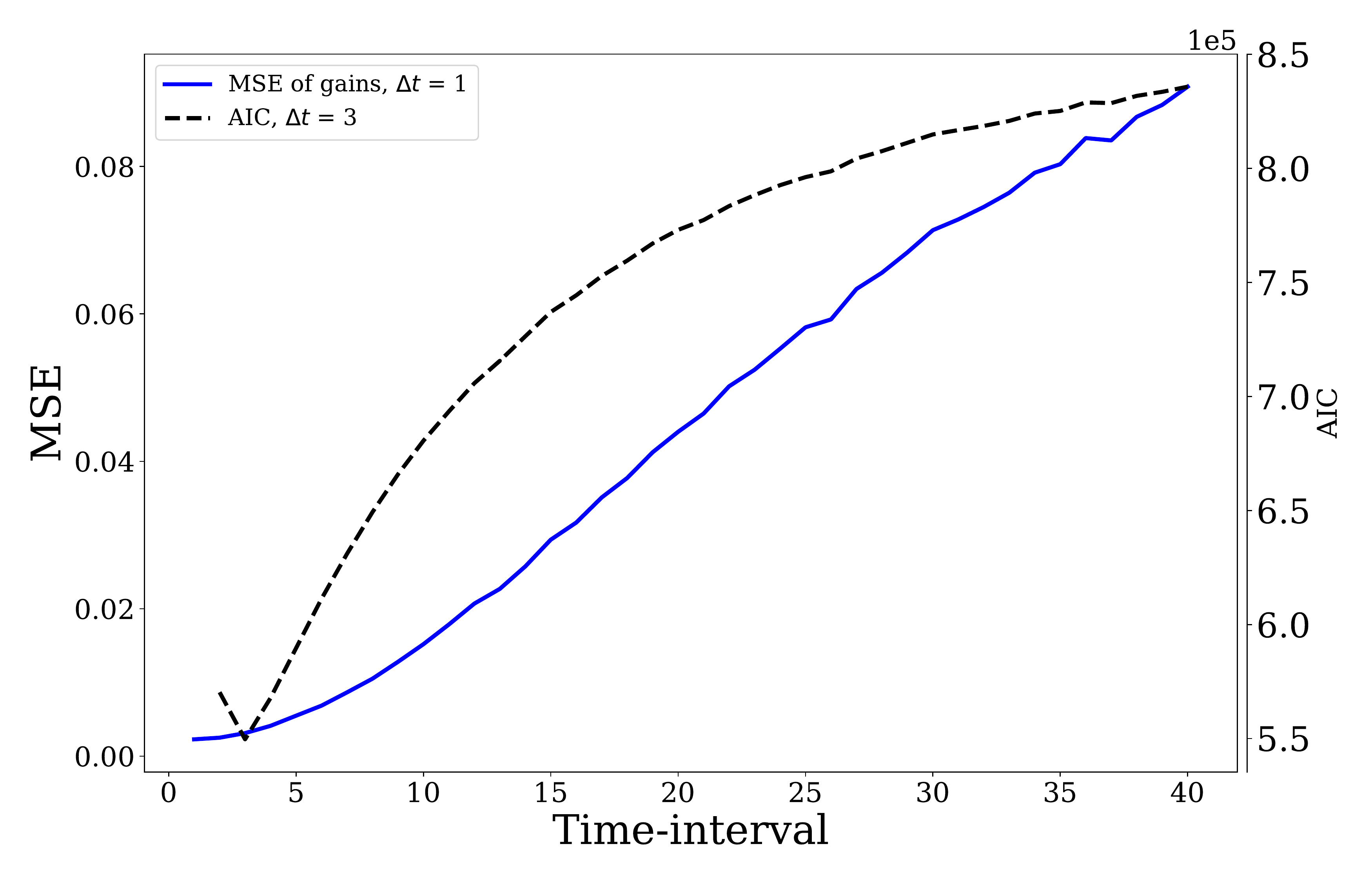}
 \caption{$\sigma_f \,=\, 0.3, l \,= \,100, \,\sigma_{\mathrm{rms}} \,=\, 0.6$ Jy}\label{fig:1dchi4}
 \end{subfigure}
 \caption[AIC plots for time only simulations]{Plots of the MSE of the estimated gains at different time intervals (blue) and the AIC of the reconstructed boxcar gains at different intervals based on the gains (black dashed lines).}
\label{fig:1d_chi_red}
\end{figure*}
Fig. \ref{fig:1d_chi_red} shows plots of the MSE of the estimated gains and the AIC of the boxcar reconstructed gains at different time intervals based on the solution at time interval 1. These plots depict the following:
\begin{enumerate}
\itemsep1em
\item{Fig. \ref{fig:1dchi1} is a high SNR simulation with rapidly varying gains. Here, the optimal interval is short, and this is well predicted by the AIC.}
\item{Fig. \ref{fig:1dchi2} is a medium SNR regime with rapidly, but not highly varying gains. A longer time interval can be used, and this is correctly predicted.}
\item{Fig. \ref{fig:1dchi3} is a case with a low SNR and slowly varying gains. Because of the low SNR, we need a long time interval in this scenario. The stability of the gains also favours this. The AIC accurately predicts the optimal interval.}
\item{Fig. \ref{fig:1dchi4} is a case of extremely high variability (this is shown in Fig. \ref{fig:1d_input_gains}). Here, $\sigma_f \, =\, 0.3$ suggests high variability in the gains and $l = 100$ (10 units since the integration time in our data is 10 seconds) implies variability at a relatively short time scale. The AIC prediction is close to the minimal MSE of the gains.}
\end{enumerate}
\bsp	
\label{lastpage}
\end{document}